\documentclass[12pt]{article}
\usepackage[margin=1in]{geometry}
\usepackage{amsmath}
\usepackage{amssymb}
\usepackage{bm}
\usepackage{mathtools} 
\usepackage{graphicx}
\usepackage{caption}
\usepackage{subcaption}
\usepackage{multirow}
\usepackage{algpseudocode}
\usepackage{enumitem}
\usepackage{booktabs}
\usepackage{makecell}
\usepackage{algorithmicx}
\usepackage{algorithm}
\usepackage{float}
\usepackage[normalem]{ulem}
\usepackage{setspace}
\usepackage[round]{natbib}
\usepackage{xr}
\usepackage{hyperref}
\begin{document}
 
\title{\textbf{Age-Specific Logistic Regression with Complex Event Time Data}}
 

\medskip


 \author{
\textbf{Haoxuan (Charlie) Zhou},
\textbf{X. Joan Hu},
\textbf{Yi Xiong}\textbf{, and}
\textbf{Yan Yuan}
}
\maketitle

 

\medskip

\begin{abstract}

In attempt to advance the current practice for assessing and predicting the primary ovarian insufficiency (POI) risk in female childhood cancer survivors, we propose two estimating function based approaches for age-specific logistic regression. Both approaches adapt the inverse probability of censoring weighting (IPCW) strategy and yield consistent estimators with asymptotic normality. The first approach modifies the IPCW weights used by \cite{im2023development} to account for doubly censoring. The second approach extends the outcome weighted IPCW approach 
to use the information of the subjects censored before the analysis time. We consider variance estimation for the estimators and explore by simulation the two approaches implemented in the situations where the conditional right-censoring time distribution required in the IPCW weighs is unknown and approximated using the survival random forest approaches, stratified empirical distribution functions, or the estimator under the Cox proportional hazards model. The numerical studies indicate that the second approach is more efficient when right-censoring is relatively heavy, whereas the first approach is preferable when the right-censoring is light. We also observe that the performance of the two approaches heavily relies on the estimation of censoring distribution in our simulation settings. The POI data from a childhood cancer survivor study are employed throughout the paper for motivation and illustration. Our data analysis provides new insight into understanding the POI risk among cancer survivors.
\end{abstract}

\noindent\textit{keywords:} Doubly Censored Event Time, Estimating Function, Estimation of Censoring Time Distribution, Inverse Probability of Censoring Weight(IPCW)

\section{Introduction}\label{sec1: Introduction}

Assessing the risk of developing a disease by a specific age with a patient's risk profile has become increasingly important in clinical decision-making \citep{chemaitilly2017premature,mostoufi2016endocrine,eshre2016eshre,touraine2024premature}. The age-specific and patient-specific risk prediction is crucial, particularly for the vulnerable population of childhood cancer survivors, as cancer treatment can subsequently put them at risk of adverse health outcomes. For example, ovarian tissue is vulnerable to gonadotoxic cancer treatments such as pelvic radiotherapy and alkylating agents, which can accelerate the age-related decline of the ovarian follicle and subsequently result in premature ovarian insufficiency (POI)\citep{chemaitilly2017premature,johnston2009normal,im2023development}. Developing a risk assessment tool for this population is challenging because it requires collecting data from a long-term follow-up study. This paper aims to model POI among female childhood cancer survivors. We use data from the Childhood Cancer Survivor Study (CCSS)\citep{robison2002study,robison2009childhood}, an ongoing multi-institutional North American cohort study of 5-year survivors of childhood cancer treated between 1970 and 1999. The large sample size and long follow-up of the study provide a unique opportunity to empirically assess disease risk among cancer survivors.

Despite its strength, observational studies like CCSS often pose challenges in statistical analysis due to censored outcomes. Survival prediction models accommodating censoring have been extensively studied, spanning from traditional approaches that posit semi-parametric assumptions (e.g., Cox proportional hazards (PH) model\citep{cox1972regression}) to machine learning algorithms (e.g., random survival forest\citep{ishwaran2008random}) to model the relationship of time-to-event outcomes and covariates. The approaches described thus far either treat the time-to-event outcome as a continuous variable or model the discrete-time hazard of the event. Although the latter approach gains flexibility in modelling complex covariate relationships and outperforms the traditional Cox PH model in the presence of high-dimensional covariates, it always requires to transform and split the continuous follow-up time into pre-specified time intervals\citep{suresh2022survival}. The specification of time intervals can yield post-selection bias and thereby substantially influence the predictive performance of the model.

In this paper, we consider quantifying the risk of POI as the cumulative incidence probability, which is the probability of experiencing POI by a specific age. We directly model it with a logistic regression model to circumvent specifying discrete time intervals for the follow-up time. The key issue in using the logistic regression model to predict cumulative incidence probability is to accommodate censored outcomes. An increasingly common technique is to apply the inverse-probability-of-censoring weighting (IPCW) to handle censored outcomes. The IPCW approach was originally proposed to correct for censoring, particularly dependent censoring\citep{robins1992recovery}, by creating a pseudo population with more weight placed on subjects who are not censored. There are two main streams in adjusting estimators by applying the IPCW with logistic regression models. One way is to adjust the estimating equation with the IPCW\citep{zheng2006application,uno2007evaluating,yuan2018threshold,vock2016adapting,im2023development} and the alternative way is to create a weighted response variable\citep{scheike2008predicting,blanche2023logistic}. Detailed discussions on these two IPCW-based methods can be found in \cite{blanche2023logistic}. It is well acknowledged that the validity of IPCW-based methods relies on correct specification about the distribution of censoring times and it is essential to examine the efficiency loss with different ways to estimate the distribution of censoring.

One challenge arising from the CCSS data is that the target population is the cancer survivors. If a prediction model is built upon defining the time origin as the diagnosis age of cancer, it lacks interpretability of risk in a given age. Therefore, we follow the prior work in \cite{im2023development} to use age as the time scale. Since POI is developed after cancer diagnosis, the prediction model needs to adjust risk sets by correctly including those who have been diagnosed with cancer. This requires inference from doubly censored event time, which is defined as the event time can only be accurately measured within a certain range \citep{cai2004semiparametric}.

In this paper, we follow \cite{betensky2015recognizing} to incorporate an additional risk set indicator in the estimating functions with an age-specific logistic regression model, and extend the existing two IPCW methods to account for doubly-censored outcomes. We provide inference procedures based on robust variance estimators with two IPCW-based methods. Furthermore, we present simulation studies and theoretical justifications that explore the performance of the proposed approaches with various estimated censoring distributions. The paper is organized as follows. Section \ref{sec2: estimation procedure} describes the model and the proposed estimation procedures. In Section \ref{sec3: real data analysis}, the proposed approaches are applied to the CCSS data that motivated this research. Section \ref{sec4: simulation} reports the simulation studies which is conducted to evaluate our findings. We conclude in Section \ref{sec5: disscusion} with final remarks.
\section{Age-Specific Logistic Regression Analysis}\label{sec2: estimation procedure}

\subsection{Notation and Modeling}
Consider a study on the risk of a particular event. Let $T$ be the age at the event, say, experiencing the aforementioned POI. Suppose the event of interest only takes place after an initial event, say, the cancer diagnosis, which occurs at age $V$. We assume the event age $T$ can be modeled by the following age-specific logistic regression model, at a prespecified age $t_0$,
\begin{align}\label{model: considered model} 
    \log\left( \frac{P(T \le t_{0}|\mathbf{Z}, T>V, V<t_{0}) }{1-P(T \le t_{0}|\mathbf{Z}, T>V, V<t_{0} )} \right)= \alpha(t_{0})+\bm{\beta}(t_{0})^{T}\mathbf{Z} 
 \end{align} 
where $\alpha(t_{0})$ represents the intercept and $\bm{\beta}(t_{0})$ is a $p\times 1$ vector of regression coefficients at $t_{0}$. This model differs from the original model considered by \cite{im2023development}, 
 \begin{align}\label{model: previous model} 
    \log\left( \frac{P(T < t_{0}|\mathbf{Z}, T>V) }{1-P(T < t_{0}|\mathbf{Z}, T>V )} \right)= \alpha(t_{0})+\bm{\beta}(t_{0})^{T}\mathbf{Z},
 \end{align} 
unless $T$ is continuous and $V<t_0$. Motivated by the CCSS study, we aim to estimate the model parameters in (\ref{model: considered model}) using the data described below. 

Suppose the study subjects are independent with the ages at the initial event and the event of interest
as $V_i$ and $T_i$ for $i=1,\ldots,n$, respectively. The event is subject to doubly censoring, with the left censoring at $V_i$ and the right censoring at $C_i$, where all $V_i$ and $C_i$ are available. For subject $i$, denote the observed age at study exit by $U_i = \min\{T_i,C_i\}$ with the indicator $\delta_i = \mathbb{I}(T_i \le C_i)$,
and the $p\times 1$ covariate vector by $\mathbf{Z_i}$. We assume that the event time and the right censoring times are independent conditional on the covariates. The available data are $O = \{O_{i}, \text{ } 1\le i\le n\}$ with $O_{i} =(U_{i},\delta_{i},V_{i},\mathbf{Z}_{i})$ for $i=1,\ldots,n$.

The model (\ref{model: considered model}) is not conventional in survival analysis. However, it enjoys the natural parameter interpretation of logistic regression. Repeating the logistic regression analysis for different $t_{0}$, one can estimate the effect size on the log-odds scale of the event risk over time. It can provide a natural dynamic risk prediction.\\

\subsection{Proposed Estimation Procedures}
When all the event ages $T_i$ are available at $t_0>V_i$, one may estimate the model parameters $\alpha(t_{0})$ and $\bm{\beta}(t_{0})$ using the standard iterative algorithms; see the book by \cite{mccullagh2019generalized}. When $T_{i}$ is subject to censoring, the IPCW strategy has been widely employed \citep{robins2006inverse}. We assume that the support of the censoring time's conditional survival function $G(c|\mathbf{Z})=P(C\ge c|\mathbf{Z})$ contains $[0, t_0]$. Following the IPCW approach, we propose two estimating function-based procedures.

\subsubsection{Approach A: IPCW with generalized linear model.}
The approach by Im et al in \cite{im2023development} can be written as 
 \begin{align}\label{eq: previous estimating equation}
 \begin{split}
\bm{U}(\alpha,\bm{\beta};t_{0}|G) =
        \displaystyle\sum_{i=1}^{n}
        \begin{bmatrix}
           1 \\
           \mathbf{Z}_{i} 
         \end{bmatrix}
        W^*_{i}(t_{0};G)
        \left[
        \mathbb{I}(T_{i} < t_{0}) -
        \frac{\exp\{\alpha(t_{0})+\bm{\beta}(t_{0})^{T}\mathbf{Z}_{i} \}}{\exp\{\alpha(t_{0})+\bm{\beta}(t_{0})^{T}\mathbf{Z}_{i} \}+1}
        \right],
\end{split}
\end{align}
 where the IPCW is
 \[
 W^*_{i}(t_{0};G)
 =\frac{\mathbb{I}(U_i < t_{0})\delta_i}{G(U_i|\mathbf{Z}_i)} + \frac{\mathbb{I}(U_i \ge t_{0})}{G(t_{0}|\mathbf{Z}_i)},
\]
for $i=1,\ldots, n$.
Adapting the estimating function in (\ref{eq: previous estimating equation}) to account for left censoring, approach A is given by
 \begin{align}\label{eq: estimating equation a}
\begin{split}
        \bm{U}_{A}(\alpha,\bm{\beta};t_{0}|G) =\displaystyle\sum_{i=1}^{n} \mathbb{I}(t_{0} \ge V_{i})
        \begin{bmatrix}
           1 \\
           \mathbf{Z}_{i}  
         \end{bmatrix}
        W_{i}(t_{0};G)
        \left[
        \mathbb{I}(T_{i} \le t_{0}) 
        -\frac{\exp\{\alpha(t_{0})+\bm{\beta}(t_{0})^{T}\mathbf{Z}_{i} \}}{\exp\{\alpha(t_{0})+\bm{\beta}(t_{0})^{T}\mathbf{Z}_{i} \}+1}
        \right],
\end{split}
\end{align}
where the IPCW is 
\begin{align}\label{eq: IPCW}
\begin{split}
    W_i(t_{0};G)=
    \frac{\mathbb{I}(U_i \le t_{0})\delta_i}{G(U_i|\mathbf{Z}_i)}+
    \frac{\mathbb{I}(U_i > t_{0})}{G(t_{0}|\mathbf{Z}_i)}.
\end{split}
\end{align}
When the selected analysis time $t_0$ is after all the initial events of the study subjects, the available data reduce to right censored event times and the proposed estimating function is then
\[
        \displaystyle\sum_{i=1}^{n}
        \begin{bmatrix}
           1 \\
           \mathbf{Z}_{i} 
         \end{bmatrix}
        W_{i}(t_{0};G)
        \left[
        \mathbb{I}(T_{i} \le t_{0})
         -\frac{\exp\{\alpha(t_{0})+\bm{\beta}(t_{0})^{T}\mathbf{Z}_{i} \}}{\exp\{\alpha(t_{0})+\bm{\beta}(t_{0})^{T}\mathbf{Z}_{i} \}+1}
        \right],
\]
which has been considered in the literature \citep{zheng2006application,blanche2023logistic}. We note that the component $\mathbb{I}(t_{0} \ge V_{i})$ in (\ref{eq: estimating equation a}) is necessary if $t_0$ is before age 21 in the aforementioned CCSS study, to account for the left censoring in the data.

Under some regularity conditions, it is straightforward to establish the consistency and asymptotic normality of the estimator with the estimation function in (\ref{eq: estimating equation a}) at a fixed $t_0>0$, provided that the event time and censoring time are independent conditional on $\mathbf{Z}$.

\subsubsection{Approach B: Outcome weighted IPCW}
As guarded by the IPCW in (\ref{eq: IPCW}), approach A excludes all the subjects censored before $t_{0}$. Appropriately using the available information of those excluded subjects is likely to yield more efficient estimation, especially at $t_0$ where the censoring is heavy. This consideration leads to approach B:
\begin{align}\label{eq: estimating equation b}
\begin{split}
        \bm{U}_{B}(\alpha,\bm{\beta};t_{0}|G) 
        &=\displaystyle\sum_{i=1}^{n} \mathbb{I}(t_{0} \ge V_{i})
        \begin{bmatrix}
           1 \\
           \mathbf{Z}_{i} 
         \end{bmatrix}
        \left[
       \mathbb{I}(T_{i} \le t_{0}) W_{i}(t_{0};G)
         -\frac{\exp\{\alpha(t_{0})+\bm{\beta}(t_{0})^{T}\mathbf{Z}_{i}\}}{\exp\{\alpha(t_{0})+\bm{\beta}(t_{0})^{T}\mathbf{Z}_{i}\}+1}
        \right].
\end{split}
\end{align}
The estimating function \ref{eq: estimating equation b} is an extension of the outcome weighted IPCW approach presented in \cite{scheike2008predicting} to account for left censoring. 

One can show that the estimating function $\bm{U}_{B}(\alpha,\bm{\beta};t_{0}|G)$ is unbiased if the event time and censoring time are independent conditional on $\mathbf{Z}$. It is also straightforward to establish, under some regularity conditions, the consistency and asymptotic normality of the estimator derived from the estimation function $\bm{U}_{B}(\alpha,\bm{\beta};t_{0}|G)$ in (\ref{eq: estimating equation b}) at a fixed $t_0>0$.

\subsubsection{Comparison of Approaches A and B in asymptotic efficiency}
We now compare the asymptotic efficiency of the two proposed approaches at a fixed time $t_0$. Let $\bm{\theta} = (\alpha(t_{0}),\bm{\beta}(t_{0})^{T})^{T}$, $\hat{\bm{\theta}}_A$ and $\hat{\bm{\theta}}_B$ denote the estimators of approaches A and B, which are the solutions to $\bm{U}_{A}(\bm{\theta};t_{0}|G)=0$ and  $\bm{U}_{B}(\bm{\theta};t_{0}|G)=0$, respectively. Viewing $\bm{U}_{A}(\bm{\theta};t_{0}|G)=\sum_{i=1}^n \mathbf{A}_i$ with i.i.d terms $\mathbf{A}_i$ with mean zero and $\bm{U}_{B}(\bm{\theta};t_{0}|G)=\sum_{i=1}^n \mathbf{B}_i$ with i.i.d. terms $\mathbf{B}_i$ with mean zero, under the conventional regularity conditions, we can show that
$\sqrt{n}(\hat{\bm{\theta}}_{A} -\bm{\theta}) \xrightarrow[]{d} N(0, AV_{A}(\bm{\theta};G))$ and $\sqrt{n}(\hat{\bm{\theta}}_{B} -\bm{\theta}) \xrightarrow[]{d} N(0, AV_{B}(\bm{\theta};G))$ as $n\rightarrow\infty$, provided that
the number of study subjects with the observed event $n^{*} = \displaystyle\sum_{i=1}^{n}I(V_{i}<t_{0})\delta_{i} \rightarrow\infty$ as $n\rightarrow\infty$. 

The difference of the asymptotic variances of the two estimators is then
\begin{align}\label{eq: diff asy var}
\begin{split}
    AV_{A}(\bm{\theta};G) - AV_{B}(\bm{\theta};G) 
    &= \Gamma_{A}^{-1}(\bm{\theta};G)\Sigma_{A}(\bm{\theta};G)\left(\Gamma_{A}^{-1}(\bm{\theta};G)\right)^{T} -
    \Gamma_{B}^{-1}(\bm{\theta};G)\Sigma_{B}(\bm{\theta};G)\left(\Gamma_{B}^{-1}(\bm{\theta};G)\right)^{T}\\
    &= \Gamma_{A}^{-1}(\bm{\theta};G) \left[\Sigma_{B}(\bm{\theta};G) - \Sigma_{A}(\bm{\theta};G) \right] \left(\Gamma_{A}^{-1}(\bm{\theta};G)\right)^{T},   
\end{split}
\end{align}
where $\Gamma_A(\bm{\theta};G)$ and $\Gamma_{B}(\bm{\theta};G)$ are the limits of $-\frac{1}{n}\frac{\partial \bm{U}_{A}(\bm{\theta}|G)}{\partial \bm{\theta}^{T}}$ and $-\frac{1}{n}\frac{\partial \bm{U}_{B}(\bm{\theta}|G)}{\partial \bm{\theta}^{T}}$, and $\Sigma_{A}(\bm{\theta};G)$ and $\Sigma_{B}(\bm{\theta};G)$ are the asymptotic variances of
$\frac{1}{\sqrt{n}}\bm{U}_{A}(\bm{\theta}|G)$ and $\frac{1}{\sqrt{n}}\bm{U}_{B}(\bm{\theta}|G)$. The last equation in (\ref{eq: diff asy var}) is due to $\Gamma_{A}(\bm{\theta};G) = \Gamma_{B}(\bm{\theta};G)$ since $E(W_{i}(t_{0};G)|T_{i},\mathbf{Z}_{i}  ) = 1$. The detailed derivation is provided in Supplementary Material \ref{supp: derv_diff_asy_var}. 

One may estimate $\Gamma_A(\bm{\theta};G)$ and $\Gamma_{B}(\bm{\theta};G)$ by $-\frac{1}{n}\frac{\partial \bm{U}_{A}(\bm{\theta}|G)}{\partial \bm{\theta}^{T}}$ and $-\frac{1}{n}\frac{\partial \bm{U}_{B}(\bm{\theta}|G)}{\partial \bm{\theta}^{T}}$, and $\frac{1}{n}\sum_{i=1}^n \mathbf{A}_{i}\mathbf{A}_{i}^{T}$ and $\frac{1}{n}\sum_{i=1}^n \mathbf{B}_{i}\mathbf{B}_{i}^{T}$ may be used to
estimate $\Sigma_{A}(\bm{\theta};G)$ and $\Sigma_{B}(\bm{\theta};G)$, respectively. The estimated  difference of the two asymptotic variances can provide efficiency comparison of the two approaches.

\subsection{Practical Operation of Approaches A and B}

The conditional distribution of the censoring time is usually unknown in practice. To implement the proposed approaches, we need to estimate $G(\cdot|\mathbf{Z})$ in (\ref{eq: IPCW}) using the available data. 

Let $\tilde{\bm{\theta}}_A$ be the parameter estimator derived from the estimating function $\bm{U}_{A}(\alpha,\bm{\beta};t_{0}|\hat{G})$. If $\hat{G}(\cdot|\mathbf{Z})$ is a uniformly consistent estimator, one may establish the consistency and asymptotic normality of $\tilde{\bm{\theta}}_A$. Express the estimating function as $\bm{U}_{A}(\alpha,\bm{\beta};t_{0}|\hat{G})=\sum_{i=1}^n \tilde{\mathbf{A}}_i$, of which $\tilde{\mathbf{A}}_i$'s are correlated through $\hat{G}(\cdot|\mathbf{Z})$ and $\mbox{E}(\tilde{\mathbf{A}}_i|\mathbf{Z}_{i},V_i)=0$. The sandwich estimator for the variance of 
$\tilde{\bm{\theta}}_A$ is 
\begin{equation}\label{eq: varestmAapp}
\widehat{\mbox{Var}}\big\{\tilde{\bm{\theta}}_A\big\}
=\Big(\frac{\partial \bm{U}_A(\bm{\theta};t_{0}|\hat{G})}{\partial \bm{\theta}^{T}}\Big)^{-1}
\hat{V}\big\{\bm{U}_A(\bm{\theta};t_{0}|\hat{G})\big\}
\left[\Big(\frac{\partial \bm{U}_A(\bm{\theta};t_{0}|\hat{G})}{\partial \bm{\theta}^{T}}\Big)^{-1}\right]^{T},
\end{equation}
where $\hat{V}\big\{\bm{U}_A(\bm{\theta};t_{0}|\hat{G})\big\}=\sum_{i=1}^n \big(\tilde{\mathbf{A}}_i - \bar{\tilde{\mathbf{A}}}\big)
\big(\tilde{\mathbf{A}}_i - \bar{\tilde{\mathbf{A}}}\big)^{T}$ with $\bar{\tilde{\mathbf{A}}}=\frac{1}{n}\sum_{i=1}^n \tilde{\mathbf{A}}_i$.

Similarly, the asymptotic properties of the estimator $\tilde{\bm{\theta}}_B$ from the estimating function $\bm{U}_B(\bm{\theta};t_{0}|\hat{G})$ can be derived. However, there appear to be no convenient ways to implement the sandwich estimator for the variance of $\tilde{\bm{\theta}}_B$ in this form 
\[
\Big(\frac{\partial \bm{U}_B(\bm{\theta};t_{0}|\hat{G})}{\partial \bm{\theta}^{T}}\Big)^{-1}
\hat{V}\big\{\bm{U}_B(\bm{\theta};t_{0}|\hat{G})\big\}
\left[\Big(\frac{\partial \bm{U}_B(\bm{\theta};t_{0}|\hat{G})}{\partial \bm{\theta}^{T}}\Big)^{-1}\right],
\]
where the middle term $\hat{V}\big\{\bm{U}_B(\bm{\theta};t_{0}|\hat{G})\big\}$ is derived from
$\mbox{Var}\big\{\bm{U}_B(\bm{\theta};t_{0}|\hat{G})\big\}$.
Write $\bm{U}_{B}(\alpha,\bm{\beta};t_{0}|\hat{G})=\sum_{i=1}^n \tilde{\mathbf{B}}_i$, of which $\tilde{\mathbf{B}}_i$'s are correlated through $\hat{G}(\cdot|\mathbf{Z})$ but $\mbox{E}(\tilde{\mathbf{B}}_i|\mathbf{Z}_{i},V_i)\neq 0$ in general. As
\[
\mbox{Var}\big\{\bm{U}_B(\bm{\theta};t_{0}|\hat{G})\big\}=\mbox{E}\big\{\mbox{Var}\big[\sum_{i=1}^n \tilde{\mathbf{B}}_i|\mathbf{Z}_{i},V_i\big]\big\}+\mbox{Var}\big\{\mbox{E}\big[\sum_{i=1}^n \tilde{\mathbf{B}}_i|\mathbf{Z}_{i},V_i\big]\big\},
\]
the first term on the right-hand side can be estimated by $\hat{V}_{app}\big\{\bm{U}_B(\bm{\theta};t_{0}|\hat{G})\big\}=\sum_{i=1}^n \big(\tilde{\mathbf{B}}_i - \bar{\tilde{\mathbf{B}}}\big)
\big(\tilde{\mathbf{B}}_i - \bar{\tilde{\mathbf{B}}}\big)^{T}$ with $\bar{\tilde{\mathbf{B}}}=\frac{1}{n}\sum_{i=1}^n \tilde{\mathbf{B}}_i$. The second term results from using $\hat{G}(\cdot|\mathbf{Z})$ in the IPCW and is generally positive. It is hard to estimate, especially when $\hat{G}(\cdot|\mathbf{Z})$ is not a parametric estimator. In fact, since
\begin{align}
\mbox{E}(\tilde{\mathbf{B}}_i|\mathbf{Z}_i,V_{i})=\mathbb{I}(t_{0} \ge V_{i})
        \begin{bmatrix}1 \\ \mathbf{Z}_{i} \end{bmatrix}
        \mbox{E}\Big\{\mathbb{I}(T_i \le t_0)\big[\frac{G(T_i|\mathbf{Z}_i)}{\hat{G}(T_i|\mathbf{Z}_i)} -1\big]\big|U_{i},\mathbf{Z}_i, V_{i}\Big\},
\end{align}   
its variance estimation depends heavily on the estimator $\hat{G}(\cdot|\mathbf{Z})$. One may employ a resampling variance estimation procedure to calculate the variance estimate in practice. Alternatively, one may ignore that second term and use the corresponding form of the variance estimator for $\tilde{\bm{\theta}}_A$ to estimate the variance of 
$\tilde{\bm{\theta}}_B$ as
\begin{equation}\label{eq: varestmBapp}
\widehat{\mbox{Var}}\big\{\tilde{\bm{\theta}}_B\big\}
=\Big(\frac{\partial \bm{U}_B(\bm{\theta};t_{0}|\hat{G})}{\partial \bm{\theta}}\Big)^{-1}
\hat{V}_{app}\big\{\bm{U}_B(\bm{\theta};t_{0}|\hat{G})\big\}
\Big(\frac{\partial \bm{U}_B(\bm{\theta};t_{0}|\hat{G})}{\partial \bm{\theta}}\Big)^{-1}.
\end{equation}
This variance estimator underestimates $\tilde{\bm{\theta}}_B$'s variance. When $\hat{G}(\cdot|\mathbf{Z})$ is sufficiently close to $G(\cdot|\mathbf{Z})$, the resulting bias from the variance underestimation may be acceptable. The sandwich estimators $\widehat{\mbox{Var}}\big\{\tilde{\bm{\theta}}_A\big\}$ and $\widehat{\mbox{Var}}\big\{\tilde{\bm{\theta}}_B\big\}$ in (\ref{eq: varestmAapp}) and (\ref{eq: varestmBapp}) are used to obtain the sandwich standard error estimates in the real data analysis reported in Section \ref{sec3: real data analysis}. We found that these estimated standard errors works well in practice since they were close to the bootstrap standard errors. Also, the bootstrap standard errors work well when the number of bootstrapping is sufficiently large (Figure \ref{fig: compare_se_all} in Supplementary Material). A detailed discussion is presented in Section \ref{sec3: real data analysis}. 

We apply the survival random forest (SRF), a machine learning technique, to estimate $G( \cdot | \mathbf{Z})$ for implementing approaches A and B. Section \ref{sec3: real data analysis} showcases how to apply approaches A and B using real data. Extensive simulations in Section \ref{sec4: simulation} demonstrate the effectiveness of estimating $G( \cdot | \mathbf{Z})$ using the SRF with appropriately chosen hyperparameters, as well as three alternative estimators for the censoring distribution: (i) a stratified empirical cumulative distribution function (ECDF), where the continuous covariate $Z_1$ is partitioned into quartiles ($[0, 0.25)$, $[0.25, 0.5)$, $[0.5, 0.75)$, and $[0.75, 1]$); (ii) a Cox PH model for the censoring time $C$ (referred to as standard Cox model in this paper), and (iii) ) a Cox PH model for the gap time $C^{*} = C - (V + 5)$ (referred to as gap-time Cox model in this paper). This latter formulation accounts for the CCSS cohort structure, in which all subjects survived at least five years after cancer diagnosis. The SRF method results in satisfactory performance for both approaches A and B in simulation studies.\\

\section{Analysis of Premature Ovarian Insufficient (POI) Data} \label{sec3: real data analysis}

This section reports the analysis of the POI data from the Childhood Cancer Survivor Study (CCSS) by applying the proposed
approaches. The real world data analysis guided the design of the simulation studies reported in Section \ref{sec4: simulation}.

\subsection{Descriptive Statistics}
The CCSS is a multi-institutional study that enrolled childhood cancer survivors from 31 centers across North America. Eligibility criteria for the cohort included a cancer diagnosis before the age of 21 between 1970 and 1999, and survival of at least five years post-diagnosis \citep{robison2002study,robison2009childhood}. For this analysis, the study sample was further restricted to female survivors who were at least 18 years old at their last follow-up and had available self-reported menstrual information \citep{im2023development}. The event of interest is POI, which is defined as the cessation of ovarian function before the age of 40. This is a particularly salient late effect for female cancer survivors, with a cumulative incidence that far exceeds that of the general population. \citep{mishra2017early,eshre2016eshre}.

Following \cite{im2023development}, we select five exposures and two interaction terms as the potential covariates. The five exposures are the rescaled age at cancer diagnosis (age divided by 21 so its range is between 0 and 1, $Z_1$), race (categorized as Caucasian, African American, or Other, which is baseline, $Z_2$ and $Z_3$, respectively), and the indicators for receiving bone marrow transplant (BMT, $Z_4$), alkylating agents ($Z_5$), and radiotherapy to the abdomen, pelvis, or whole body ($Z_6$). The two interactions are the rescaled age at cancer diagnosis with the indicator for BMT ($Z_1*Z_4$) and it with the indicator for radiotherapy ($Z_1*Z_6$). 

The event POI is defined as occurring after the age of cancer diagnosis and prior to age 40 in the CCSS study. The right-censoring age is the minimum of the age at surgical premature menopause (SPM), the age at the second malignant neoplasm, and the age at last follow-up. The censoring ages of all the study subjects are available. Assuming data is missing at random, we used a study group comprised of 6,961 subjects with complete data for the analysis. Table \ref{tab:desc_stat} presents descriptive statistics of the six exposure variables and the POI status. Among those subjects, $11.47\%$ experienced the POI before the censoring. The majority of the subjects were Caucasian. More than half of them received alkylating agents, while $22.63\%$ received radiation therapy and only $4.28\%$ received BMT.

\begin{table}[htbp]
\centering
\captionof{table}{Descriptive statistics of the POI data used in the analysis}\label{tab:desc_stat}
\begin{tabular}{|lc|r|}
\hline
\multicolumn{2}{|l|}{}                                                                                & \multicolumn{1}{c|}{\begin{tabular}[c]{@{}c@{}}Number of Subjects \\ (total n = 6961)\end{tabular}} \\ \hline
\multicolumn{1}{|l|}{\multirow{2}{*}{Status of POI Development}}                                   & Yes             & 798                                                                                            \\ \cline{2-3} 
\multicolumn{1}{|l|}{}                                                              & No              & 6163                                                                                           \\ \hline
\multicolumn{1}{|l|}{Age at Cancer Diagnosis (\textit{$21\times Z_{1}$})}                    & Median (Q1, Q3) & 7.43 (3.27, 13.86)                                                                             \\ \hline
\multicolumn{1}{|l|}{\multirow{3}{*}{Race (vs Caucasian)}} & African American (\textit{$Z_{2}$}=1)          & 389                                                                                            \\ \cline{2-3} 
\multicolumn{1}{|l|}{}                                                              & Other (\textit{$Z_{3}$}=1)          & 469                                                                                            \\ \cline{2-3} 
\multicolumn{1}{|l|}{}                                                              & Caucasian (\textit{$Z_{2}$}=\textit{$Z_{3}$}=0)         & 6103                                                                                           \\ \hline
\multicolumn{1}{|l|}{\multirow{2}{*}{Having Received BMT }}                & Yes (\textit{$Z_{4}$}=1)           & 298                                                                                            \\ \cline{2-3} 
\multicolumn{1}{|l|}{}                                                              & No (\textit{$Z_{4}$}=0)             & 6663                                                                                           \\ \hline
\multicolumn{1}{|l|}{\multirow{2}{*}{Having Received Alkylating agents}}  & Yes (\textit{$Z_{5}$}=1)            & 3516                                                                                           \\ \cline{2-3} 
\multicolumn{1}{|l|}{}                                                              & No (\textit{$Z_{5}$}=0)             & 3445                                                                                           \\ \hline
\multicolumn{1}{|l|}{\multirow{2}{*}{Having Received Radiotherapy}}       & Yes (\textit{$Z_{6}$}=1)             & 1575                                                                                           \\ \cline{2-3} 
\multicolumn{1}{|l|}{}                                                              & No (\textit{$Z_{6}$}=0)              & 5386                                                                                           \\ \hline
\end{tabular}
\end{table}

\subsection{Inferential Analysis}
The study by \cite{im2023development} focused on estimating the probability of POI between ages 21 and 40 conditional on covariates, referred to as `Approach by Im et al' below. Replacing $I (T_{i} < t_{0})$, their response variable at the analysis time $t_0$, by $I (T_{i}\le t_{0})$, the response variable in our analysis, approach by Im et al is then the same as the proposed approach A without accounting for the left censoring due to the cancer diagnosis, an IPCW based GLM (the generalized linear model) approach. We conduct an analysis at ages from 17 to 40 by applying the three approaches, approach by Im et al and approaches A and B. The $G(\cdot|\mathbf{Z})$ in the IPCW estimated by the SRF, where both the number of trees and the minimal node size to split at (referred to as node size in this paper) are set to 100. All analyses are conducted in R. The R packages \texttt{ranger} and \texttt{survival} were employed to estimate $\hat{G}(\cdot\mid\mathbf{Z})$.

\begin{figure}[htbp]
    \centering
    \begin{subfigure}[t]{0.9\textwidth}
        \centering
         \includegraphics[width=1\textwidth]{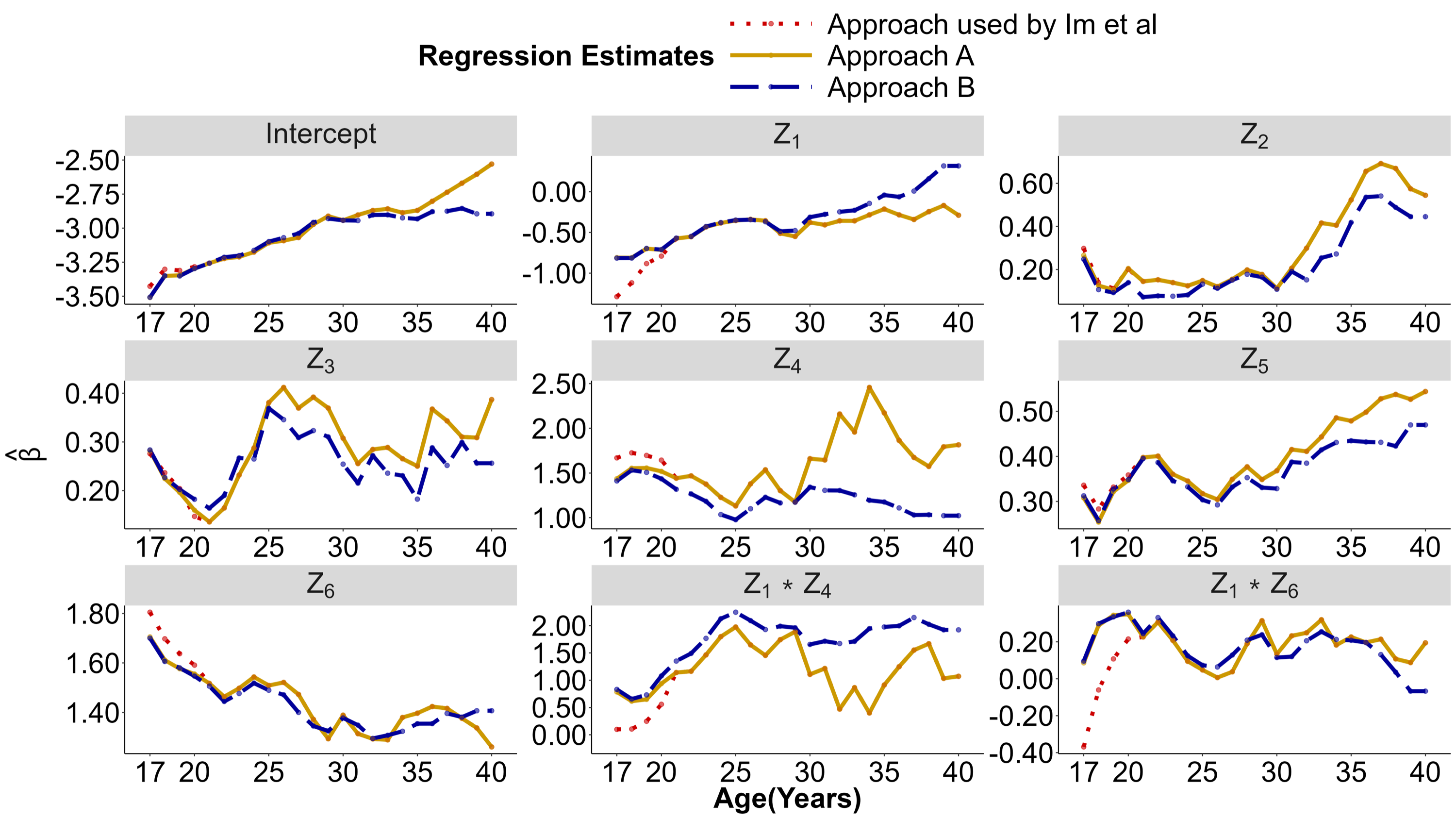}
         \caption{Estimated Coefficients}
        \label{fig: real_srf_comp_coef}
    \end{subfigure}%
        \vfill
    \begin{subfigure}[t]{0.9\textwidth}
        \centering
            \includegraphics[width=1\textwidth]{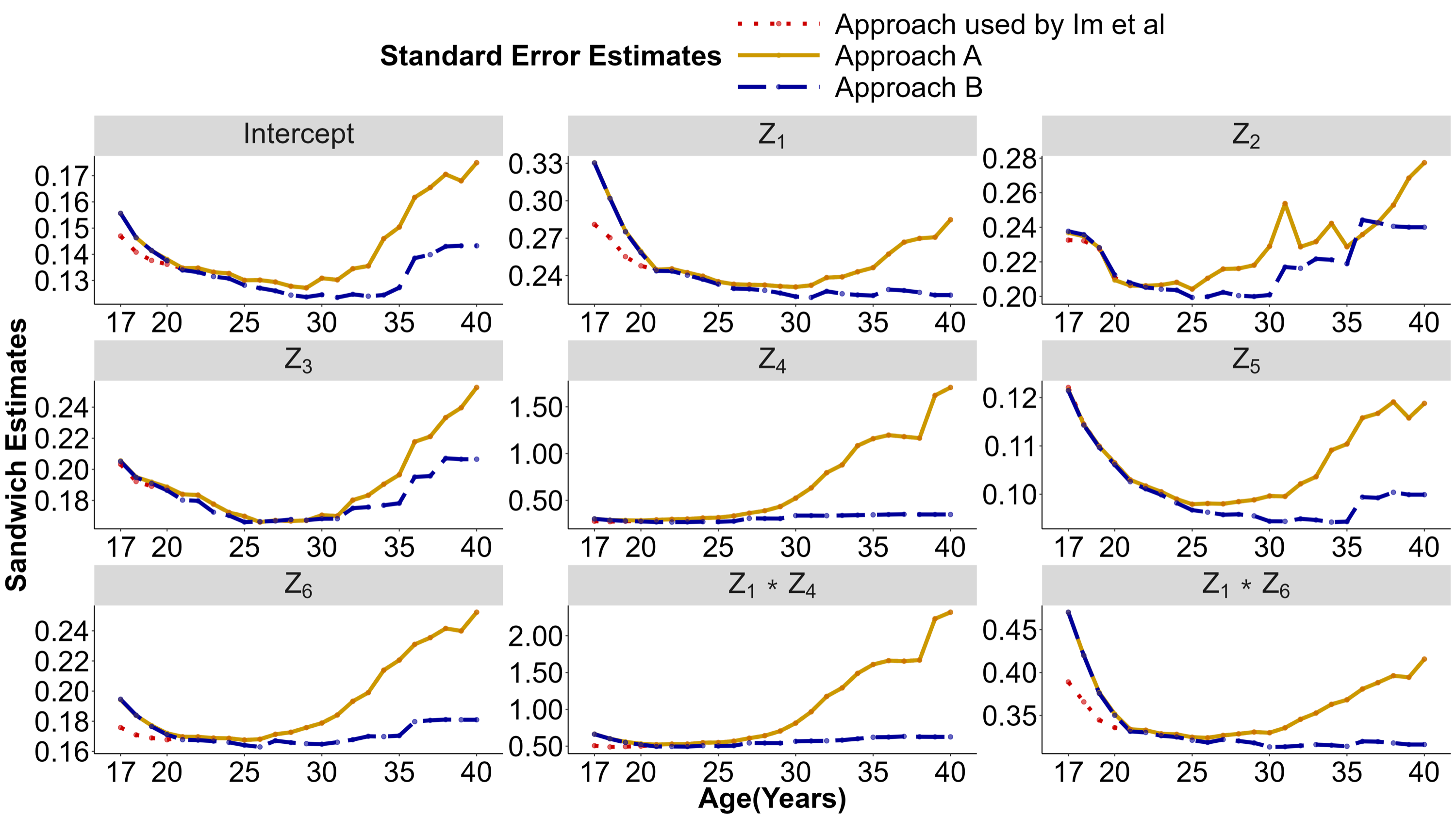}
            \caption{Sandwich Standard Error Estimates}
        \label{fig: real_srf_comp_se}
    \end{subfigure}
    \caption{POI data analysis outcomes by three approaches aided by SRF estimate $\hat{G}(\cdot|\mathbf{Z})$ with the number of trees and the node size in SRF are set to 100. The sandwich standard errors are calculated using Equation \ref{eq: varestmAapp} and \ref{eq: varestmBapp}, receptively.
             $Z_{1}\text{: Rescaled age at cancer diagnosis}$,
             $Z_{2}\text{: Race-African American}$,
             $Z_{3}\text{: Race-Other}$,
             $Z_{4}\text{: Receipt of BMT}$,
             $Z_{5}\text{: Receipt of Alkylating agents}$
             $Z_{6}\text{: Receipt of radiation to the abdomen/pelvis/total body}$} \label{fig: real_srf_comp_coef_se}
\end{figure}
The estimated age-specific intercept and regression coefficients between age 21 and 40 are presented in Figure \ref{fig: real_srf_comp_coef}. The estimates associated with approach by Im et al are different from the ones using Approaches A and B before age 21. The difference is apparent in the estimates for the coefficients of $Z_{1}$, the rescaled age at cancer diagnosis. It verifies that the estimates of approach by Im et al before $t_0=21$ are biased since they do not adjust for the at risk set. When $t_{0} \ge 21$, approach by Im et al is equivalent to approach A and thus yields the same estimates. The estimates by approaches A and B are similar except for the coefficients of $Z_{4}$ (BMT) and the interactions $Z_{1}*Z_4$, of which the estimates by approach A fluctuate after age 30. It is likely because that only 4.28\% of the study subjects received BMT overall. 

The estimated standard errors (SEs) using the sandwich variance estimators (\ref{eq: varestmAapp}) and (\ref{eq: varestmBapp}) are presented in Figure \ref{fig: real_srf_comp_se}. In general, SEs associated with Approach B are smaller than the ones associated with Approach A, particularly after age 30. The gaps increase as the $t_0$ increases. The magnitudes of the estimated SEs associated with the estimated coefficients of $Z_4$ and interaction $Z_{1}*Z_{4}$ are relatively larger compared to the other estimated SEs. It reflects again little information available in the data on the effect of $Z_4$. 

We constructed approximate $95\%$ confidence intervals (CIs) for all the model parameters of Approaches A and B (Figure \ref{fig: real_srf_comp_coef_with_se} in Supplementary Material). We observe the covariate effects on POI risk as follows: The younger the cancer diagnosis age ($Z_1$), the lower the risk. But the age at cancer diagnosis effect diminished as $t_{0}$ increases. Non-Caucasian subjects consistently show a higher risk than Caucasian subjects. However, these effects were not statistically significant at most ages. The estimated effect of BMT is influenced by the approach used. In Approach B, subjects exposed to BMT ($Z_4$) have a significantly higher risk than those not exposed . The age at cancer diagnosis modifies the effect BMT, the BMT exposure effect increases as the age at cancer diagnosis increases. The BMT exposure effect on the POI risk increases as attained age ($t_0$) increases to 25 and then remains at the same level. The alkylating agents ($Z_5$) significantly increase POI risk. Its effect shows an increasing trend over $t_{0}$. The radiation  exposure ($Z_6$) increases the POI risk, showing an increasing trend over $t_{0}$. The cancer diagnosis age does not modify the effect of radiation exposure ($Z_{1}*Z_{6}$).
\begin{figure}[htbp]
    \centering
    \includegraphics[width=1\textwidth]{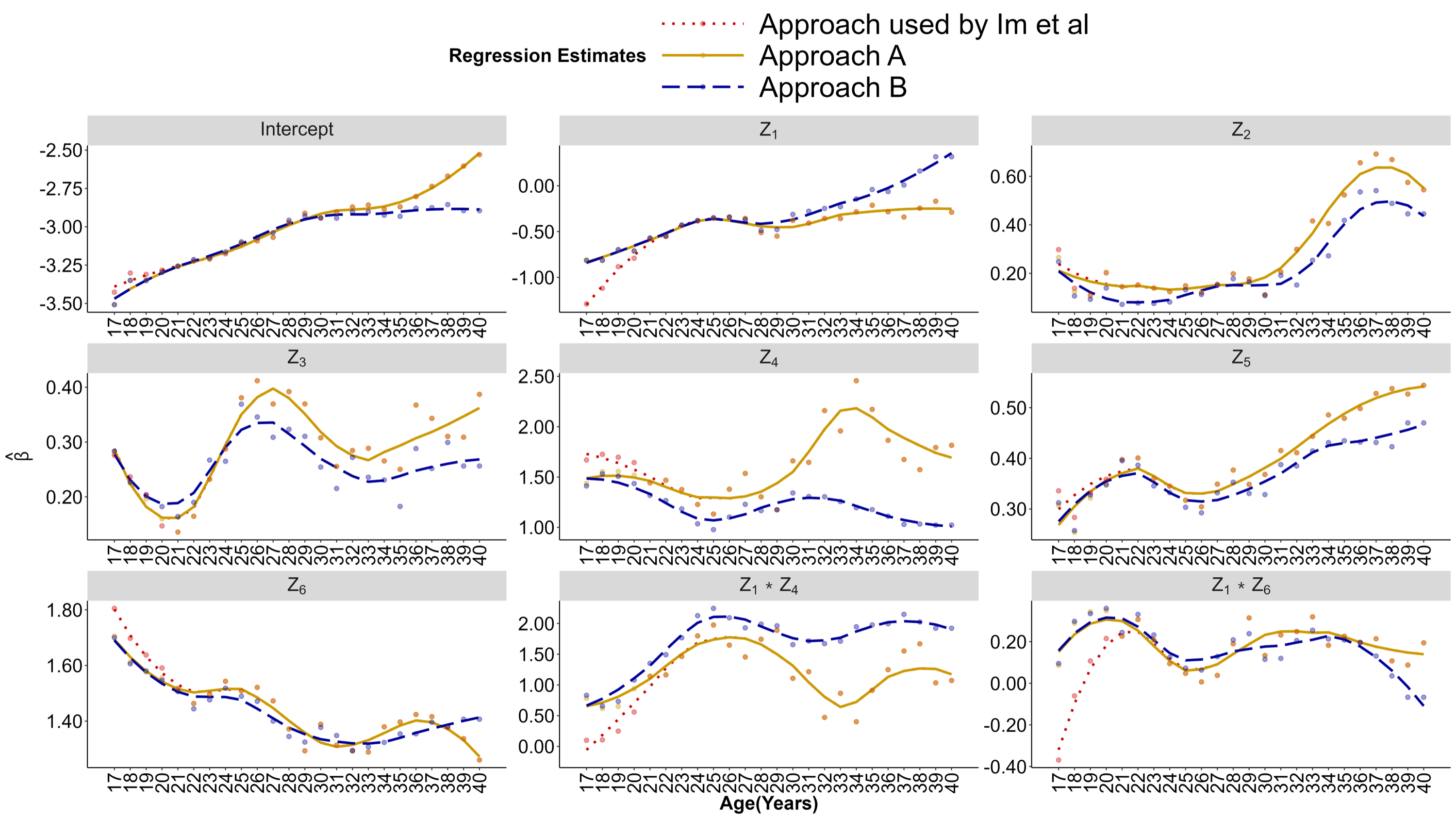}
    \captionof{figure}{The estimated coefficients were obtained using three approaches. The LOESS method was applied when span value is set to 0.5. The $\hat{G}(\cdot|\mathbf{Z})$ is obtained using SRF. The number of trees and the node size in SRF are set to 100.
             $Z_{1}\text{: Rescaled age at cancer diagnosis}$,
             $Z_{2}\text{: Race-African American}$,
             $Z_{3}\text{: Race-Other}$,
             $Z_{4}\text{: Receipt of BMT}$,
             $Z_{5}\text{: Receipt of Alkylating agents}$
             $Z_{6}\text{: Receipt of radiation to the abdomen/pelvis/total body}$}\label{fig: real_srf_comp_coef_loess_one}
\end{figure}
We estimated the conditional distribution of censoring time $G(\cdot|\mathbf{Z})$ using SRF, the stratified ECDF, the standard Cox model, and the gap-time Cox model.  These estimates are presented in Figures \ref{fig: real_appa_comp_coef} and \ref{fig: real_appb_comp_coef} (Supplementary Material). We observe that the estimation method for the censoring distribution has a substantial impact on the coefficient estimates. On the other hand, its effect on standard error estimation is negligible across all approaches (Figures \ref{fig: real_appa_comp_coef_se} and \ref{fig: real_appb_comp_coef_se}, Supplementary Material).

In addition, locally estimated scatterplot smoothing (LOESS) was applied to each collection of the estimated coefficients using the three approaches. The LOESS curves with the span value at $0.5$ are presented in Figure \ref{fig: real_srf_comp_coef_loess_one}. It shows the trends of the covariate effects over $t_{0}$. The trends by Approaches A and B are similar, except for those associated with $Z_4$ (BMT receipt) and its interaction with $Z_1$ (rescaled cancer diagnosis age). A more comprehensive comparison, with the span values at $0.3, 0.5$ and $0.8$, is presented in Figure \ref{fig: real_srf_comp_coef_loess}, and the same conclusion can be drawn.

\section{Simulation}\label{sec4: simulation}


We conducted three simulation studies to examine the consistency, efficiency, and robustness of the proposed estimators with four different methods to estimate the censoring distribution: the standard Cox model, the gap-time Cox model, the stratified ECDF, and SRF. We also report the estimated result of using the true CDF of the censoring time, $G(\cdot|\mathbf{Z})$, as a benchmark.

Across the simulation studies, each dataset consisted of $n=7000$ subjects and included two covariates: a continuous variable $Z_{1}$ and a binary variable $Z_{2}$. All results are based on 1000 replications. \textit{R} was used to conduct the simulation. 

\subsection{Data Generation}\label{sec4: simulation data generation}

The main difference among the three simulation studies is in the generation of the event time $T|\mathbf{Z}, T > V$. The data generation procedure for subject $i$ in Simulations 1 and 3 is reported in Algorithm \ref{alg:data_sim}. In Simulation 2, an intermediate step, \textit{Step 1b}, is introduced after Step 1. In this step, subjects are stratified into two groups based on $V_{i}$. The event indicator $\mathbb{I}(T_{i}|\mathbf{Z}_{i}, T_{i} > V_{i})$ is then simulated from different logistic distributions depending on $V_{i}$ .
\begin{algorithm}[ht]
    \small
    \singlespacing
    \vspace{-20pt}
    \caption{Simulation Process for one repetition}\label{alg:data_sim}
    \begin{flushleft}
        \textbf{For} subject $i = 1, \ldots, n$ \textbf{do}:
        \vspace{-5pt}
        \begin{enumerate}[itemsep=2pt, topsep=2pt, parsep=0pt]
            \item \textbf{Generate} $Z_{1i} \sim \text{Beta}(a_{1},a_{2})$ and $Z_{2i} \sim \text{Bern}(p)$
            \item \textbf{Calculate} $V_{i} = 21 \times Z_{1i}$
            \item \textbf{Generate Censoring Time:} 
            \begin{itemize}[itemsep=1pt, topsep=1pt, parsep=0pt]
                \item Sample $C^{*}_{i} \sim \text{Weibull}(\psi_{3i}, \psi_{4i})$
                \item Set $C_{i} = C^{*}_{i} + V_{i} + 5$
            \end{itemize}
        \end{enumerate}
        
        \vspace{5pt}
        \textbf{For} subject $i = 1, \ldots, n$, let $Y_{i}(t_{0}) \coloneqq \mathbb{I}(T_{i}|\mathbf{Z}_{i}, T_{i} > V_{i})$, and for each $t_{0}$ \textbf{do}:
        \vspace{-5pt}
        \begin{enumerate}[itemsep=2pt, topsep=2pt, parsep=0pt]
            \setcounter{enumi}{3}
            \item \textbf{Determine Event at $t_0$:} \\
            Calculate $\pi_{i}(t_{0}) = P(T_{i}\le t_{0} \mid \mathbf{Z}_{i}, T_{i} > V_{i}, V_{i} < t_{0})$ \\
            Sample $Y_{i}(t_{0}) \sim \text{Bern}(\pi_{i}(t_{0}))$
            \item \textbf{If} $Y_{i}(t_{0}) = 1$ \textbf{then}:
            \begin{itemize}[itemsep=1pt, topsep=1pt, parsep=0pt]
                \item Set $q \gets 1$
                \item \textbf{While} $Y_{i}(t_{0}-(q-1)s) == 1$ \textbf{do}:
                \begin{itemize}[itemsep=1pt, topsep=1pt, parsep=0pt]
                    \item Calculate $\pi_{i}(t_{0}-qs) = P(T_{i}\le t_{0}-qs \mid \mathbf{Z}_{i}, T_{i} > V_{i}, V_{i} < t_{0}-qs)$
                    \item \textbf{Check Threshold:} \textbf{If} $\pi_{i}(t_{0}-qs) < 0.005$ \textbf{then} \\
                    \hspace*{1em} Set $Y_{i}(t_{0}-qs) = 0$ and \textbf{break}
                    \item Calculate $\rho_{cond} = \frac{\pi_{i}(t_{0}-qs)}{\pi_{i}(t_{0}-(q-1)s)}$ 
                    \item Sample $Y_{i}(t_{0}-qs) \sim \text{Bern}(\rho_{cond})$
                    \item \textbf{If} $Y_{i}(t_{0}-qs)=1$, set $q \gets q + 1$. \textbf{Else} break
                \end{itemize}
                \item Set $T_{i} = t_{0}-(q-1)s$
            \end{itemize}
            \item Let $\delta_{i} = \mathbb{I}(T_i \le C_i)$ and $U_{i} = \min\{T_{i}, C_{i}\}$
            \item Store generated data is $\{ Y_{i}(t_{0}), U_{i}, \delta_{i},V_{i}, C_{i}, \mathbf{Z}_{i} \}$
        \end{enumerate}
    \end{flushleft}
\end{algorithm}
In all the simulation studies, the time unit $s$ is set to $1/12$. The continuous covariate $Z_{1}$ was sampled from a $\text{Beta}(a_1, a_2)$ distribution, with the age at diagnosis defined as $V = 21\times Z_1$. We specified $(a_1, a_2) = (0.94, 1.06)$ for Simulations 1 and 3, and $a_1 = a_2 = 2$ for Simulation 2. The binary covariate $Z_2$ followed a $\text{Bernoulli}(0.40)$ distribution. The event time $T|\mathbf{Z}, T > V, V<t_{0}$ was only generated for subjects whose $\mathbb{I}(T_{i}|\mathbf{Z}_{i}, T_{i} > V_{i})=1$ at each given $t_{0}$ from its assumed distribution. To mimic the POI data where subjects survive at least 5 years post-diagnosis, the censoring time was defined as $C = C^{*} + V + 5$. The component $C^{*}$ was generated from a Weibull distribution with shape parameter $\psi_{3i}$ and scale parameter $\psi_{4i}$. These parameters varied by scenario: $\psi_{3i} = 3.34- 0.10 \times Z_{2i}$ and $\psi_{4i} = 21.00-2.00 \times Z_{2i}$ in Simulation 1.1; $\psi_{3i} = 6.00 - 1.00 \times Z_{2i}$ and $\psi_{4i} = 31.00 - 2.00 \times Z_{2i}$ in Simulation 1.2; $\psi_{3i} = 3.34 - 0.10 \times  Z_{2i}$ and $\psi_{4i} = 22.00 - 2.00 \times Z_{2i}$ in Simulation 2; and $\psi_{3i} = 3.34 - 2.00 \times Z_{2i}$ and $\psi_{4i} = 20.00$ in Simulation 3. Descriptive statistics for the simulated studies are summarized in Table \ref{table: desc_sim_data}.

\begin{table}[htbp]
\centering
\setlength{\tabcolsep}{3pt} 
\captionof{table}{The descriptive statistics of censoring rate and age at cancer diagnosis in different simulation studies.}
\label{table: desc_sim_data}
\begin{tabular}{@{}lcccccclccc@{}}
\toprule
& \multicolumn{5}{c}{\textbf{Censoring Rate}} & & \multicolumn{4}{c}{\textbf{Age at Cancer Diagnosis ($\boldsymbol{V}$)}} \\
\cmidrule(lr){2-6} \cmidrule(lr){8-11}
& \makecell{Cleaned\\Data} & \makecell{Sim 1.1} & \makecell{Sim 1.2} & \makecell{Sim 2} & \makecell{Sim 3} & & & \makecell{Cleaned\\Data} & \makecell{Sim 1\\and 3} & \makecell{Sim 2} \\
\midrule
Age 13 & 0.10\% & - & - & 0.11\% & - & & Min & 0.00 & 0.00 & 0.13 \\
Age 14 & 0.14\% & - & - & 0.20\% & - & & Q1 & 3.27 & 4.56 & 6.86 \\
Age 15 & 0.22\% & - & - & 0.33\% & 0.94\% & & Median & 7.43 & 9.62 & 10.50 \\
Age 20 & - & - & - & 2.29\% & 4.87\% & & Q3 & 13.86 & 15.01 & 14.15 \\
Age 21 & 5.57\% & 8.21\% & 0.35\% & - & - & & Max & 21.00 & 20.99 & 20.87 \\
Age 25 & - & - & - & 7.39\% & 14.38\% & & Mean & 8.59 & 9.87 & 10.50 \\
Age 30 & 31.38\% & 35.86\% & 3.41\% & 14.96\% & 29.94\% & & SD & 5.95 & 6.05 & 4.69 \\
Age 35 & 50.51\% & 55.00\% & 6.83\% & 22.12\% & 48.58\% & & & & & \\
Age 40 & 66.46\% & 71.12\% & 10.58\% & 26.55\% & 65.82\% & & & & & \\
\bottomrule
\end{tabular}
\end{table}
\subsection{Simulation Outcomes}
We assessed the consistency and efficiency of the proposed approaches using four metrics: the sample mean of the estimates (SMEAN), the sample standard deviation of the estimates (SSD), the sample mean of the estimated standard errors (SMESE), and the root sample mean squared error of the estimates (RSMSE) in Simulations 1 and 2. We assessed the robustness of the proposed approaches based on a comparison of the estimated conditional survival probabilities with their true values in Simulations 2 and 3.

\subsubsection{Simulation 1: Examining consistency and efficiency}
We consider two different censoring rates.  \medskip

\noindent{\sf \underline{Simulation 1.1: Heavy Censoring}} We generated data with censoring rate comparable to the aforementioned real data. Specifically, the data were generated from a logistic regression. 
\begin{align}\label{model sim1: true model}
\begin{split}
\log\left( \frac{P(T \le t|\mathbf{Z}, T > V)}{1-P(T \le t|\mathbf{Z}, T > V)} \right)= \alpha_{1}(t)+\beta_1(t) Z_1 + \beta_2(t) Z_2,
\end{split}
\end{align}
where the intercept $\alpha_{0}(t)= \gamma_{0} + \gamma_{1}t$, with $\gamma_{0} = -7.5$ and $\gamma_{1}=0.23$. The regression coefficients $\bm{\beta}(t)= \{\beta_{1}(t), \beta_{2}(t)\}^{T}$ are held constant over time with $\beta_{1}(t) = \beta_{1} = -4.83$ and $\beta_{2}(t) = \beta_{2} = -1.00$. All the estimates from the three approaches were evaluated at four distinct time points: 21, 30, 35, and 40 using the simulated data. 

The three hyperparameters of SRF (the number of candidate variables drawn in each split; the node size; the number of trees) are set to 2, 200, and 100, respectively. We choose the combination of hyperparameters based on the lowest value of RSMSE. The estimated coefficients by Approach B always have smaller RSMSEs than those by Approach A at ages 35 and 40.
\begin{table}[htbp]
\centering
\footnotesize 
\setlength{\tabcolsep}{2pt} 
\renewcommand{\arraystretch}{1.4} 
\captionof{table}{The result of simulation 1.1. We report the sample mean of the estimates (SMEAN), the sample standard deviation of the estimates (SSD), the sample mean of the estimated standard errors  (SMESE), and the root sample mean squared error of the estimates (RSMSE) of each approach. In SRF, the number of tree is set to 100 and the node size is set to 200.}\label{table: sim1set11:result}
\resizebox{0.82\textwidth}{!}{%
\begin{tabular}{|c|c|ccccc|ccccc|ccccc|}
\hline
 &  & \multicolumn{5}{c|}{$\alpha(21) = -2.670 $} & \multicolumn{5}{c|}{$\beta_{1}(21) = -4.830$} & \multicolumn{5}{c|}{$\beta_{2}(21) = -1.000$} \\ \hline
Age 21 &  & SRF & Cox $C$ &  Cox $C^{*}$ & ECDF & True & SRF & Cox $C$ &  Cox $C^{*}$ & ECDF & True & SRF & Cox $C$ &  Cox $C^{*}$ & ECDF & True \\ \hline
\multirow{4}{*}{\begin{tabular}[c]{@{}c@{}}Approach used\\ by Im et al/\\ Approach A\end{tabular}} 
 & SMEAN & -2.666 & -2.615 & -2.670 & -2.637 & -2.668 & -4.916 & -5.037 & -4.912 & -5.013 & -4.913 & -1.027 & -1.021 & -1.026 & -1.023 & -1.030 \\
 & SSD   & 0.189 & 0.190 & 0.189 & 0.190 & 0.189 & 0.691 & 0.702 & 0.692 & 0.708 & 0.692 & 0.309 & 0.307 & 0.309 & 0.307 & 0.309 \\
 & SMESE & 0.187 & 0.187 & 0.187 & 0.188 & 0.187 & 0.653 & 0.663 & 0.653 & 0.670 & 0.653 & 0.311 & 0.310 & 0.311 & 0.310 & 0.311 \\
 & RSMSE & 0.189 & 0.197 & 0.189 & 0.193 & 0.189 & 0.696 & 0.732 & 0.696 & 0.731 & 0.697 & 0.310 & 0.307 & 0.310 & 0.308 & 0.310 \\ \hline
\multirow{4}{*}{Approach B} 
 & SMEAN & -2.670 & -2.730 & -2.664 & -2.702 & -2.668 & -4.908 & -4.713 & -4.916 & -4.792 & -4.913 & -1.033 & -1.047 & -1.038 & -1.033 & -1.030 \\
 & SSD   & 0.189 & 0.186 & 0.190 & 0.185 & 0.190 & 0.691 & 0.668 & 0.692 & 0.664 & 0.691 & 0.309 & 0.307 & 0.309 & 0.307 & 0.309 \\
 & SMESE & 0.187 & 0.183 & 0.187 & 0.183 & 0.186 & 0.653 & 0.632 & 0.653 & 0.628 & 0.653 & 0.311 & 0.309 & 0.311 & 0.310 & 0.311 \\
 & RSMSE & 0.189 & 0.195 & 0.190 & 0.188 & 0.189 & 0.695 & 0.678 & 0.697 & 0.665 & 0.695 & 0.311 & 0.310 & 0.311 & 0.309 & 0.310 \\ \hline
 &  & \multicolumn{5}{c|}{$\alpha(30) = -0.600 $} & \multicolumn{5}{c|}{$\beta_{1}(30) = -4.830$} & \multicolumn{5}{c|}{$\beta_{2}(30) = -1.000$} \\ \hline
Age 30 &  & SRF & Cox $C$ &  Cox $C^{*}$ & ECDF & True & SRF & Cox $C$ &  Cox $C^{*}$ & ECDF & True & SRF & Cox $C$ &  Cox $C^{*}$ & ECDF & True \\ \hline
\multirow{4}{*}{\begin{tabular}[c]{@{}c@{}}Approach used\\ by Im et al/\\ Approach A\end{tabular}} 
 & SMEAN & -0.578 & -0.586 & -0.593 & -0.524 & -0.595 & -4.890 & -4.819 & -4.859 & -5.065 & -4.857 & -0.993 & -1.026 & -1.013 & -0.997 & -1.004 \\
 & SSD   & 0.123 & 0.121 & 0.125 & 0.117 & 0.127 & 0.327 & 0.317 & 0.331 & 0.325 & 0.333 & 0.193 & 0.188 & 0.195 & 0.180 & 0.195 \\
 & SMESE & 0.131 & 0.126 & 0.129 & 0.124 & 0.129 & 0.346 & 0.325 & 0.342 & 0.337 & 0.342 & 0.190 & 0.183 & 0.190 & 0.179 & 0.189 \\
 & RSMSE & 0.125 & 0.122 & 0.126 & 0.139 & 0.127 & 0.332 & 0.317 & 0.333 & 0.401 & 0.334 & 0.193 & 0.190 & 0.195 & 0.180 & 0.195 \\ \hline
\multirow{4}{*}{Approach B} 
 & SMEAN & -0.616 & -0.805 & -0.594 & -0.742 & -0.600 & -4.813 & -4.366 & -4.848 & -4.420 & -4.845 & -1.023 & -1.016 & -1.017 & -1.020 & -1.006 \\
 & SSD   & 0.121 & 0.112 & 0.122 & 0.108 & 0.125 & 0.322 & 0.296 & 0.326 & 0.271 & 0.331 & 0.193 & 0.184 & 0.195 & 0.180 & 0.196 \\
 & SMESE & 0.130 & 0.119 & 0.129 & 0.115 & 0.129 & 0.341 & 0.308 & 0.340 & 0.283 & 0.340 & 0.188 & 0.178 & 0.188 & 0.178 & 0.188 \\
 & RSMSE & 0.122 & 0.234 & 0.122 & 0.178 & 0.125 & 0.322 & 0.551 & 0.326 & 0.491 & 0.331 & 0.194 & 0.185 & 0.195 & 0.181 & 0.196 \\ \hline
 &  & \multicolumn{5}{c|}{$\alpha(35) = 0.550 $} & \multicolumn{5}{c|}{$\beta_{1}(35) = -4.830$} & \multicolumn{5}{c|}{$\beta_{2}(35) = -1.000$} \\ \hline
Age 35 &  & SRF & Cox $C$ &  Cox $C^{*}$ & ECDF & True & SRF & Cox $C$ &  Cox $C^{*}$ & ECDF & True & SRF & Cox $C$ &  Cox $C^{*}$ & ECDF & True \\ \hline
\multirow{4}{*}{\begin{tabular}[c]{@{}c@{}}Approach used\\ by Im et al/\\ Approach A\end{tabular}} 
 & SMEAN & 0.704 & 0.229 & 0.575 & 0.644 & 0.567 & -5.175 & -4.148 & -4.886 & -5.089 & -4.886 & -0.925 & -1.136 & -1.040 & -1.003 & -1.006 \\
 & SSD   & 0.188 & 0.225 & 0.197 & 0.160 & 0.202 & 0.421 & 0.500 & 0.450 & 0.356 & 0.455 & 0.219 & 0.333 & 0.264 & 0.212 & 0.258 \\
 & SMESE & 0.198 & 0.220 & 0.197 & 0.171 & 0.197 & 0.421 & 0.461 & 0.429 & 0.357 & 0.426 & 0.227 & 0.306 & 0.251 & 0.217 & 0.245 \\
 & RSMSE & 0.243 & 0.392 & 0.198 & 0.186 & 0.202 & 0.544 & 0.846 & 0.453 & 0.440 & 0.459 & 0.231 & 0.360 & 0.267 & 0.212 & 0.258 \\ \hline
\multirow{4}{*}{Approach B} 
 & SMEAN & 0.448 & 0.436 & 0.550 & 0.295 & 0.548 & -4.640 & -4.679 & -4.845 & -4.183 & -4.841 & -1.063 & -0.911 & -1.000 & -1.017 & -1.011 \\
 & SSD   & 0.159 & 0.209 & 0.181 & 0.132 & 0.193 & 0.351 & 0.520 & 0.417 & 0.270 & 0.432 & 0.212 & 0.296 & 0.246 & 0.205 & 0.247 \\
 & SMESE & 0.184 & 0.217 & 0.193 & 0.150 & 0.193 & 0.379 & 0.504 & 0.419 & 0.280 & 0.416 & 0.219 & 0.277 & 0.238 & 0.205 & 0.234 \\
 & RSMSE & 0.189 & 0.238 & 0.180 & 0.287 & 0.193 & 0.399 & 0.541 & 0.417 & 0.701 & 0.431 & 0.221 & 0.309 & 0.246 & 0.206 & 0.247 \\ \hline
 &  & \multicolumn{5}{c|}{$\alpha(40) = 1.700 $} & \multicolumn{5}{c|}{$\beta_{1}(40) = -4.830$} & \multicolumn{5}{c|}{$\beta_{2}(40) = -1.000$} \\ \hline
Age 40 &  & SRF & Cox $C$ &  Cox $C^{*}$ & ECDF & True & SRF & Cox $C$ &  Cox $C^{*}$ & ECDF & True & SRF & Cox $C$ &  Cox $C^{*}$ & ECDF & True \\ \hline
\multirow{4}{*}{\begin{tabular}[c]{@{}c@{}}Approach used\\ by Im et al/\\ Approach A\end{tabular}} 
 & SMEAN & 2.422 & 0.542 & 1.861 & 1.858 & 1.845 & -6.103 & -3.044 & -5.147 & -5.186 & -5.139 & -0.818 & -1.312 & -1.053 & -0.979 & -0.993 \\
 & SSD   & 0.314 & 0.986 & 0.505 & 0.387 & 0.508 & 0.530 & 1.695 & 0.955 & 0.712 & 0.943 & 0.244 & 1.298 & 0.519 & 0.439 & 0.497 \\
 & SMESE & 0.328 & 0.610 & 0.429 & 0.409 & 0.430 & 0.540 & 0.955 & 0.742 & 0.693 & 0.740 & 0.263 & 0.671 & 0.411 & 0.416 & 0.399 \\
 & RSMSE & 0.787 & 1.520 & 0.530 & 0.418 & 0.528 & 1.378 & 2.461 & 1.006 & 0.796 & 0.991 & 0.304 & 1.334 & 0.521 & 0.439 & 0.497 \\ \hline
\multirow{4}{*}{Approach B} 
 & SMEAN & 1.140 & 2.433 & 1.675 & 1.289 & 1.678 & -3.859 & -5.729 & -4.762 & -3.928 & -4.772 & -1.173 & -1.053 & -1.011 & -1.021 & -1.074 \\
 & SSD   & 0.222 & 0.839 & 0.451 & 0.257 & 0.447 & 0.364 & 1.201 & 0.805 & 0.430 & 0.817 & 0.226 & 0.857 & 0.449 & 0.339 & 0.428 \\
 & SMESE & 0.270 & 1.294 & 0.504 & 0.321 & 0.488 & 0.433 & 1.834 & 0.879 & 0.515 & 0.868 & 0.256 & 1.069 & 0.440 & 0.358 & 0.415 \\
 & RSMSE & 0.602 & 1.114 & 0.452 & 0.485 & 0.448 & 1.037 & 1.499 & 0.807 & 0.999 & 0.818 & 0.284 & 0.858 & 0.449 & 0.339 & 0.434 \\ \hline
\end{tabular}%
}
\end{table}

From Table ~\ref{table: sim1set11:result} and Figure ~\ref{fig: smean_compre_method_sim11} (Supplementary Material), we found that consistency was held when censoring distribution was estimated by methods other than directly using standard Cox model. The standard Cox model yielded biased estimates for both $\beta_{1}(t)$ and $\beta_{2}(t)$ across both approaches as censoring rates increased at ages 30, 35, and 40, likely due to a violation of the proportional hazards assumption for the censoring process.


The efficiency of the estimated coefficients is contingent upon the censoring rate at $t_{0}$, as well as the method employed to estimate the censoring distribution (Table ~\ref{table: sim1set11:result} and Figure~\ref{fig: ssd_compre_method_sim11}). The SE estimation is valid when the censoring distribution is estimated by methods other than the standard Cox model, as the SMESE and SSD are in close agreement. While the SMESE and SSD are also similar when using the standard Cox model, the presence of bias in the coefficient estimates, $\beta_{1}(t)$ and $\beta_{2}(t)$, makes the discussion of SE validity for that model unnecessary.

\begin{figure}[htbp]
    \centering
    \includegraphics[width=\textwidth]{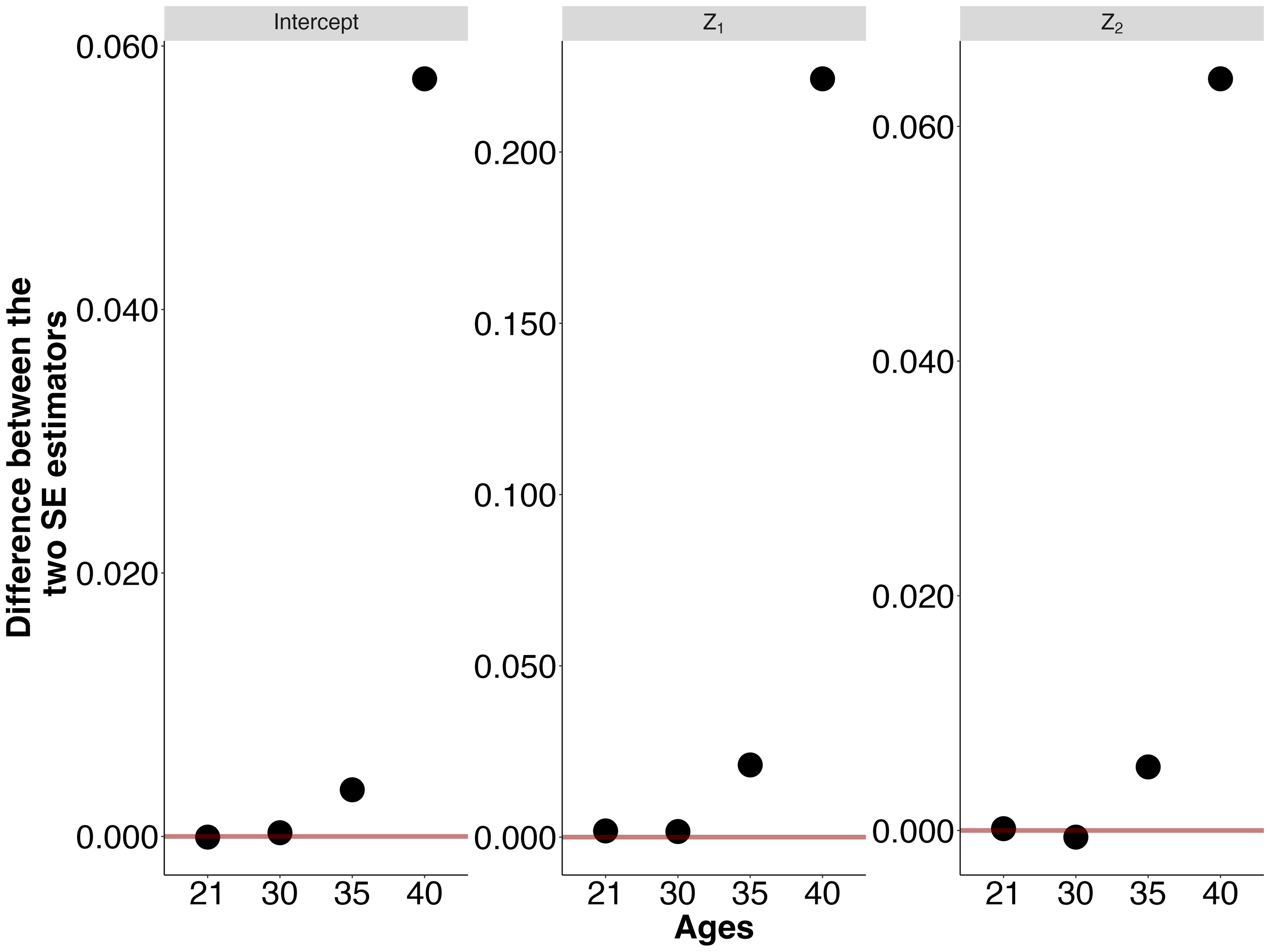}
    \captionof{figure}{Differences between estimated variances of the estimators by Approach A and Approach B when $G(\cdot|\mathbf{Z})$ is known in Simulation 1.1. The red line in each plot represents a difference of 0.}
    \label{fig:sim11_combined_settings}
\end{figure}

Figure \ref{fig:sim11_combined_settings} presents the differences between the sample standard deviations of the estimators by approach A and approach B when $G(\cdot|\mathbf{Z})$ is known. The red horizontal line at zero serves as a reference for equal efficiency, and positive values indicate that Approach B is more efficient than Approach A as $t_{0}$ rises, corresponding to a higher censoring rate. 

Next, we compared the RSMSE of the four methods used to estimate the censoring distribution in Table ~\ref{table: sim1set11:result} and Figure~\ref{fig: sim11_all_comp_srf}, which considers both bias and variance at the same time. The true CDF serves as a benchmark for the other methods. With Approach A, the well-tuned SRF and the gap-time Cox model consistently perform well, closely aligning with the True CDF at earlier ages (21, 30, and 35). The standard Cox model also performs adequately in early ages; however, it performs poorly at age 40 across all coefficients, where its RSMSE spikes dramatically. This shows that the standard Cox model is highly unstable at later time points compared to the SRF and the gap-time Cox model, which is due to the failure of the proportional hazards assumption. When using Approach B, the SRF and gap-time Cox model have good performance, almost perfectly mirroring the True CDF for $\alpha_{1}(t)$ and $\beta_{1}(t)$ across all ages. Across both approaches, the stratified ECDF demonstrates notably strong performance for $Z_2(t)$, achieving the lowest or near-lowest RMSE at most ages. However, its performance for the intercept and $Z_1$ is approach-dependent. Overall, these results suggest that the gap-time Cox model and the well-tuned SRF provide reliable estimation across most settings, while the standard Cox model should be used with caution at later time points where proportional hazards assumptions are most likely violated.

In conclusion, both approaches provide a consistent estimator when the censoring distribution is estimated well. When the censoring rate is high, Approach B is more efficient than Approach A when using a stratified ECDF, a well-tuned SRF, a standard Cox model, or a gap-time Cox model to estimate the censoring distribution. The results underscore the importance of modeling the censoring distribution well to achieve valid and efficient inference when analyzing doubly censored data.

\medskip
\noindent{\sf \underline{Simulation 1.2: Light Censoring}} We simulate the outcome data from the same model as in Simulation 1.1. We changed the values of $\psi_{3i}$ and $\psi_{4i}$ to yield a much lower censoring rate. The simulation results are presented in Supplementary \ref{supp: restul_of_simulation_set12}. When the censoring distribution is estimated well, both approaches A and B provide consistent estimates, while Approach A was more efficient. Furthermore, the RMSE for $\beta_{2}(t)$ is high when the censoring distribution is estimated using either the standard or the gap-time Cox model. This is due to the dependence of the shape parameter $\psi_{3i}$ (for the $C^{*}$ distribution) on the covariate $Z_{2}$, which results in a violation of the proportional hazards assumption even for the gap-time Cox model.

Simulations 1.1 and 1.2 indicate that the relative performance of the two proposed approaches depends on the censoring distribution which is consistent with previous findings for right censored data \citep{blanche2023logistic}. 

\subsubsection{Simulation 2: Examining Proposed Approaches under a Mixture of Logistic Distributions}

This simulation study was conducted to serve two purposes: (i) to verify the necessity of accounting for left-censoring, and (ii) to investigate the robustness of the proposed approaches against outcome model misspecification. 

The event time was generated from a mixture of two logistic regression models depending on the age of cancer diagnosis:
\begin{align}\label{model sim2: true model}
\begin{split}
\log\left( \frac{P(T\le t|\mathbf{Z}, T > V)}{1-P(T \le t|\mathbf{Z}, T > V,)} \right)=
\begin{cases}
    \alpha_{1}(t)+\beta_1(t)Z_1+\beta_{21}(t)Z_2, \quad\quad\quad\text{ when $0 \le V<16$},\\
    \alpha_{2}(t)+\beta_1(t)Z_1+\beta_{22}(t)Z_2, \quad\quad\quad\text{when $16 \le V\le 21$},
\end{cases}
\end{split}
\end{align}
where $\alpha_{1}(t)= \gamma_{01} + \gamma_{1}t$ with $\gamma_{01} = -6.3$, $\gamma_{1}=0.30$, $\beta_{1}(t) = \beta_{1}$, and $\beta_{21}(t) = \beta_{21}$. We set $\beta_{1}=-0.36$ and $\beta_{21}=1.00$. In addition, $\alpha_{2}(t)= \gamma_{02} + \gamma_{1} t$. We set $\gamma_{02}=-6.90$, $\beta_{22}(t) = \beta_{22} = 1.60$.

In this simulation, we had an intermediate step, denoted \textit{Step 1b}, after Step 1 of algorithm \ref{alg:data_sim} for the necessity of adjusting the risk set by comparing the Approach by Im et al and approaches A and B at $t_{0}$ = 13, 14, and 15. When employing SRF, both the number of trees and the node size are set to 100 in simulation 2.1, and the node size changes to 500 in simulation 2.2.

\medskip
\noindent{\sf \underline{Simulation 2.1: Risk Set Adjustment}} Table~\ref{table: sim2:result} presents the results using approaches A and B and the Approach by Im et al. It demonstrates that both approaches A and B yield unbiased estimates. The RSMSEs for both approaches are consistently low and comparable across all methods used to estimate the censoring distribution. This is due to the very low censoring rates at these ages, resulting in $\hat{G}(\cdot|\mathbf{Z})/\hat{G}(\cdot)$ close to one for all subjects. 

In contrast, the Approach by Im et al exhibits substantial bias in the estimation of $\beta_1(t_{0})$, as reflected by larger deviations of SMEANs from the true values and higher RSMSEs. The large bias in $\beta_{1}(t_{0})$ occurs because the method includes subjects into the analysis before they are actually at risk. Since a subject's cancer diagnosis age equals $21 \times Z_{1}$, this mistake specifically adds too many people with high $Z_{1}$ values to the data. Because these wrongly included subjects cannot yet experience the event, the model interprets the association between high $Z_{1}$ value and the absence of events as evidence that $Z_{1}$ is more protective than it truly is, producing a substantial negative bias in $\hat{\beta}_1(t_{0})$. On the other hand, $Z_{2}$ is independent of $Z_{1}$. Consequently, the estimate of $\beta_2$ remains unbiased. The intercept $\alpha_1$ shows moderate bias, as the misincluded subjects contribute additional zero outcomes to the estimating equation, but the effect is less pronounced than for $\beta_1$

\medskip
\noindent{\sf \underline{Simulation 2.2. Robustness against model misspecification}} 
When SRF, gap-time Cox model, or stratified ECDF is used to estimate the censoring distribution (Figure~\ref{fig: sim22a_srf_ci}, \ref{fig: sim22a_coxcplus_ci}, and\ref{fig: sim22a_ecdf_ci} in Supplementary Materials), both approaches A and B yield survival probability estimates that closely align with the true survival probability across all time points, indicating that both approaches are consistent and efficient in this setting. When the standard Cox model is used to estimate the censoring distribution (Figure~\ref{fig: sim22a_cox_ci}), Approach B shows an increasing bias as $t_0$ increases, while Approach A remains comparatively robust and estimates are closer to the true survival probability.

Overall, the results in Table \ref{table: sim2:result} demonstrate that proper risk set adjustment is essential in doubly censored data. Otherwise, as in the Approach by Im et al, it may lead to substantial bias in parameter estimation. Both approaches A and B, which incorporate risk set adjustment, yield unbiased and efficient estimates under low censoring rates. The choice of method for estimating the censoring distribution has minimal impact on estimates of regression coefficients when censoring rate is low. However, Figures~\ref{fig: sim22a_srf_ci}--\ref{fig: sim22a_tcdf_ci} show the necessity of correctly specifying the censoring model in IPCW-based methods when the outcome model is mis-specified.


\subsubsection{Simulation 3: Robustness of Proposed Approaches when Age at POI Follows a Weibull distribution}
Lastly, we generated event times from a Weibull distribution using Cox PH model, with the hazard function taking the form $\lambda\nu t^{\nu-1}\exp\{\bm{\beta}(t)^{T}\mathbf{Z}\} $\citep{bender2005generating,austin2012generating}. We set $\lambda = 4.50\times 10^{-9}$, $\nu=5.00$, and coefficients $\beta_{1}(t)=\beta_{1}= 2.00$ and $\beta_{2}(t)=\beta_{2} =-0.30$. 

The estimated survival probability using either Approach A or B was close to the true survival curve across all $t_{0}$ when SRF, the gap-time Cox model, or the stratified ECDF is used to estimate the censoring distribution (Figures~\ref{fig: sim31_srf_ci}, \ref{fig: sim31_coxcplus_ci}, and \ref{fig: sim31_ecdf_ci} in Supplementary Material). The number of trees is set to 100 and the node size is set to 500 in SRF. This result demonstrates that these three methods provide reliable $\hat{G}(\cdot|\mathbf{Z})$ in this context, yielding consistent estimators for both approaches. In contrast, when the standard Cox model was used to estimate the censoring distribution (Figure \ref{fig: sim31_cox_ci}), Approach A showed bias at age 40 where the censoring rate was higher.

Furthermore, the initial data-generating algorithm only generates data at each selected $t_{0}$, where the true event time is only generated for subjects satisfying $I(T_i \le t_0) = 1$. To validate this approach, we consider an alternative algorithm that generates $T_i$ for all subjects, as detailed in Algorithm \ref{alg: another_data_sim}. We applied this alternative algorithm to replicate the scenarios in Simulation 3 and computed the estimated survival probabilities for all methods across both approaches. As shown in Figure \ref{fig: sim32_comapre}, the differences between the two algorithms are negligible.

In conclusion, simulations 2.2 and 3 show that estimating the censoring distribution becomes more important when event times follow a non-logistic distribution. Poor estimation of the censoring distribution can easily lead to biased results. Finally, we confirm the validity of our proposed data-generating process.

\section{Final Remarks} \label{sec5: disscusion}

This paper proposed two IPCW-based estimating function approaches for age-specific logistic regression in order to handle doubly censored event times. Left censoring is accommodated by modifying the risk sets at analysis time $t_{0}$, while right censoring is addressed using IPCW. Our primary contributions are the adaptation and evaluation of the two proposed approaches and assessing the impact of different censoring distribution estimation methods on the coefficient estimation through simulation studies.

There are four major findings. Firstly, proper risk set adjustment is essential for valid inference; approaches that fail to do so will yield biased coefficient estimates. Secondly, when the censoring rate is high, approach B is more efficient than approach A when the censoring distribution is well estimated (e.g., well-tuned SRF). When the censoring rate is low, approach A is more efficient when the true censoring distribution is well estimated. The superior performance of approach B when censoring rate is high is due to its use of more information. This pattern suggests that relative efficiency is dynamic and depends on the censoring rate at the analysis age $t_0$. Thirdly, non-parametric methods, such as SRF, are preferred for estimating the censoring distribution as they do not rely on the proportional hazards assumption. While the stratified ECDF is an alternative, it becomes impractical as the number of continuous covariates increases and requires a sufficient sample size within each stratum to ensure estimation accuracy. Conversely, although SRF avoids these issues, it necessitates careful hyperparameter tuning; suboptimal parameter selection can lead to biased coefficient estimates and unreliable standard errors. Lastly, both the coefficients and their standard error estimates are affected by $\hat{G}(\cdot|\mathbf{Z})$. The sandwich standard error estimate is reasonable when $G(\cdot\mid\mathbf{Z})$ is estimated well.


There are several limitations in this study that present avenues for future investigation. Firstly, our approaches did not consider competing risks. The age at SPM is treated as part of the censoring age in the current study; however, subjects experiencing SPM first cannot have POI afterwards. Secondly, the estimation of the coefficient functions, $\alpha(t_{0})$ and $\mathbf{\beta}(t_{0})$, were performed in two steps by smoothing the point-wise estimated coefficients at a set of chosen $t_{0}$. To address these two issues, future research could focus on extending these approaches to a competing risks framework and developing an integrated, one-step procedure for estimating the coefficient function. One possible approach would be to incorporate penalized splines or kernel-based generalized estimating equation. Thirdly, exploring the extension of the proposed approaches to a continuous-time scale would be valuable. Fourthly, machine learning models such as the survival support vector machine or the super learner for survival prediction could be used to estimate the censoring distribution. Furthermore, more investigation on the use of SRF is needed. Previous research has identified the importance of the loss function used for tuning hyperparameters in SRF \citep{berkowitz2025targeted}. We did not investigate any other loss functions besides RSMSE. Lastly, we did not evaluate the performance of the five considered methods for estimating the censoring distribution in high-dimensional settings; however, the advantages of the RSF approach are expected to be more pronounced in scenarios involving a larger number of covariates. In summary, this study provides a framework for modeling doubly censored event times and offers guidance on selecting the proper approach and method to estimate censoring distribution.


\section*{Acknowledgments}
The authors are grateful to Yunshan (Daisy) Dai for proofreading the manuscript.

\section*{Financial disclosure}
This work was supported in part by Simon Fraser University through Graduate Fellowships and a PhD Research Scholarship awarded to Haoxuan (Charlie) Zhou. Research utilizing the Childhood Cancer Survivor Study data under Professor Yan Yuan is supported by the National Cancer Institute (U24CA55727), a resource that facilitates research on long-term childhood and adolescent cancer survivors.

\section*{Conflict of interest}

The authors declare no potential conflicts of interest.

\bibliography{wileyNJD-AMA}

@article{mostoufi2016endocrine,
  title={Endocrine abnormalities in aging survivors of childhood cancer: a report from the Childhood Cancer Survivor Study},
  author={Mostoufi-Moab, Sogol and Seidel, Kristy and Leisenring, Wendy M and Armstrong, Gregory T and Oeffinger, Kevin C and Stovall, Marilyn and Meacham, Lillian R and Green, Daniel M and Weathers, Rita and Ginsberg, Jill P and others},
  journal={Journal of Clinical Oncology},
  volume={34},
  number={27},
  pages={3240--3247},
  year={2016},
  publisher={American Society of Clinical Oncology}
}

@article{touraine2024premature,
  title={Premature ovarian insufficiency},
  author={Touraine, Philippe and Chabbert-Buffet, Nathalie and Plu-Bureau, Genevieve and Duranteau, Lise and Sinclair, Andrew H and Tucker, Elena J},
  journal={Nature Reviews Disease Primers},
  volume={10},
  number={1},
  pages={63},
  year={2024},
  publisher={Nature Publishing Group UK London}
}

@article{im2023development,
  title={Development and validation of age-specific risk prediction models for primary ovarian insufficiency in long-term survivors of childhood cancer: a report from the Childhood Cancer Survivor Study and St Jude Lifetime Cohort},
  author={Im, Cindy and Lu, Zhe and Mostoufi-Moab, Sogol and Delaney, Angela and Yu, Lin and Baedke, Jessica L and Han, Yutong and Sapkota, Yadav and Yasui, Yutaka and Chow, Eric J and others},
  journal={The Lancet Oncology},
  volume={24},
  number={12},
  pages={1434--1442},
  year={2023},
  publisher={Elsevier}
}

@article{robison2002study,
  title={Study design and cohort characteristics of the Childhood Cancer Survivor Study: a multi-institutional collaborative project},
  author={Robison, Leslie L and Mertens, Ann C and Boice, John D and Breslow, Norman E and Donaldson, Sarah S and Green, Daniel M and Li, Frederic P and Meadows, Anna T and Mulvihill, John J and Neglia, Joseph P and others},
  journal={Medical and Pediatric Oncology},
  volume={38},
  number={4},
  pages={229--239},
  year={2002},
  publisher={Wiley Online Library}
}

@article{robison2009childhood,
  title={The Childhood Cancer Survivor Study: a National Cancer Institute--supported resource for outcome and intervention research},
  author={Robison, Leslie L and Armstrong, Gregory T and Boice, John D and Chow, Eric J and Davies, Stella M and Donaldson, Sarah S and Green, Daniel M and Hammond, Sue and Meadows, Anna T and Mertens, Ann C and others},
  journal={Journal of Clinical Oncology},
  volume={27},
  number={14},
  pages={2308--2318},
  year={2009},
  publisher={American Society of Clinical Oncology}
}

@article{eshre2016eshre,
issn = {0268-1161},
journal = {Human reproduction (Oxford)},
pages = {926--937},
volume = {31},
publisher = {Oxford University Press},
number = {5},
year = {2016},
title = {ESHRE Guideline: management of women with premature ovarian insufficiency},
copyright = {The Author 2014. Published by Oxford University Press on behalf of the European Society of Human Reproduction and Embryology. All rights reserved. For Permissions, please email: journals.permissions@oup.com 2014},
language = {eng},
address = {England},
author = {Webber, L. and Davies, M. and Anderson, R. and Bartlett, J. and Braat, D. and Cartwright, B. and Cifkova, R. and de Muinck Keizer-Schrama, S. and Hogervorst, E. and Janse, F. and Liao, L. and Vlaisavljevic, V. and Zillikens, C. and Vermeulen, N.},
organization = {European Society for Human Reproduction and Embryology (ESHRE) Guideline Group on POI},
}

@article{mishra2017early,
  title={Early menarche, nulliparity and the risk for premature and early natural menopause},
  author={Mishra, Gita D and Pandeya, Nirmala and Dobson, Annette J and Chung, Hsin-Fang and Anderson, Debra and Kuh, Diana and Sandin, Sven and Giles, Graham G and Bruinsma, Fiona and Hayashi, Kunihiko and others},
  journal={Human Reproduction},
  volume={32},
  number={3},
  pages={679--686},
  year={2017},
  publisher={Oxford University Press}
}

@article{robins2006inverse,
  title={Inverse probability weighting in survival analysis},
  author={Robins, JM and Rotnitzky, A},
  journal={Survival and event history analysis. Chichester. UK: Wiley},
  pages={266--71},
  year={2006}
}

@book{mccullagh2019generalized,
publisher = {Taylor and Francis, an imprint of Routledge},
isbn = {9780412317606},
year = {2019},
title = {Generalized Linear Models / by P. McCullagh.},
edition = {Second},
language = {eng},
author = {McCullagh, P.},
}

@article{zheng2006application,
  title={Application of the time-dependent ROC curves for prognostic accuracy with multiple biomarkers},
  author={Zheng, Yingye and Cai, Tianxi and Feng, Ziding},
  journal={Biometrics},
  volume={62},
  number={1},
  pages={279--287},
  year={2006},
  publisher={Oxford University Press}
}

@article{uno2007evaluating,
  title={Evaluating prediction rules for t-year survivors with censored regression models},
  author={Uno, Hajime and Cai, Tianxi and Tian, Lu and Wei, Lee-Jen},
  journal={Journal of the American Statistical Association},
  volume={102},
  number={478},
  pages={527--537},
  year={2007},
  publisher={Taylor \& Francis}
}

@article{scheike2008predicting,
  title={Predicting cumulative incidence probability by direct binomial regression},
  author={Scheike, Thomas H and Zhang, Mei-Jie and Gerds, Thomas A},
  journal={Biometrika},
  volume={95},
  number={1},
  pages={205--220},
  year={2008},
  publisher={Oxford University Press}
}

@article{blanche2023logistic,
  title={On logistic regression with right censored data, with or without competing risks, and its use for estimating treatment effects},
  author={Blanche, Paul Fr{\'e}d{\'e}ric and Holt, Anders and Scheike, Thomas},
  journal={Lifetime Data Analysis},
  volume={29},
  number={2},
  pages={441--482},
  year={2023},
  publisher={Springer}
}

@article{vock2016adapting,
  title={Adapting machine learning techniques to censored time-to-event health record data: A general-purpose approach using inverse probability of censoring weighting},
  author={Vock, David M and Wolfson, Julian and Bandyopadhyay, Sunayan and Adomavicius, Gediminas and Johnson, Paul E and Vazquez-Benitez, Gabriela and O’Connor, Patrick J},
  journal={Journal of Biomedical Informatics},
  volume={61},
  pages={119--131},
  year={2016},
  publisher={Elsevier}
}

@article{cai2004semiparametric,
  title={Semiparametric regression analysis for doubly censored data},
  author={Cai, T and Cheng, S},
  journal={Biometrika},
  volume={91},
  number={2},
  pages={277--290},
  year={2004},
  publisher={Oxford University Press}
}

@article{yuan2018threshold,
  title={A threshold-free summary index of prediction accuracy for censored time to event data},
  author={Yuan, Yan and Zhou, Qian M and Li, Bingying and Cai, Hengrui and Chow, Eric J and Armstrong, Gregory T},
  journal={Statistics in Medicine},
  volume={37},
  number={10},
  pages={1671--1681},
  year={2018},
  publisher={Wiley Online Library}
}

@article{bender2005generating,
  title={Generating survival times to simulate Cox proportional hazards models},
  author={Bender, Ralf and Augustin, Thomas and Blettner, Maria},
  journal={Statistics in Medicine},
  volume={24},
  number={11},
  pages={1713--1723},
  year={2005},
  publisher={Wiley Online Library}
}

@article{austin2012generating,
  title={Generating survival times to simulate Cox proportional hazards models with time-varying covariates},
  author={Austin, Peter C},
  journal={Statistics in Medicine},
  volume={31},
  number={29},
  pages={3946--3958},
  year={2012},
  publisher={Wiley Online Library}
}

@article{berkowitz2025targeted,
  title={Targeted tuning of random forests for quantile estimation and prediction intervals},
  author={Berkowitz, Matthew and Altman, Rachel MacKay and Loughin, Thomas M},
  journal={arXiv preprint arXiv:2507.01430},
  year={2025}
}

@article{chemaitilly2017premature,
  title={Premature ovarian insufficiency in childhood cancer survivors: a report from the St. Jude Lifetime Cohort},
  author={Chemaitilly, Wassim and Li, Zhenghong and Krasin, Matthew J and Brooke, Rachel J and Wilson, Carmen L and Green, Daniel M and Klosky, James L and Barnes, Nicole and Clark, Kerri L and Farr, Jennifer B and others},
  journal={The Journal of Clinical Endocrinology \& Metabolism},
  volume={102},
  number={7},
  pages={2242--2250},
  year={2017},
  publisher={Oxford University Press}
}

@article{johnston2009normal,
  title={Normal ovarian function and assessment of ovarian reserve in the survivor of childhood cancer},
  author={Johnston, Richard J and Wallace, W Hamish B},
  journal={Pediatric blood \& cancer},
  volume={53},
  number={2},
  pages={296--302},
  year={2009},
  publisher={Wiley Online Library}
}

@article{cox1972regression,
  title={Regression models and life-tables},
  author={Cox, David R},
  journal={Journal of the Royal Statistical Society: Series B (Methodological)},
  volume={34},
  number={2},
  pages={187--202},
  year={1972},
  publisher={Wiley Online Library}
}

@article{ishwaran2008random,
issn = {1932-6157},
journal = {The Annals of Applied Statistics},
pages = {841--860},
volume = {2},
publisher = {Institute of Mathematical Statistics},
number = {3},
year = {2008},
title = {Random Survival Forests},
copyright = {Copyright 2008 Institute of Mathematical Statistics},
language = {eng},
address = {Cleveland, OH},
author = {Ishwaran, Hemant and Kogalur, Udaya B. and Blackstone, Eugene H. and Lauer, Michael S.},
}

@article{suresh2022survival,
  title={Survival prediction models: an introduction to discrete-time modeling},
  author={Suresh, Krithika and Severn, Cameron and Ghosh, Debashis},
  journal={BMC Medical Research Methodology},
  volume={22},
  number={1},
  pages={207},
  year={2022},
  publisher={Springer}
}

@incollection{robins1992recovery,
  title={Recovery of information and adjustment for dependent censoring using surrogate markers},
  author={Robins, James M and Rotnitzky, Andrea},
  booktitle={AIDS Epidemiology},
  pages={297--331},
  year={1992},
  publisher={Springer}
}

@article{betensky2015recognizing,
  title={Recognizing the problem of delayed entry in time-to-event studies: better late than never for clinical neuroscientists},
  author={Betensky, Rebecca A and Mandel, Micha},
  journal={Annals of Neurology},
  volume={78},
  number={6},
  pages={839--844},
  year={2015},
  publisher={Wiley Online Library}
}






\end{document}


\maketitle
\tableofcontents
\newpage
\renewcommand{\thealgorithm}{S\arabic{algorithm}}

Section \ref{supp: derv_diff_asy_var} presents the derivation of the difference of the asymptotic variances of the two estimators when the conditional survival function of the censoring time $G(\cdot\mid\mathbf{Z})$ is known. The additional real data analysis results are presented in Section \ref{supp: real_data}. The additional simulation results of Simulations 1, 2, and 3 are reported in Section \ref{supp: restul_of_simulation_set1}, \ref{supp: restul_of_simulation_set2}, and \ref{supp: restul_of_simulation_set3}, respectively.
%
%
%
%
\section{Derivation of difference in asymptotic variance between the two approaches}\label{supp: derv_diff_asy_var}
Given $G(\cdot\mid\mathbf{Z})$ is known, we first derive the asymptotic variances of the estimators obtained using approaches A and B. We then show the difference between these two asymptotic variances in Equation \ref{eq: diff asy var full}.
\begin{align}\label{eq: diff two estimation functions}
\begin{split}
        \Delta(\bm{\theta};t_{0}|G) 
        &=\bm{U}_{A}(\alpha,\bm{\beta};t_{0}|G) - \bm{U}_{B}(\alpha,\bm{\beta};t_{0}|G) \\
        & = \displaystyle\sum_{i=1}^{n} \mathbb{I}(t_{0} \ge V_{i})
        \begin{bmatrix}
           1 \\
           \mathbf{Z}_{i} 
        \end{bmatrix}
        \left(1- W_{i}(t_{0};G)\right)
        \frac{\exp\{\alpha(t_{0})+\bm{\beta}(t_{0})^{T}\mathbf{Z}_{i}\}}{\exp\{\alpha(t_{0})+\bm{\beta}(t_{0})^{T}\mathbf{Z}_{i}\}+1}
\end{split}
\end{align}
We first show the derivation of $\Sigma_{A}(\bm{\theta};G)$, $\Sigma_{B}(\bm{\theta};G)$, and $\Sigma_{\Delta}$ below
\begin{align}
\begin{split}
\Sigma_{A}(\bm{\theta};G) 
&= Var\left( \mathbb{I}(t_{0} \ge V_{i})
\begin{bmatrix}
           1 \\
           \mathbf{Z}_{i}  
\end{bmatrix}
W_{i}(t_{0};G)
\left[
\mathbb{I}(T_{i} \le t_{0}) -
\frac{\exp\{\alpha(t_{0})+\bm{\beta}(t_{0})^{T}\mathbf{Z}_{i} \}}{\exp\{\alpha(t_{0})+\bm{\beta}(t_{0})^{T}\mathbf{Z}_{i} \}+1}
\right]\right)\\
&= E\left( \mathbb{I}(t_{0} \ge V_{i})
\begin{bmatrix}
           1 \\
           \mathbf{Z}_{i}  
\end{bmatrix}
\begin{bmatrix}
           1 \\
           \mathbf{Z}_{i}  
\end{bmatrix}^{T}
E\left\{
\left(
W_{i}(t_{0};G)
\left[
\mathbb{I}(T_{i} \le t_{0}) -
\frac{\exp\{\alpha(t_{0})+\bm{\beta}(t_{0})^{T}\mathbf{Z}_{i} \}}{\exp\{\alpha(t_{0})+\bm{\beta}(t_{0})^{T}\mathbf{Z}_{i} \}+1}
\right]
\right)^{2}
| \mathbf{Z}_{i} , V_{i}
\right\}
\right)
\\
\Sigma_{B}(\bm{\theta};G)  
&= Var\left(
\mathbb{I}(t_{0} \ge V_{i})
        \begin{bmatrix}
           1 \\
           \mathbf{Z}_{i} 
         \end{bmatrix}
        \left[
       \mathbb{I}(T_{i} \le t_{0}) W_{i}(t_{0};G)
         -
        \frac{\exp\{\alpha(t_{0})+\bm{\beta}(t_{0})^{T}\mathbf{Z}_{i}\}}{\exp\{\alpha(t_{0})+\bm{\beta}(t_{0})^{T}\mathbf{Z}_{i}\}+1}
        \right]
\right)\\
&= E\left( \mathbb{I}(t_{0} \ge V_{i})
\begin{bmatrix}
           1 \\
           \mathbf{Z}_{i}  
\end{bmatrix}
\begin{bmatrix}
           1 \\
           \mathbf{Z}_{i}  
\end{bmatrix}^{T}
E\left\{
\left(
       \mathbb{I}(T_{i} \le t_{0}) W_{i}(t_{0};G)
         -
        \frac{\exp\{\alpha(t_{0})+\bm{\beta}(t_{0})^{T}\mathbf{Z}_{i}\}}{\exp\{\alpha(t_{0})+\bm{\beta}(t_{0})^{T}\mathbf{Z}_{i}\}+1}
\right)^{2}|\mathbf{Z}_{i} , V_{i}
\right\}
\right)\\
\Sigma_{\Delta}(\bm{\theta};G)  
&= Var\left(
\mathbb{I}(t_{0} \ge V_{i})
        \begin{bmatrix}
           1 \\
           \mathbf{Z}_{i} 
        \end{bmatrix}
        \left(1- W_{i}(t_{0};G)\right)
        \frac{\exp\{\alpha(t_{0})+\bm{\beta}(t_{0})^{T}\mathbf{Z}_{i}\}}{\exp\{\alpha(t_{0})+\bm{\beta}(t_{0})^{T}\mathbf{Z}_{i}\}+1}
\right)\\
&= E\left( \mathbb{I}(t_{0} \ge V_{i})
\begin{bmatrix}
           1 \\
           \mathbf{Z}_{i}  
\end{bmatrix}
\begin{bmatrix}
           1 \\
           \mathbf{Z}_{i}  
\end{bmatrix}^{T}
\left(
       \frac{\exp\{\alpha(t_{0})+\bm{\beta}(t_{0})^{T}\mathbf{Z}_{i}\}}{\exp\{\alpha(t_{0})+\bm{\beta}(t_{0})^{T}\mathbf{Z}_{i}\}+1}
\right)^{2}
E\left\{
       (1- W_{i}(t_{0};G))^{2}|\mathbf{Z}_{i} , V_{i}
\right\}
\right)
\end{split}
\end{align}
Next, we can show $\Gamma_{A}(\bm{\theta};G)$ and $\Gamma_{B}(\bm{\theta};G)$ as below
\begin{align}
\begin{split}
\frac{1}{n}\frac{\partial U_{A}(\bm{\theta};t_{0}|G) }{\partial \bm{\theta}^{T}}
= 
&\frac{1}{n}
\left\{
- \displaystyle\sum_{i=1}^{n}
\mathbb{I}(t_{0} \ge V_{i})
\begin{bmatrix}
           1 \\
           \mathbf{Z}_{i}  
\end{bmatrix}
\begin{bmatrix}
           1 \\
           \mathbf{Z}_{i}  
\end{bmatrix}^{T}
W_{i}(t_{0};G)
\frac{\exp\{\alpha(t_{0})+\bm{\beta}(t_{0})^{T}\mathbf{Z}_{i}\}}{(\exp\{\alpha(t_{0})+\bm{\beta}(t_{0})^{T}\mathbf{Z}_{i}\}+1)^{2}}
\right\}
\xrightarrow{a.s.}\\
&-E\left(
\mathbb{I}(t_{0} \ge V_{i})
\begin{bmatrix}
           1 \\
           \mathbf{Z}_{i}  
\end{bmatrix}
\begin{bmatrix}
           1 \\
           \mathbf{Z}_{i}  
\end{bmatrix}^{T}
W_{i}(t_{0};G)
\frac{\exp\{\alpha(t_{0})+\bm{\beta}(t_{0})^{T}\mathbf{Z}_{i}\}}{(\exp\{\alpha(t_{0})+\bm{\beta}(t_{0})^{T}\mathbf{Z}_{i}\}+1)^{2}}
\right)
\triangleq -\Gamma_{A}(\bm{\theta};G)
\end{split}
\end{align}

\begin{align}
\begin{split}
\frac{1}{n}\frac{\partial U_{B}(\bm{\theta};t_{0}|G) }{\partial \bm{\theta}^{T}}
= 
&\frac{1}{n}
\left\{
- \displaystyle\sum_{i=1}^{n}
\mathbb{I}(t_{0} \ge V_{i})
\begin{bmatrix}
           1 \\
           \mathbf{Z}_{i}  
\end{bmatrix}
\begin{bmatrix}
           1 \\
           \mathbf{Z}_{i}  
\end{bmatrix}^{T}
\frac{\exp\{\alpha(t_{0})+\bm{\beta}(t_{0})^{T}\mathbf{Z}_{i}\}}{(\exp\{\alpha(t_{0})+\bm{\beta}(t_{0})^{T}\mathbf{Z}_{i}\}+1)^{2}}
\right\}
\xrightarrow{a.s.}\\
&-E\left(
\mathbb{I}(t_{0} \ge V_{i})
\begin{bmatrix}
           1 \\
           \mathbf{Z}_{i}  
\end{bmatrix}
\begin{bmatrix}
           1 \\
           \mathbf{Z}_{i}  
\end{bmatrix}^{T}
\frac{\exp\{\alpha(t_{0})+\bm{\beta}(t_{0})^{T}\mathbf{Z}_{i}\}}{(\exp\{\alpha(t_{0})+\bm{\beta}(t_{0})^{T}\mathbf{Z}_{i}\}+1)^{2}}
\right)
\triangleq -\Gamma_{B}(\bm{\theta};G)
\end{split}
\end{align}
The $\Gamma_{A}(\bm{\theta};G)= \Gamma_{B}(\bm{\theta};G)$ since $E(W_{i}(t_{0};G)|T_{i},\mathbf{Z}_{i}  ) = 1$.
We then shown
\begin{align}
\begin{split}
\frac{1}{n}\frac{\partial \Delta(\bm{\theta};t_{0}|G) }{\partial \bm{\theta}^{T}}
= 
&\frac{1}{n}
\left\{
\displaystyle\sum_{i=1}^{n}
\mathbb{I}(t_{0} \ge V_{i})
\begin{bmatrix}
           1 \\
           \mathbf{Z}_{i}  
\end{bmatrix}
\begin{bmatrix}
           1 \\
           \mathbf{Z}_{i}  
\end{bmatrix}^{T}
(1-W_{i}(t_{0};G))
\frac{\exp\{\alpha(t_{0})+\bm{\beta}(t_{0})^{T}\mathbf{Z}_{i}\}}{(\exp\{\alpha(t_{0})+\bm{\beta}(t_{0})^{T}\mathbf{Z}_{i}\}+1)^{2}}
\right\}
\xrightarrow{a.s.}\\
&E\left(
\mathbb{I}(t_{0} \ge V_{i})
\begin{bmatrix}
           1 \\
           \mathbf{Z}_{i}  
\end{bmatrix}
\begin{bmatrix}
           1 \\
           \mathbf{Z}_{i}  
\end{bmatrix}^{T}
(1-W_{i}(t_{0};G))
\frac{\exp\{\alpha(t_{0})+\bm{\beta}(t_{0})^{T}\mathbf{Z}_{i}\}}{(\exp\{\alpha(t_{0})+\bm{\beta}(t_{0})^{T}\mathbf{Z}_{i}\}+1)^{2}}
\right)
=0
\end{split}
\end{align}
As a result, we can show 
\begin{align}
\begin{split}\label{eq: diff asy var full}
    &AV_{A}(\bm{\theta};G) - AV_{B}(\bm{\theta};G) 
     = \Gamma_{A}^{-1}(\bm{\theta};G)\Sigma_{A}(\bm{\theta};G)\left(\Gamma_{A}^{-1}(\bm{\theta};G)\right)^{T} -
    \Gamma_{B}^{-1}(\bm{\theta};G)\Sigma_{B}(\bm{\theta};G)\left(\Gamma_{B}^{-1}(\bm{\theta};G)\right)^{T}\\
    &= \Gamma_{A}^{-1}(\bm{\theta};G) 
       \left(\Sigma_{\Delta}(\bm{\theta}; G) + 
        E\left[
        2\mathbb{I}(t_{0} \ge V_{i})
        \begin{bmatrix}
           1 \\
           \mathbf{Z}_{i}  
        \end{bmatrix}
        \begin{bmatrix}
           1 \\
           \mathbf{Z}_{i}  
        \end{bmatrix}^{T} 
        \frac{\exp\{\alpha(t_{0})+\bm{\beta}(t_{0})^{T}\mathbf{Z}_{i}\}}{\exp\{\alpha(t_{0})+\bm{\beta}(t_{0})^{T}\mathbf{Z}_{i}\}+1}\right. \right. \\
        &\quad \left. \left. 
        \left\{\frac{\exp\{\alpha(t_{0})+\bm{\beta}(t_{0})^{T}\mathbf{Z}_{i}\}}{\exp\{\alpha(t_{0})+\bm{\beta}(t_{0})^{T}\mathbf{Z}_{i}\}+1}-
        E\left(
        W_{i}(t_{0};G)^{2}I(T_{i}<t_{0})|\mathbf{Z}_{i},V_{i}
        \right)
        \right\}\right] \right) 
       \left(\Gamma_{A}^{-1}(\bm{\theta};G)\right)^{T}    
\end{split}
\end{align}

While $AV_{A}(\bm{\theta};G) - AV_{B}(\bm{\theta};G)$ can be approximated via numerical integration, we instead illustrate the difference between the two approaches empirically. Specifically, we compare their sample standard deviations of estimates, as given in (\ref{eq: approx diff asy var full}), using data generated under Simulations 1.1 and 1.2; results are shown in Figures 3 and \ref{fig:sim12_combined_settings}, respectively.

\begin{align}
\begin{split}\label{eq: approx diff asy var full}
\frac{1}{n}\left\{
\left(\frac{1}{K-1}\sum_{k=1}^{K}(\tilde{\bm{\theta}}_{A}^{(k)}-\bar{\tilde{\bm{\theta}}}_{A})^{2}\right)
-
\left(\frac{1}{K-1}\sum_{k=1}^{K}(\tilde{\bm{\theta}}_{B}^{(k)}-\bar{\tilde{\bm{\theta}}}_{B})^{2}\right)
\right\}
\end{split}
\end{align}
Recall that there are 1000 repetitions in Simulations 1.1 and 1.2 (i.e., $1\le k \le K$, where $K=1000$), and there is 7000 subjects (i.e.,$n=7000$) in each repetition. The $\tilde{\bm{\theta}}_{A}^{(k)}$ and $\tilde{\bm{\theta}}_{B}^{(k)}$  are the estimated coefficients using approaches A and B at $t_{0}$, repetitively. The $\bar{\tilde{\bm{\theta}}}_{A} = \displaystyle\frac{1}{K}\sum_{k=1}^{K}\tilde{\bm{\theta}}_{A}^{(k)}$ and $\bar{\tilde{\bm{\theta}}}_{B} = \displaystyle\frac{1}{K}\sum_{k=1}^{K}\tilde{\bm{\theta}}_{B}^{(k)}$.

In the simulation repetition $k$, the $W^{(k)}_{i}(t_{0};G)
 =\frac{\mathbb{I}(U_i \le t_{0})\delta_i}{G(U_i|\mathbf{Z}_i^{(k)})} + \frac{\mathbb{I}(U_i > t_{0})}{G(t_{0}|\mathbf{Z}_i^{(k)})} $,  where $G(t_{0}|\mathbf{Z}_i^{(k)})$ and $\mathbf{Z}_{i}^{(k)}$ are the true IPCW and covariates,respectively.

\newpage
%
%
%
%
\section{Additional Result of POI Analysis}\label{supp: real_data}
\begin{figure}[H]
    \centering
    \begin{subfigure}[t]{0.6\textwidth}
        \centering
        \includegraphics[width=1\textwidth]{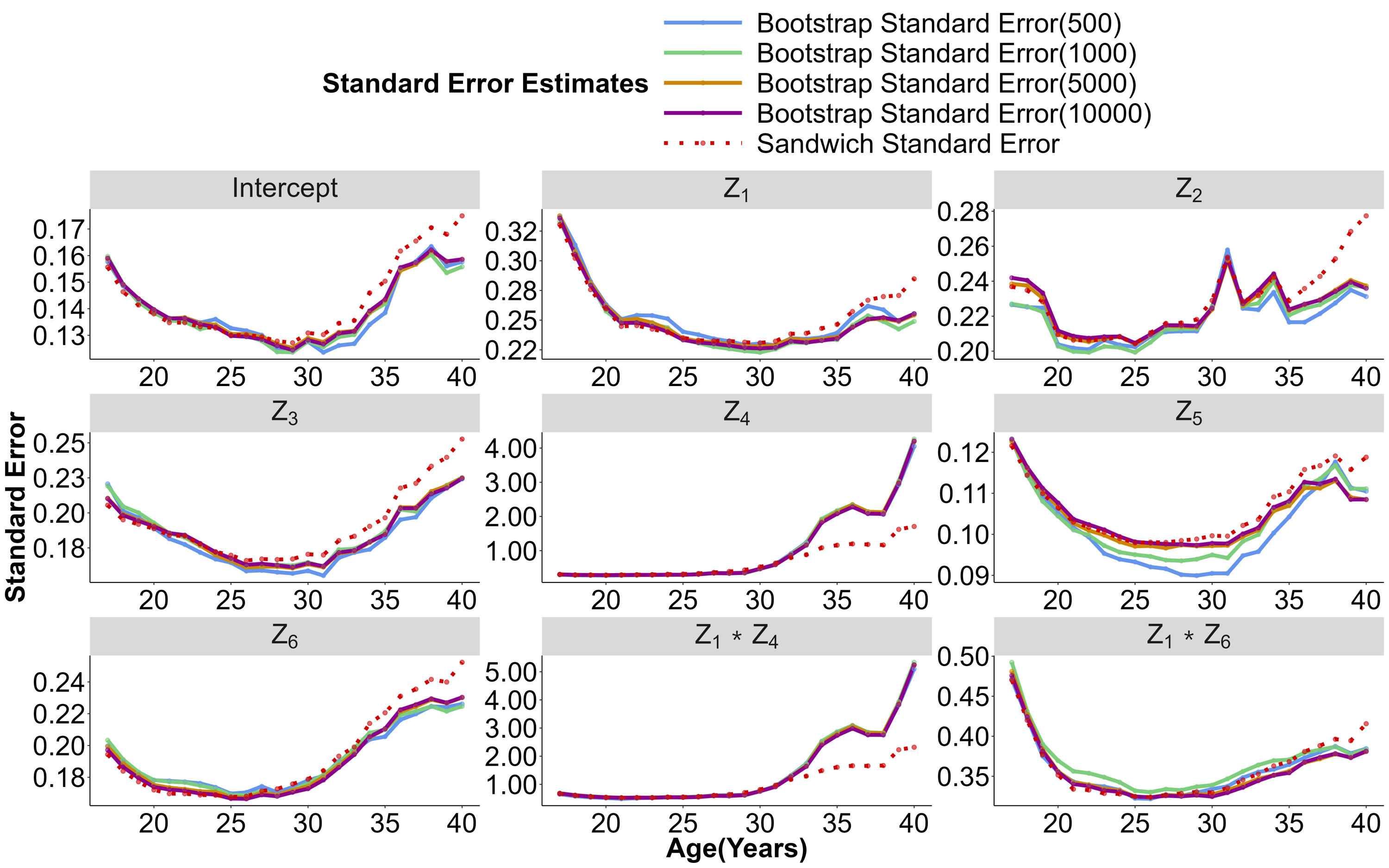}
        \caption{Approach A}
        \label{fig: compare_se_appa}
    \end{subfigure}%
    \vfill
    \begin{subfigure}[t]{0.6\textwidth}
        \centering
        \includegraphics[width=1\textwidth]{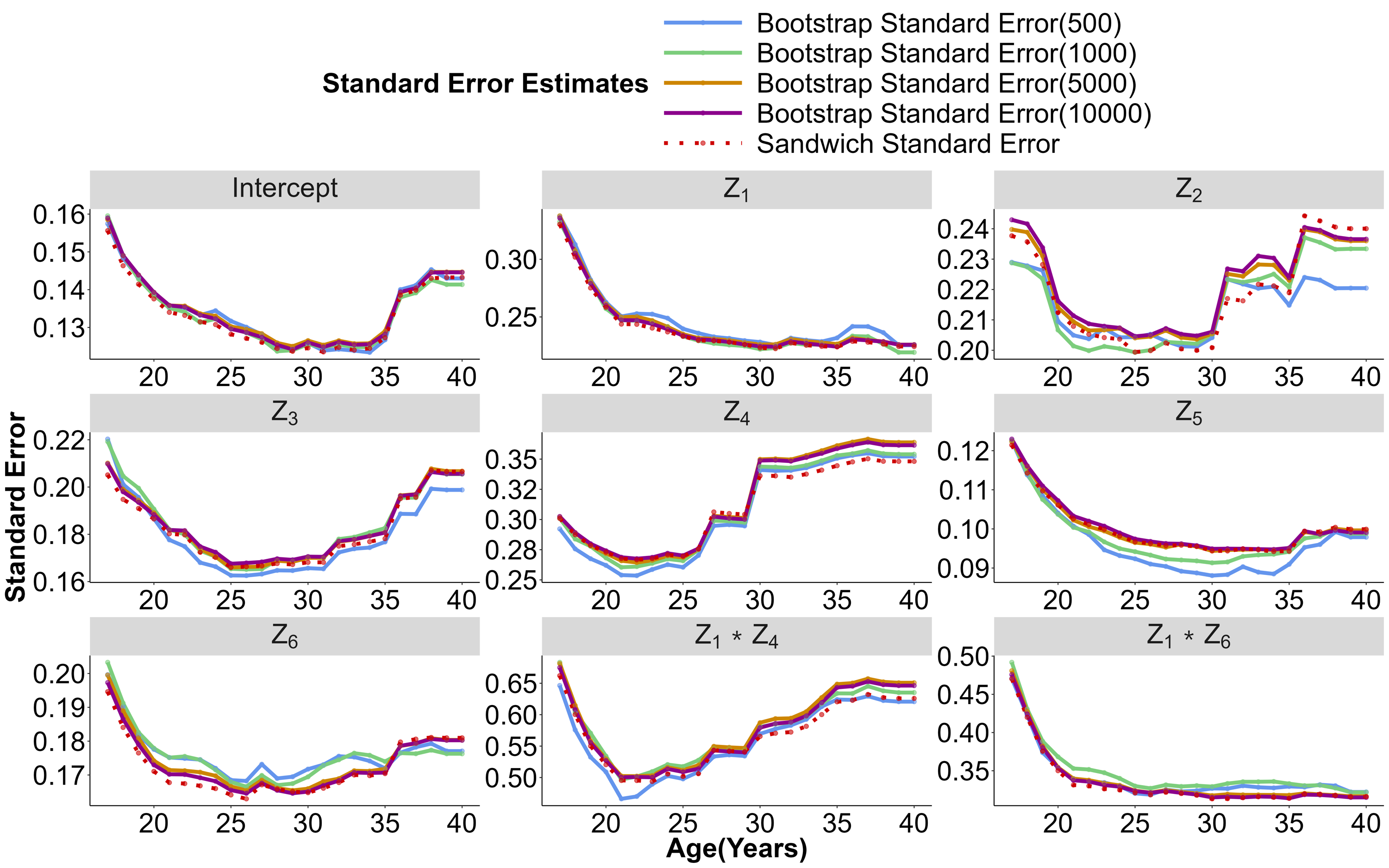}
        \caption{Approach B}
        \label{fig: compare_se_appb}
    \end{subfigure}%
    \caption{Comparison of sandwich standard error and bootstrap standard error of estimated coefficients when using both approaches. The number of bootstrap is 500,1000, 5000, and 10000. The $G(\cdot|\mathbf{Z})$ is estimated using SRF with the number of trees and the node size both set to 100. 
             $Z_{1}\text{: Rescaled age at cancer diagnosis}$,
             $Z_{2}\text{: Race-Africa American}$,
             $Z_{3}\text{: Race-Other}$,
             $Z_{4}\text{: Receipt of BMT}$,
             $Z_{5}\text{: Receipt of Alkylating agents}$
             $Z_{6}\text{: Receipt of radiation to the abdomen/pelvis/total body}$}\label{fig: compare_se_all}
\end{figure}

Figure \ref{fig: compare_se_all} compares the sandwich and bootstrap standard errors under approaches A and B for bootstrap sample sizes of 500, 1,000, 5,000, and 10,000. Under approach B, the bootstrap SEs are close to and slightly larger than the corresponding sandwich SEs, except for $Z_{2}$ when $t_{0} \ge 35$; this discrepancy will diminish as the number of bootstrap resamples increases (e.g., 50,000). Under approach A, the bootstrap and sandwich SEs are similar and slightly larger for $t_{0} \le 30$, but at later ages the bootstrap SEs tend to be slightly smaller for most coefficients. This divergence is due to the high and increasing censoring rate beyond age 30 (e.g., 32.5\% at age 30), which causes some bootstrap resamples to yield unstable estimates under approach A, particularly when the data are imbalanced (e.g., $Z_{2}$).

\begin{figure}[H]
    \centering
    \includegraphics[width=1\textwidth]{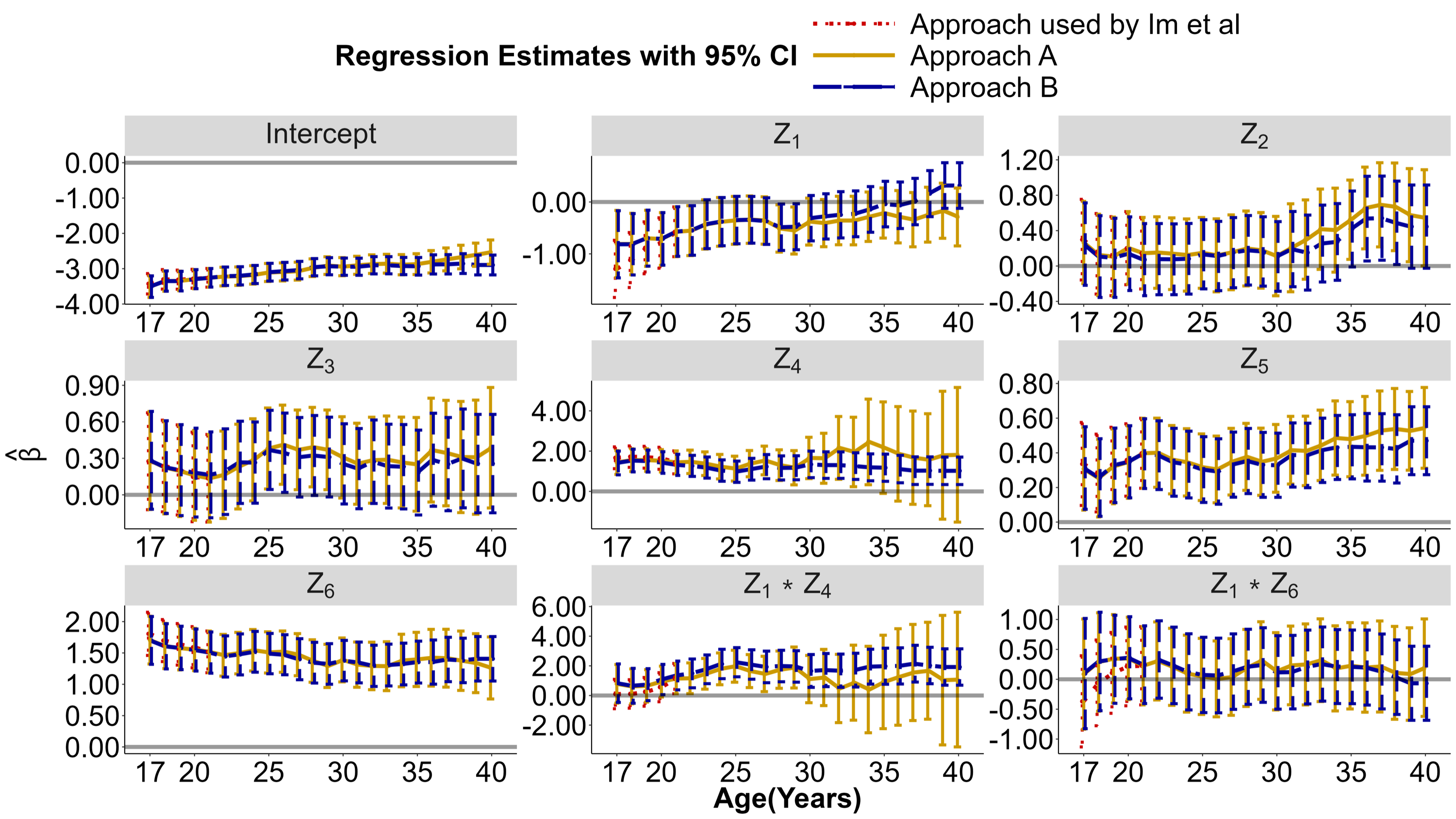}
    \caption{The estimated coefficient with $95\%$ CI, and $\hat{G}(\cdot|\mathbf{Z})$ is obtained using SRF. Following Im et al., the number of trees and the node size in SRF are set to 100. The black line denotes a horizontal line of 0. $Z_{1}\text{: Rescaled age at cancer diagnosis}$,
             $Z_{2}\text{: Race-Africa American}$,
             $Z_{3}\text{: Race-Other}$,
             $Z_{4}\text{: Receipt of BMT}$,
             $Z_{5}\text{: Receipt of Alkylating agents}$
             $Z_{6}\text{: Receipt of radiation to the abdomen/pelvis/total body}$}
             \label{fig: real_srf_comp_coef_with_se}
\end{figure}

\begin{figure}[H]
    \centering
    \includegraphics[width=0.8\textwidth]{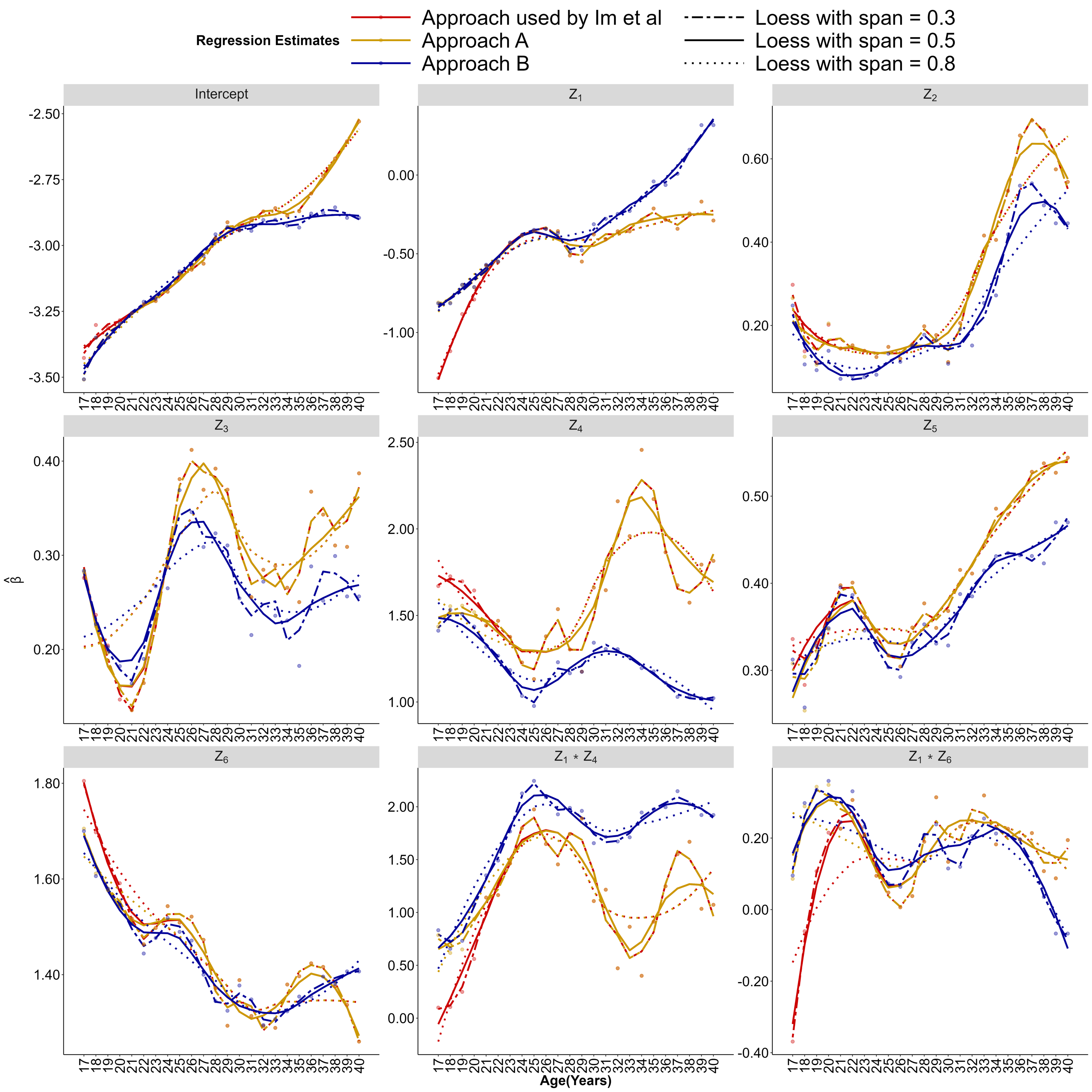}
    \captionof{figure}{The estimated coefficients were obtained using three approaches. The LOESS method was applied using three different span values: 0.3 (dash-dot line), 0.5 (solid line) and 0.8 (dotted line). The $\hat{G}(\cdot|\mathbf{Z})$ is obtained using SRF. Following Im et al., the number of trees and the node size in SRF are set to 100.
             $Z_{1}\text{: Rescaled age at cancer diagnosis}$,
             $Z_{2}\text{: Race-Africa American}$,
             $Z_{3}\text{: Race-Other}$,
             $Z_{4}\text{: Receipt of BMT}$,
             $Z_{5}\text{: Receipt of Alkylating agents}$
             $Z_{6}\text{: Receipt of radiation to the abdomen/pelvis/total body}$}\label{fig: real_srf_comp_coef_loess}
\end{figure}

\newpage
The estimated coefficients along with their $95\%$ CI using both approaches A and B is shown in Figure \ref{fig: real_srf_comp_coef_with_se}. 
\newpage
In addition, a two-step procedure is employed to estimate the coefficient functions, where the LOESS method is fitted on top of the estimated coefficient at each $t_0$. We chose span values equal to 0.3, 0.5, and 0.8. The result is shown in Figure \ref{fig: real_srf_comp_coef_loess}

\begin{figure}[H]
    \centering
    \begin{subfigure}[t]{0.9\textwidth}
        \centering
        \includegraphics[width=1\textwidth]{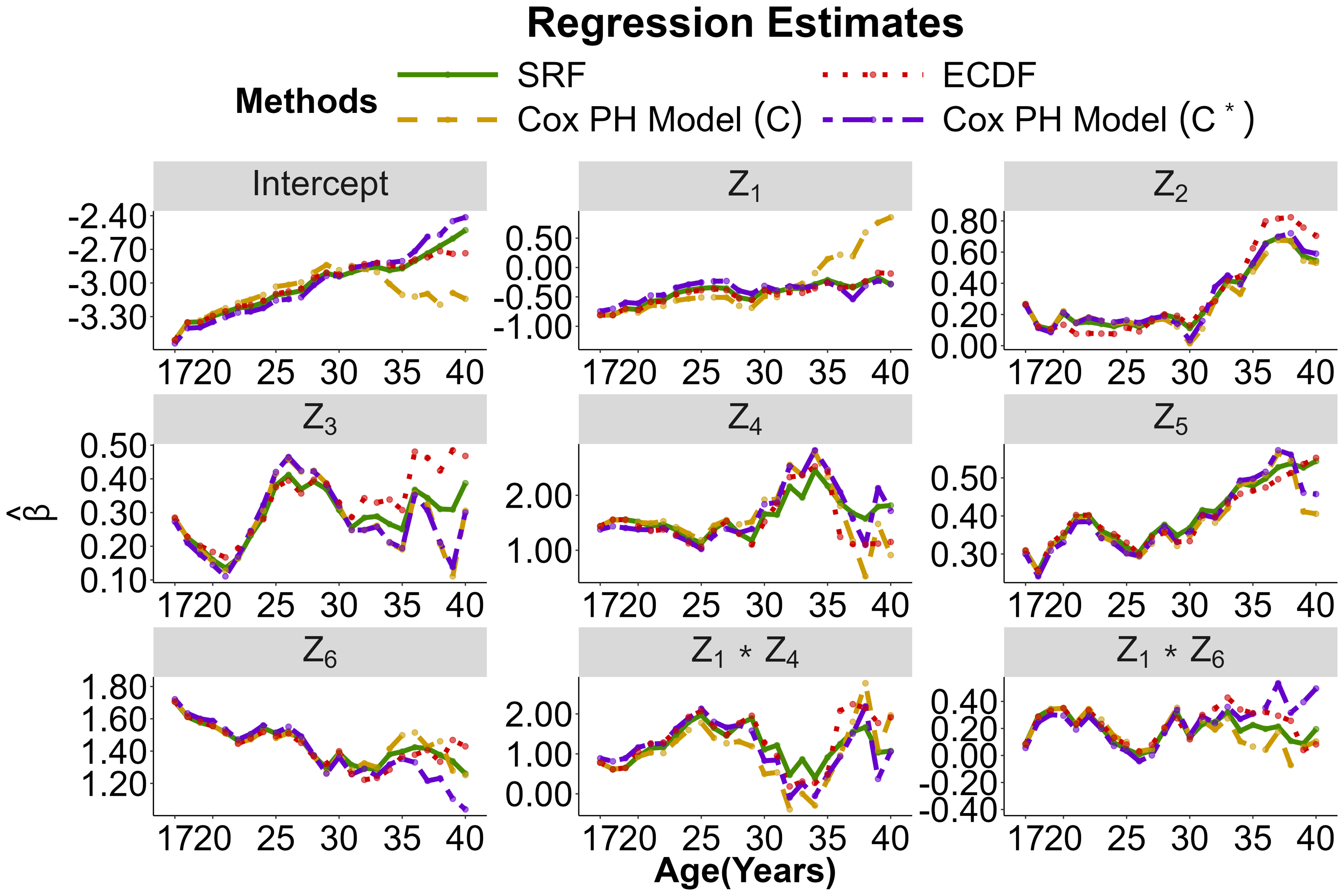}
        \caption{Estimated Coefficients using Approach A}
        \label{fig: real_appa_comp_coef}
    \end{subfigure}%
    \vfill
    \begin{subfigure}[t]{0.9\textwidth}
        \centering
        \includegraphics[width=1\textwidth]{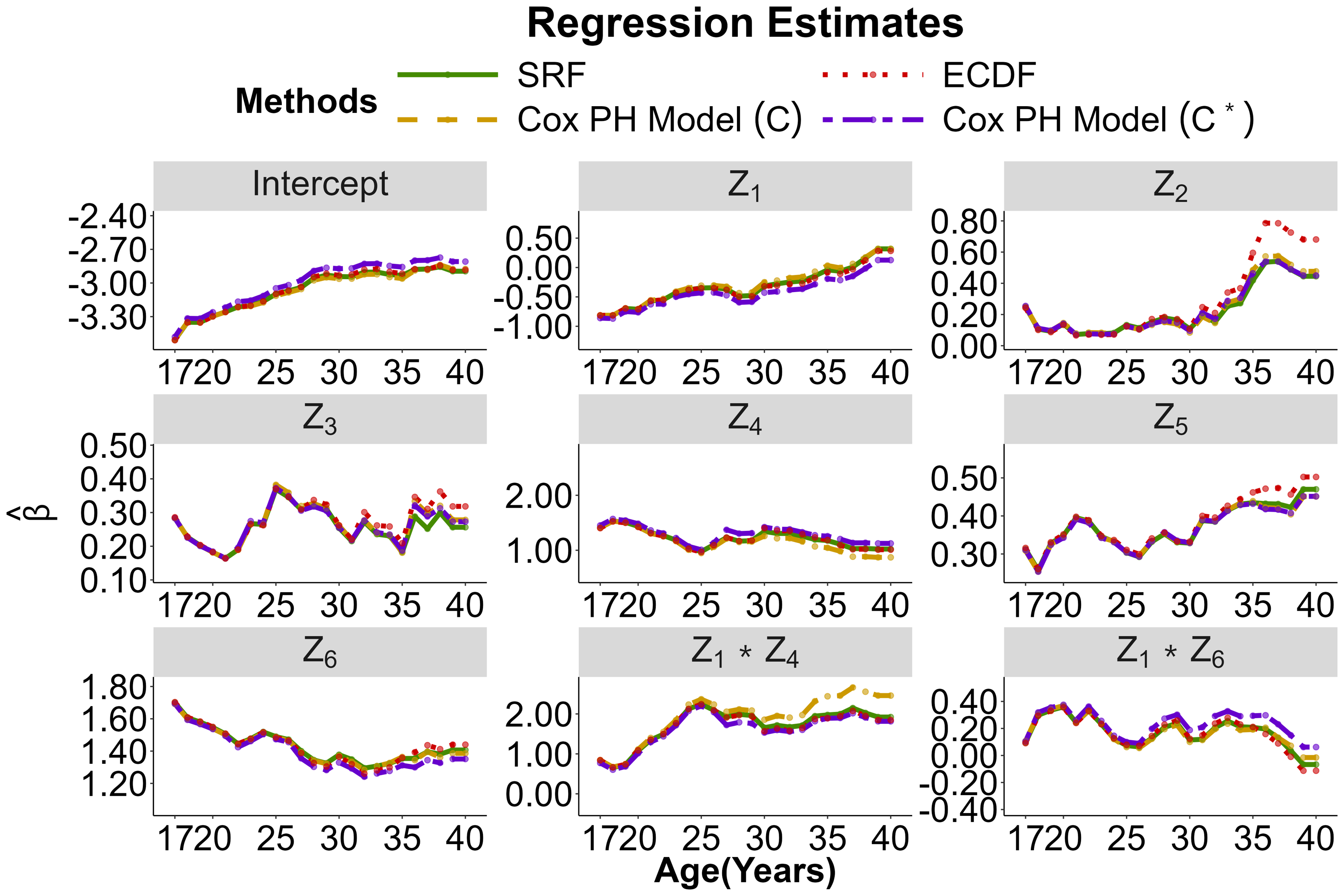}
        \caption{Estimated Coefficients using Approach B}
        \label{fig: real_appb_comp_coef}
    \end{subfigure}%
    \caption{The estimated coefficient using approach A and B with different methods to obtained $\hat{G}(\cdot|\mathbf{Z})$. The number of trees and the node size in SRF are set to 100.
             $Z_{1}\text{: Rescaled age at cancer diagnosis}$,
             $Z_{2}\text{: Race-Africa American}$,
             $Z_{3}\text{: Race-Other}$,
             $Z_{4}\text{: Receipt of BMT}$,
             $Z_{5}\text{: Receipt of Alkylating agents}$
             $Z_{6}\text{: Receipt of radiation to the abdomen/pelvis/total body}$}
\end{figure}

\begin{figure}[H]
\centering
        \begin{subfigure}[b]{0.8\textwidth}
            \centering
            \includegraphics[width=1\textwidth]{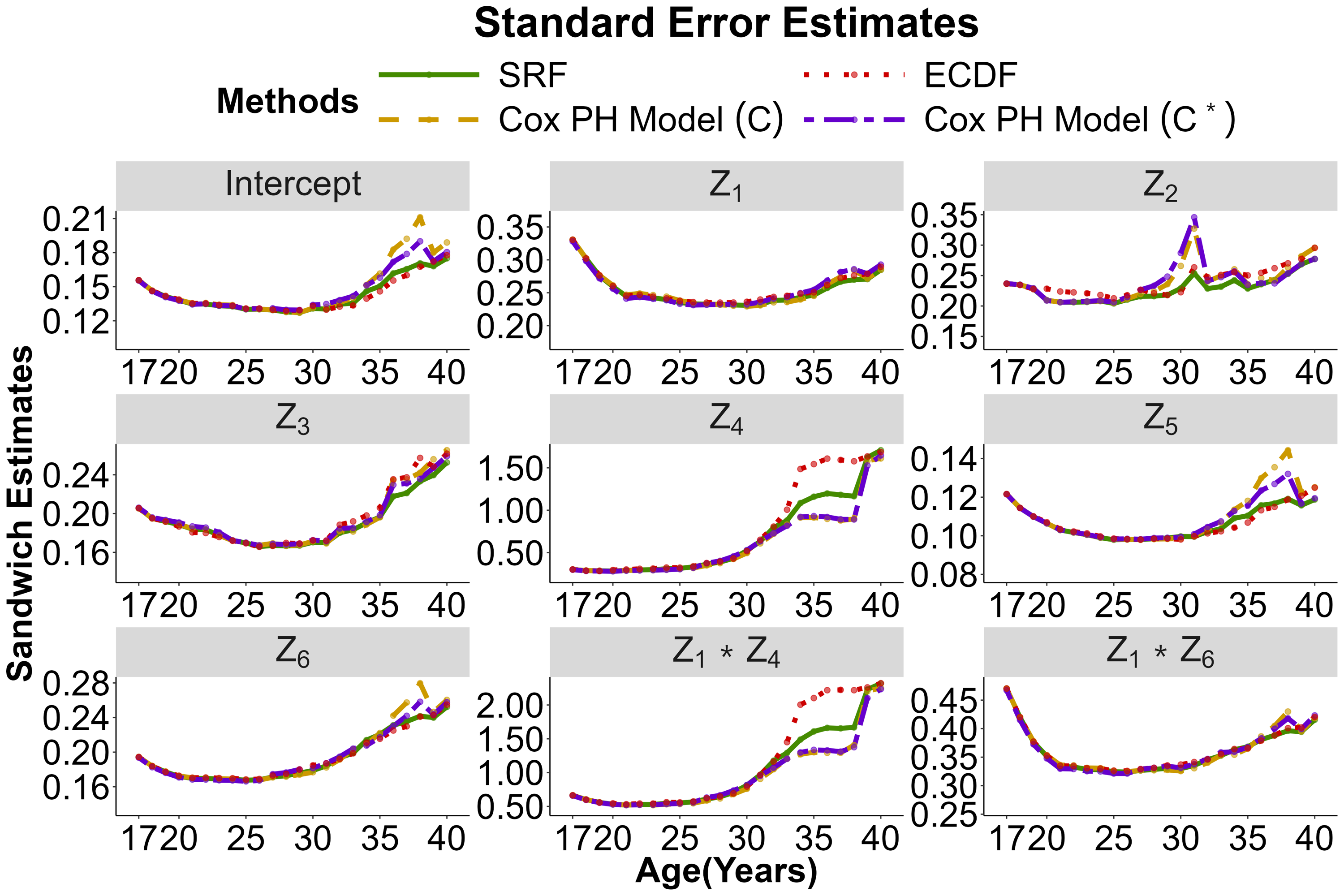}
        \caption{Standard Error Estimates using Approach A}
        \label{fig: real_appa_comp_coef_se}
        \end{subfigure}
        \vfill 
        \begin{subfigure}[b]{0.8\textwidth}
            \centering
            \includegraphics[width=1\textwidth]{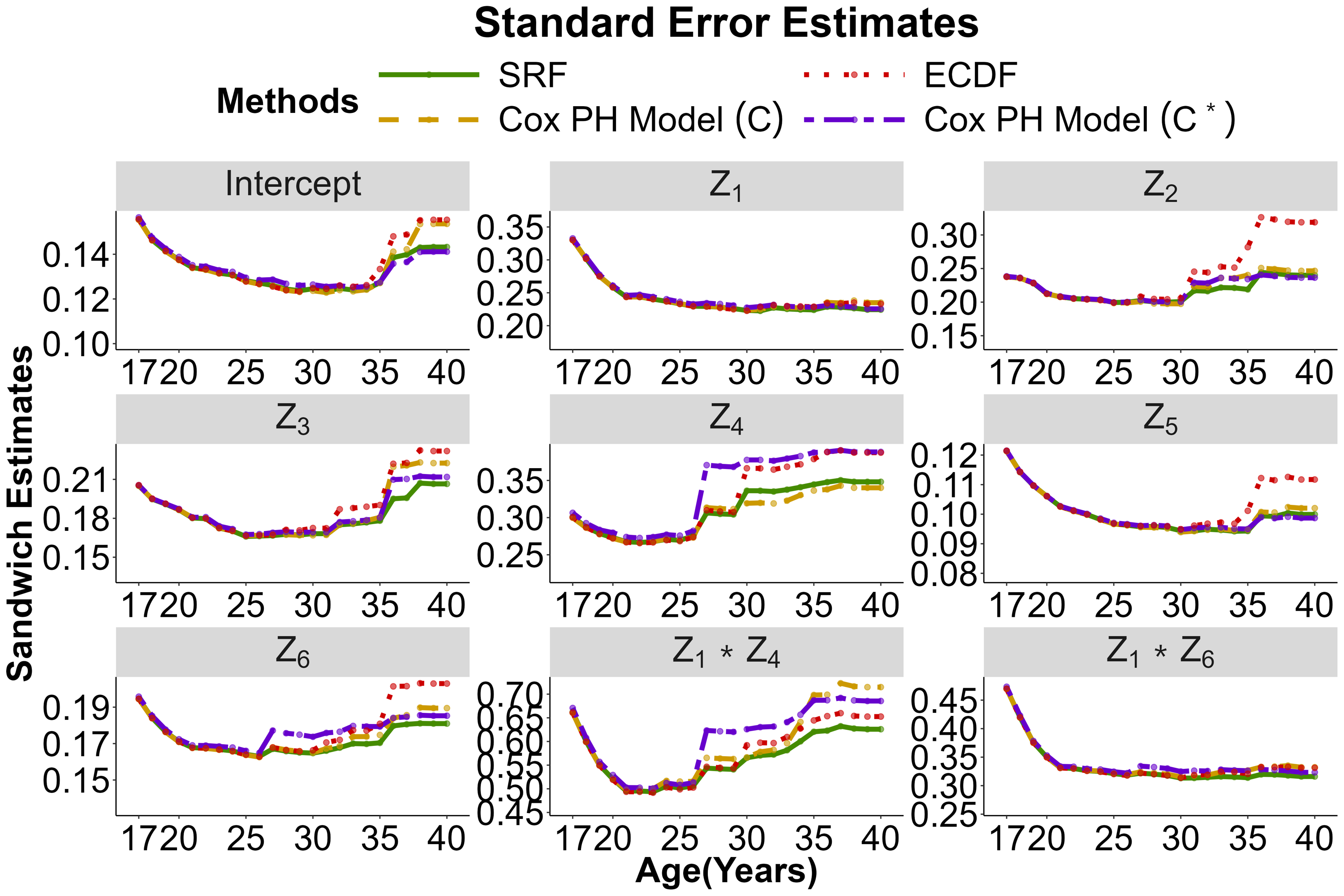}
        \caption{Standard Error Estimates using Approach B}
        \label{fig: real_appb_comp_coef_se}
        \end{subfigure}
        \caption{The standard error of estimated coefficient using approach A and B with different models to obtained $\hat{G}(\cdot|\mathbf{Z})$. The number of trees and the node size in SRF are set to 100.
             $Z_{1}\text{: Rescaled age at cancer diagnosis}$,
             $Z_{2}\text{: Race-Africa American}$,
             $Z_{3}\text{: Race-Other}$,
             $Z_{4}\text{: Receipt of BMT}$,
             $Z_{5}\text{: Receipt of Alkylating agents}$
             $Z_{6}\text{: Receipt of radiation to the abdomen/pelvis/total body}$.}
\end{figure}

The estimated coefficients obtained by five different methods for calculating $G(\cdot\mid\mathbf{Z})$ are in Figure \ref{fig: real_appa_comp_coef}. When with Approach A, and Figure \ref{fig: real_appa_comp_coef} with Approach B is used. We conducted the same comparison for the corresponding sandwich standard errors, which are shown in Figures \ref{fig: real_appa_comp_coef_se} and \ref{fig: real_appb_comp_coef_se}.
%
%
%
%
\section{Additional Outcome of Simulation 1}\label{supp: restul_of_simulation_set1}
We summarized the efficiency of different methods and approaches in simulation 1.1 as follows. When using a well-tuned SRF, Approach B demonstrates greater efficiency under high censoring rates but is slightly less efficient when the censoring rate is low at the earliest age (age 21). When $G(\cdot|\mathbf{Z})$ is estimated with the stratified ECDF or Cox PH model on $C^{*}$, Approach B generally yields a smaller SSD. A similar pattern is shown when the Cox PH model on $C$ is used at ages 21 and 30. However, the Cox PH model on $C$ appears to lose efficiency at ages 35 and 40 compared to other methods , particularly when using Approach B (Figure \ref{fig: ssd_compre_method_sim11}).

We summarized the efficiency of different methods and approaches in simulation 1.2 as follows. When using a well-tuned SRF, Approach A demonstrates greater efficiency than Approach B. A similar trend is observed when $G(\cdot|\mathbf{Z})$ is estimated with the stratified ECDF or Cox PH model on $C^{*}$, where Approach A consistently yields a smaller SSD than Approach B. At early ages (21 and 30), the performance across both approaches remains nearly identical, primarily due to the negligible censoring rate ($0.34\%$). However, a notable divergence occurs at age 40 when the censoring rate is low ($10.7\%$), as illustrated in Figure \ref{fig: ssd_compre_method_sim12}.

%
%
%
%
\subsection{Simulation 1.1: Regarding Performance of Different Methods and Approaches}\label{supp: restul_of_simulation_set11}
We compare the root sample mean square error (RSMSE) of estimates when using a survival random forest (SRF) with different sets of input parameters to obtain $\hat{G}(\cdot\mid\mathbf{Z})$. Since we only have two covariates, we use both of them in each split. We select the number of trees from 100 to 1000 for some of the simulation datasets. Since a higher number of trees provides a similar RSMSE but takes much longer to run, we have decided to stick with 100 trees. We used both variables at each split, and the minimal node size to split at (i.e., node size) is selected from 15, 50, 100, 200, 500, and 3500. We refer to these as situations 1 to 6 and denote them as S1 to S6, respectively. A lower RSMSE is considered to be better. We observe that small node sizes perform well when the covariate is a binary variable, while large node sizes perform well when the covariate is a continuous variable. We select node size equal to 200 (i.e., S4) when using SRF in both approaches.

\begin{figure}[htbp]
    \centering 
    \includegraphics[width=0.9\textwidth]{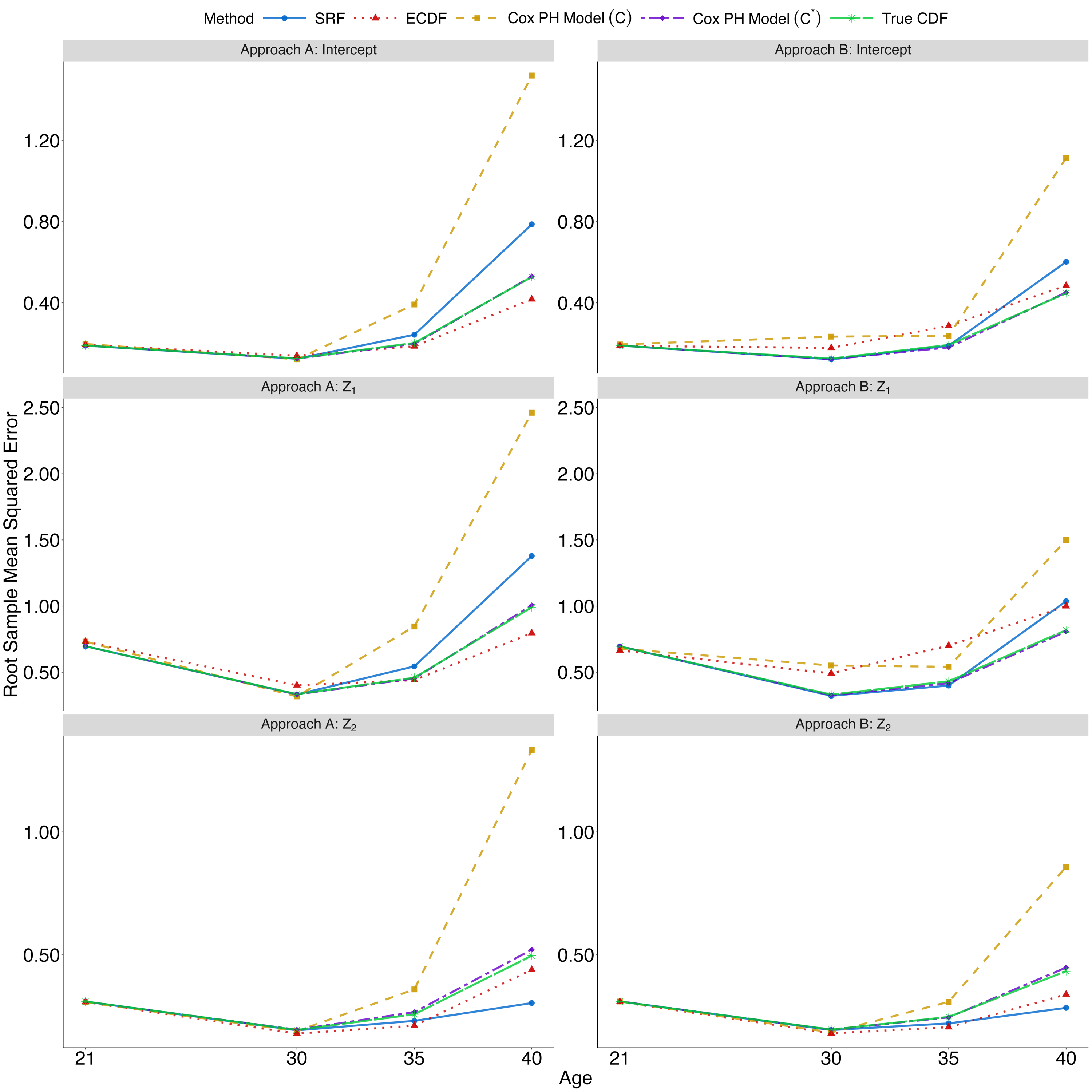}
    \caption{Comparison of Root Sample Mean Square Error (RSMSE) for coefficient estimates in Simulation 1.1. Results are shown for two approaches across five $\hat{G}(\cdot \mid \mathbf{Z})$ which is obtained using: (i) SRF with 100 trees and node size of 200; (ii) stratified ECDF; (iii) Cox PH model on $C$; (iv) Cox PH model on $C^{*}$; and (v) the true CDF.  }
    \label{fig: sim11_all_comp_srf}
\end{figure}

To find which method has the best performance when using each approach to estimate coefficients, we compare the RSMSE of estimated coefficients when using SRF with the selected hyperparameter value, stratified ECDF , the Cox PH model on $C$, the Cox PH model on $C^{*}$, and the true CDF to obtain $\hat{G}(\cdot\mid\mathbf{Z})$. For the stratified ECDF, the covariate $Z_{1}$ was partitioned into four levels: $[0, 0.25)$, $[0.25, 0.5)$, $[0.5, 0.75)$, and $[0.75, 1]$. Results for both estimation frameworks are illustrated in Figure \ref{fig: sim11_all_comp_srf}.

\begin{figure}[H]
    \centering
    \includegraphics[width=1\textwidth]{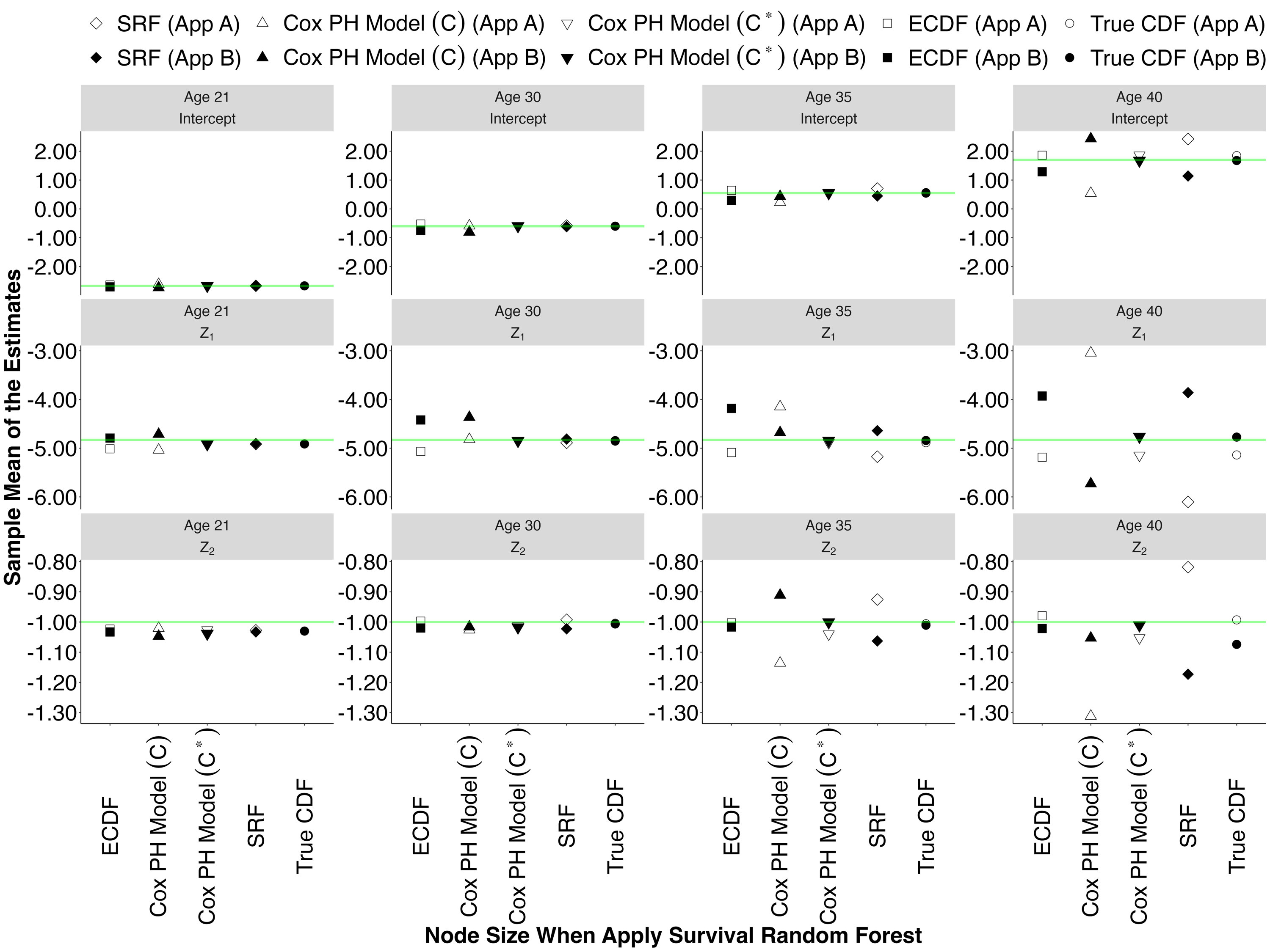}
    \caption{Comparison of sample mean of the estimates (SMEAN) for coefficient estimates in Simulation 1.1. Results are shown for two approaches across five $\hat{G}(\cdot \mid \mathbf{Z})$ which is obtained using: (i) SRF with 100 trees and node size of 200; (ii) stratified ECDF; (iii) Cox PH model on $C$; (iv) Cox PH model on $C^{*}$; and (v) the true CDF.}\label{fig: smean_compre_method_sim11}
\end{figure}

\newpage
The Figure \ref{fig: smean_compre_method_sim11} compares the estimated coefficients obtained using two approaches, while the censoring distribution is estimated using different methods alongside their true values.

\begin{figure}[H]
    \centering
    \includegraphics[width=1\textwidth]{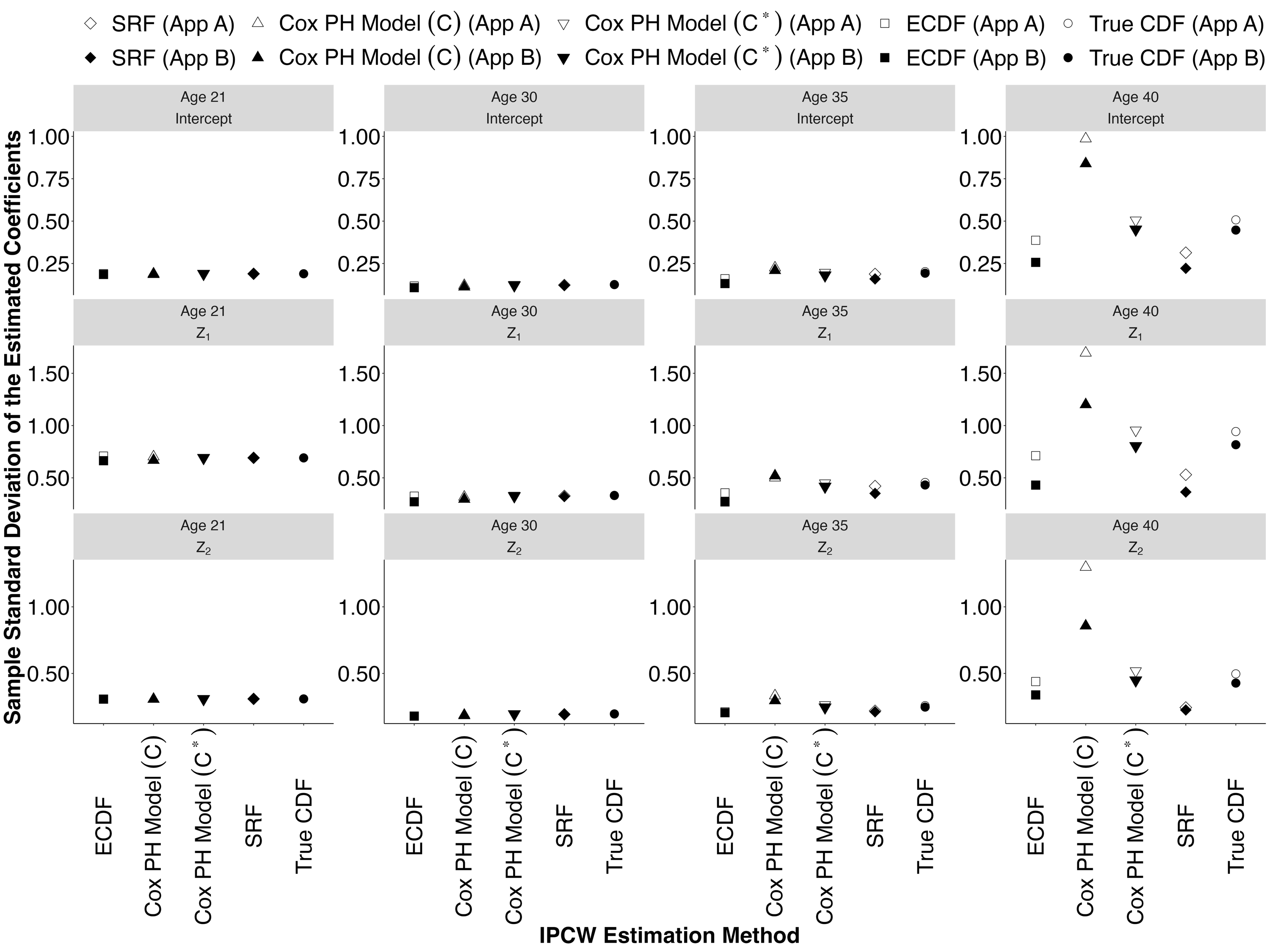}
    \caption{Comparison of sample standard deviation of the estimates (SSD) for coefficient estimates in Simulation 1.1. Results are shown for two approaches across five $\hat{G}(\cdot \mid \mathbf{Z})$ which is obtained using: (i) SRF with 100 trees and node size of 200; (ii) stratified ECDF; (iii) Cox PH model on $C$; (iv) Cox PH model on $C^{*}$; and (v) the true CDF.}\label{fig: ssd_compre_method_sim11}
\end{figure}

\newpage
The Figure \ref{fig: ssd_compre_method_sim11} compares the sample standard deviation of the estimated coefficients using both approaches when the censoring distribution is estimated using different methods.

\begin{figure}[H]
    \centering
    \includegraphics[width=1\textwidth]{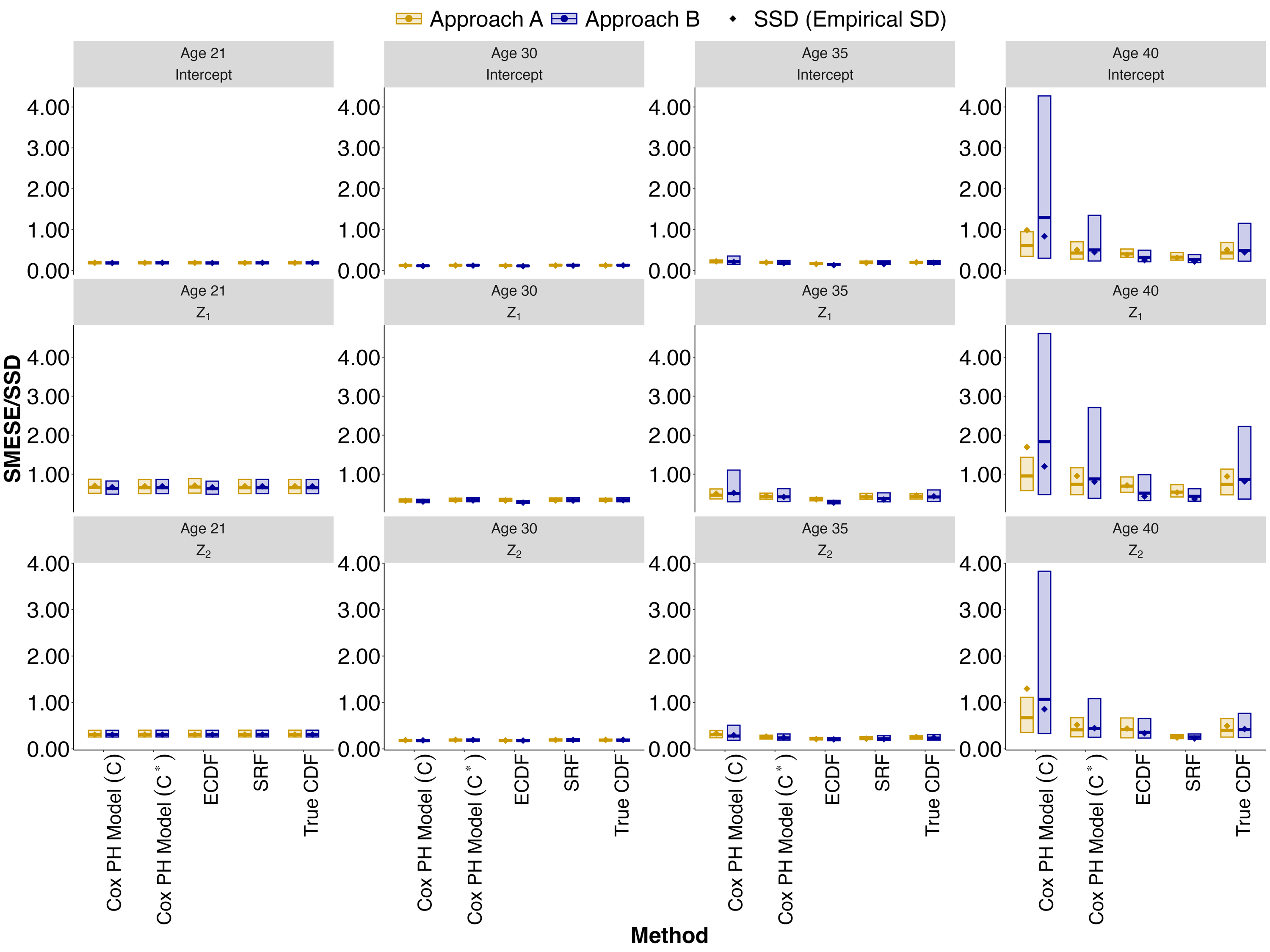}
    \caption{Evaluation of standard error (SE) estimator performance in Simulation 1.1. The plot displays the sample mean of the estimated SEs (SMESE; horizontal lines) and the corresponding 2.5\%–97.5\% empirical quantiles (shaded regions) relative to the sample standard deviation of the coefficient estimates (SSD; diamonds). Results are shown for two approaches across five $\hat{G}(\cdot \mid \mathbf{Z})$ which is obtained using: (i) SRF with 100 trees and node size of 200; (ii) stratified ECDF; (iii) Cox PH model on $C$; (iv) Cox PH model on $C^{*}$; and (v) the true CDF.}\label{fig: smese_ssd_compre_method_sim11}
\end{figure}

\newpage
The Figure \ref{fig: smese_ssd_compre_method_sim11} compares the SMESE to the SSD for assessing the standard error estimation 

\clearpage
%
%
%
%
\subsection{Simulation 1.2: Regarding Performance of Different Methods and Approaches}\label{supp: restul_of_simulation_set12}
Following the same procedure mentioned in section \ref{supp: restul_of_simulation_set11}, we select the number of trees equal to 100 and the node sizes equal to 50 when using SRF, as this set of hyperparameters shows good performance in terms of RSMSE at all ages using either approach.

\begin{figure}[htbp]
    \centering 
    \includegraphics[width=0.9\textwidth]{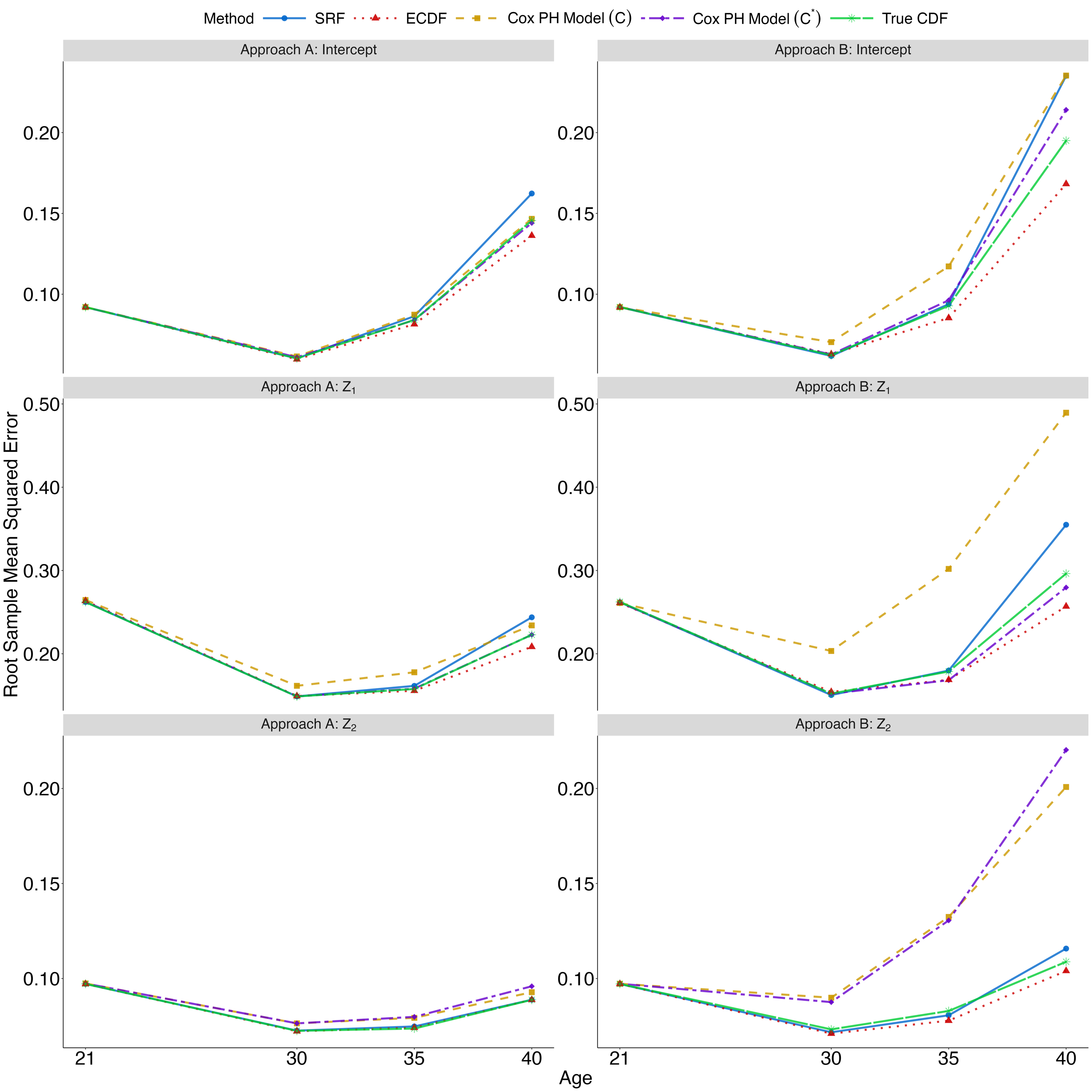}
    \caption{Comparison of Root Sample Mean Square Error (RSMSE) for coefficient estimates in Simulation 1.2. Results are shown for two approaches across five $\hat{G}(\cdot \mid \mathbf{Z})$ which is obtained using: (i) SRF with 100 trees and a minimum node size of 50; (ii) stratified ECDF; (iii) Cox PH model on $C$; (iv) Cox PH model on $C^{*}$; and (v) the true CDF.  }
    \label{fig: sim12_all_comp_srf}
\end{figure}

To find which method has the best performance when using each approach to estimate coefficients, we compare the RSMSE of estimated coefficients when using SRF with the selected hyperparameter value, stratified ECDF , the Cox PH model on $C$, the Cox PH model on $C^{*}$, and the true CDF to obtain $\hat{G}(\cdot\mid\mathbf{Z})$. For the stratified ECDF, the covariate $Z_{1}$ was partitioned into four levels: $[0, 0.25)$, $[0.25, 0.5)$, $[0.5, 0.75)$, and $[0.75, 1]$. Results for both estimation frameworks are illustrated in Figure \ref{fig: sim12_all_comp_srf}.

The Table \ref{table: sim1set12:result} summarizes the estimates of $\alpha_{1}(t)$ and $\beta(t)$ (which equal $\beta$) in simulation 1.2, using both approaches and all five methods to obtain $\hat{G}(\cdot|\mathbf{Z})$. We set the number of trees equal to 100 and the node size equal to 50, as mentioned above.
\begin{table}[H]
\centering
\fontsize{4pt}{5pt}\selectfont 

\setlength{\tabcolsep}{1.8pt} 
\renewcommand{\arraystretch}{1.4} 

\caption{The result of simulation 1.2. We report the sample mean of the estimates (SMEAN), the sample standard deviation of the estimates (SSD), the sample mean of the estimated standard errors  (SMESE), and the root mean squared error of the estimates (RSMSE) of each approach. In SRF, the number of tree is set to 100 and the node size is set to 50.}\label{table: sim1set12:result}
\resizebox{1\textwidth}{!}{%
\begin{tabular}{|c|c|ccccc|ccccc|ccccc|}
\hline
 &  & \multicolumn{5}{c|}{$\alpha(21) = -1.440 $} & \multicolumn{5}{c|}{$\beta_{1}(21) = -5.460$} & \multicolumn{5}{c|}{$\beta_{2}(21) = 1.500$} \\ \hline
Age 21 &  & SRF & Cox $C$ &  Cox $C^{*}$ & ECDF & True & SRF & Cox $C$ &  Cox $C^{*}$ & ECDF & True & SRF & Cox $C$ &  Cox $C^{*}$ & ECDF & True \\ \hline
\multirow{4}{*}{\begin{tabular}[c]{@{}c@{}}Approach used\\ by Im et al/\\ Approach A\end{tabular}} 
 & SMEAN & -1.439 & -1.436 & -1.441 & -1.436 & -1.439 & -5.472 & -5.492 & -5.477 & -5.482 & -5.471 & 1.500 & 1.506 & 1.505 & 1.500 & 1.499 \\
 & SSD   & 0.092 & 0.092 & 0.092 & 0.092 & 0.092 & 0.262 & 0.263 & 0.262 & 0.263 & 0.262 & 0.097 & 0.097 & 0.097 & 0.097 & 0.097 \\
 & SMESE & 0.092 & 0.092 & 0.092 & 0.092 & 0.092 & 0.257 & 0.258 & 0.257 & 0.257 & 0.257 & 0.099 & 0.099 & 0.099 & 0.099 & 0.099 \\
 & RSMSE & 0.092 & 0.092 & 0.092 & 0.092 & 0.092 & 0.262 & 0.265 & 0.263 & 0.263 & 0.262 & 0.097 & 0.097 & 0.097 & 0.097 & 0.097 \\ \hline
\multirow{4}{*}{Approach B} 
 & SMEAN & -1.439 & -1.442 & -1.438 & -1.442 & -1.439 & -5.471 & -5.453 & -5.465 & -5.461 & -5.471 & 1.499 & 1.494 & 1.494 & 1.499 & 1.499 \\
 & SSD   & 0.092 & 0.092 & 0.092 & 0.092 & 0.092 & 0.262 & 0.261 & 0.262 & 0.261 & 0.262 & 0.097 & 0.097 & 0.097 & 0.097 & 0.097 \\
 & SMESE & 0.092 & 0.092 & 0.092 & 0.092 & 0.092 & 0.257 & 0.256 & 0.256 & 0.256 & 0.257 & 0.099 & 0.099 & 0.099 & 0.099 & 0.099 \\
 & RSMSE & 0.092 & 0.092 & 0.092 & 0.092 & 0.092 & 0.262 & 0.261 & 0.262 & 0.261 & 0.262 & 0.097 & 0.097 & 0.097 & 0.097 & 0.097 \\ \hline
 &  & \multicolumn{5}{c|}{$\alpha(30) = 0.900 $} & \multicolumn{5}{c|}{$\beta_{1}(30) = -5.460$} & \multicolumn{5}{c|}{$\beta_{2}(30) = 1.500$} \\ \hline
Age 30 &  & SRF & Cox $C$ &  Cox $C^{*}$ & ECDF & True & SRF & Cox $C$ &  Cox $C^{*}$ & ECDF & True & SRF & Cox $C$ &  Cox $C^{*}$ & ECDF & True \\ \hline
\multirow{4}{*}{\begin{tabular}[c]{@{}c@{}}Approach used\\ by Im et al/\\ Approach A\end{tabular}} 
 & SMEAN & 0.899 & 0.915 & 0.890 & 0.906 & 0.897 & -5.462 & -5.524 & -5.466 & -5.480 & -5.456 & 1.506 & 1.526 & 1.525 & 1.501 & 1.501 \\
 & SSD   & 0.060 & 0.060 & 0.060 & 0.059 & 0.060 & 0.148 & 0.149 & 0.149 & 0.148 & 0.149 & 0.072 & 0.072 & 0.072 & 0.072 & 0.073 \\
 & SMESE & 0.062 & 0.062 & 0.062 & 0.062 & 0.062 & 0.150 & 0.149 & 0.149 & 0.149 & 0.150 & 0.070 & 0.069 & 0.070 & 0.070 & 0.070 \\
 & RSMSE & 0.060 & 0.062 & 0.061 & 0.060 & 0.060 & 0.148 & 0.162 & 0.149 & 0.149 & 0.149 & 0.073 & 0.076 & 0.076 & 0.072 & 0.072 \\ \hline
\multirow{4}{*}{Approach B} 
 & SMEAN & 0.897 & 0.863 & 0.909 & 0.881 & 0.897 & -5.446 & -5.315 & -5.426 & -5.409 & -5.456 & 1.492 & 1.445 & 1.450 & 1.500 & 1.501 \\
 & SSD   & 0.061 & 0.060 & 0.062 & 0.060 & 0.062 & 0.148 & 0.143 & 0.148 & 0.146 & 0.152 & 0.071 & 0.071 & 0.072 & 0.071 & 0.073 \\
 & SMESE & 0.063 & 0.062 & 0.064 & 0.063 & 0.063 & 0.152 & 0.146 & 0.151 & 0.151 & 0.153 & 0.071 & 0.069 & 0.070 & 0.071 & 0.071 \\
 & RSMSE & 0.061 & 0.070 & 0.063 & 0.063 & 0.062 & 0.148 & 0.203 & 0.152 & 0.154 & 0.152 & 0.072 & 0.090 & 0.088 & 0.071 & 0.073 \\ \hline
 &  & \multicolumn{5}{c|}{$\alpha(35) = 2.200 $} & \multicolumn{5}{c|}{$\beta_{1}(35) = -5.460$} & \multicolumn{5}{c|}{$\beta_{2}(35) = 1.500$} \\ \hline
Age 35 &  & SRF & Cox $C$ &  Cox $C^{*}$ & ECDF & True & SRF & Cox $C$ &  Cox $C^{*}$ & ECDF & True & SRF & Cox $C$ &  Cox $C^{*}$ & ECDF & True \\ \hline
\multirow{4}{*}{\begin{tabular}[c]{@{}c@{}}Approach used\\ by Im et al/\\ Approach A\end{tabular}} 
 & SMEAN & 2.207 & 2.231 & 2.194 & 2.206 & 2.203 & -5.478 & -5.548 & -5.475 & -5.485 & -5.467 & 1.509 & 1.531 & 1.531 & 1.503 & 1.502 \\
 & SSD   & 0.083 & 0.082 & 0.084 & 0.081 & 0.084 & 0.156 & 0.154 & 0.158 & 0.154 & 0.158 & 0.074 & 0.073 & 0.073 & 0.074 & 0.074 \\
 & SMESE & 0.082 & 0.080 & 0.081 & 0.080 & 0.081 & 0.154 & 0.151 & 0.153 & 0.151 & 0.154 & 0.073 & 0.073 & 0.073 & 0.074 & 0.074 \\
 & RSMSE & 0.083 & 0.087 & 0.084 & 0.081 & 0.084 & 0.157 & 0.178 & 0.158 & 0.156 & 0.158 & 0.074 & 0.079 & 0.080 & 0.074 & 0.074 \\ \hline
\multirow{4}{*}{Approach B} 
 & SMEAN & 2.194 & 2.115 & 2.236 & 2.181 & 2.203 & -5.440 & -5.199 & -5.441 & -5.406 & -5.467 & 1.481 & 1.391 & 1.395 & 1.505 & 1.502 \\
 & SSD   & 0.085 & 0.080 & 0.089 & 0.083 & 0.093 & 0.163 & 0.151 & 0.167 & 0.160 & 0.179 & 0.077 & 0.076 & 0.078 & 0.078 & 0.083 \\
 & SMESE & 0.090 & 0.085 & 0.091 & 0.089 & 0.090 & 0.171 & 0.157 & 0.170 & 0.168 & 0.173 & 0.079 & 0.075 & 0.077 & 0.080 & 0.080 \\
 & RSMSE & 0.085 & 0.117 & 0.096 & 0.085 & 0.093 & 0.164 & 0.302 & 0.168 & 0.169 & 0.179 & 0.079 & 0.132 & 0.131 & 0.078 & 0.083 \\ \hline
 &  & \multicolumn{5}{c|}{$\alpha(40) = 3.500 $} & \multicolumn{5}{c|}{$\beta_{1}(40) = -5.460$} & \multicolumn{5}{c|}{$\beta_{2}(40) = 1.500$} \\ \hline
Age 40 &  & SRF & Cox $C$ &  Cox $C^{*}$ & ECDF & True & SRF & Cox $C$ &  Cox $C^{*}$ & ECDF & True & SRF & Cox $C$ &  Cox $C^{*}$ & ECDF & True \\ \hline
\multirow{4}{*}{\begin{tabular}[c]{@{}c@{}}Approach used\\ by Im et al/\\ Approach A\end{tabular}} 
 & SMEAN & 3.524 & 3.542 & 3.502 & 3.490 & 3.504 & -5.499 & -5.550 & -5.483 & -5.466 & -5.468 & 1.512 & 1.529 & 1.534 & 1.500 & 1.502 \\
 & SSD   & 0.144 & 0.140 & 0.144 & 0.136 & 0.146 & 0.221 & 0.216 & 0.222 & 0.208 & 0.223 & 0.088 & 0.088 & 0.090 & 0.089 & 0.089 \\
 & SMESE & 0.149 & 0.145 & 0.148 & 0.142 & 0.149 & 0.223 & 0.219 & 0.223 & 0.212 & 0.223 & 0.089 & 0.089 & 0.090 & 0.090 & 0.090 \\
 & RSMSE & 0.146 & 0.147 & 0.144 & 0.136 & 0.146 & 0.224 & 0.234 & 0.223 & 0.208 & 0.223 & 0.089 & 0.093 & 0.096 & 0.089 & 0.089 \\ \hline
\multirow{4}{*}{Approach B} 
 & SMEAN & 3.466 & 3.322 & 3.605 & 3.514 & 3.506 & -5.397 & -5.027 & -5.503 & -5.442 & -5.471 & 1.456 & 1.323 & 1.302 & 1.523 & 1.502 \\
 & SSD   & 0.168 & 0.153 & 0.187 & 0.168 & 0.195 & 0.257 & 0.229 & 0.276 & 0.256 & 0.296 & 0.097 & 0.094 & 0.097 & 0.101 & 0.109 \\
 & SMESE & 0.193 & 0.172 & 0.203 & 0.195 & 0.197 & 0.291 & 0.255 & 0.300 & 0.293 & 0.298 & 0.107 & 0.099 & 0.102 & 0.113 & 0.110 \\
 & RSMSE & 0.172 & 0.235 & 0.214 & 0.168 & 0.195 & 0.264 & 0.489 & 0.280 & 0.257 & 0.296 & 0.107 & 0.201 & 0.220 & 0.104 & 0.109 \\ \hline
\end{tabular}%
}
\end{table}

\begin{figure}[H]
    \centering
    \includegraphics[width=1\textwidth]{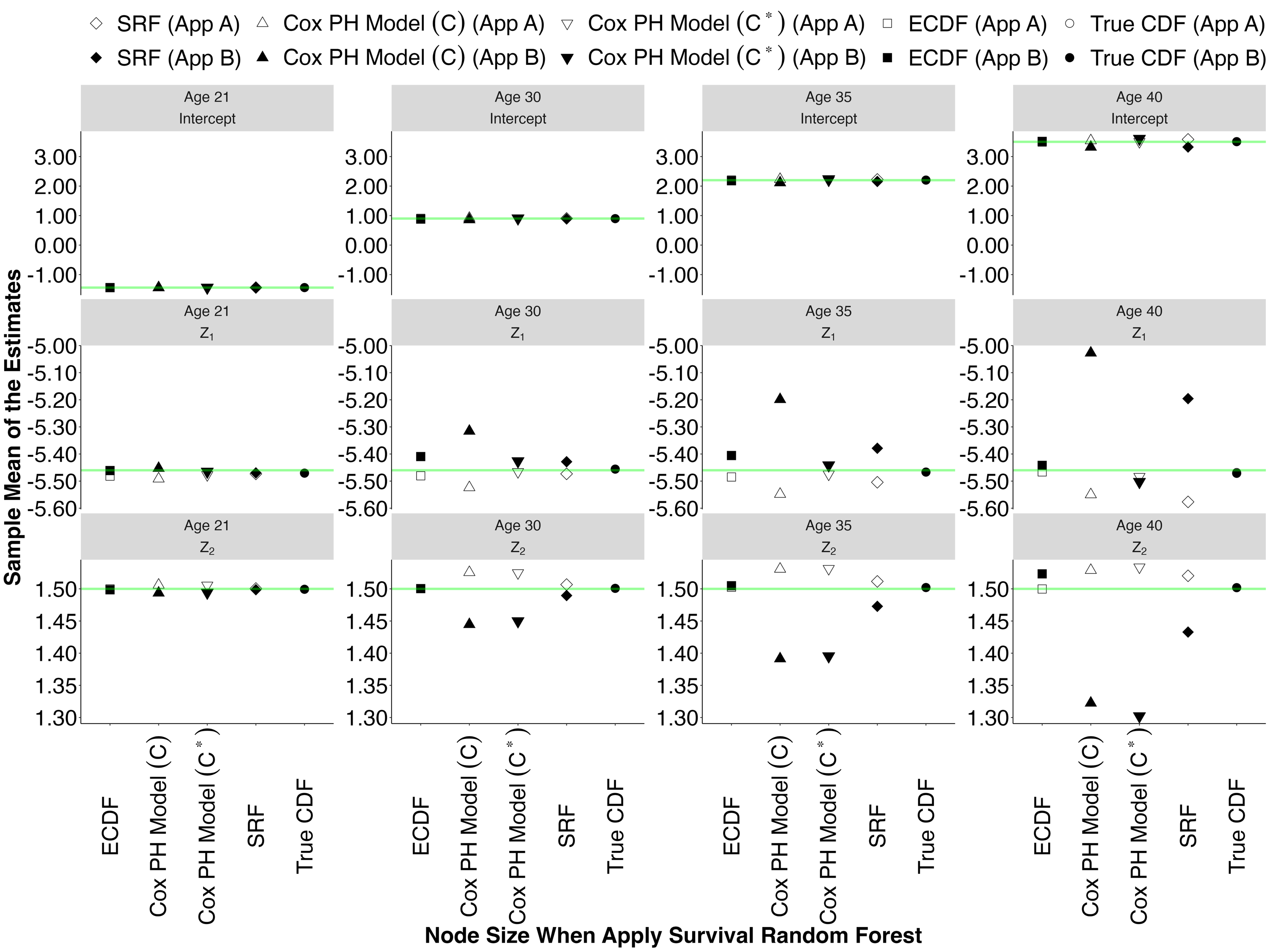}
    \caption{Comparison of sample mean of the estimates (SMEAN) for coefficient estimates in Simulation 1.2. Results are shown for two approaches across five $\hat{G}(\cdot \mid \mathbf{Z})$ which is obtained using: (i) SRF with 100 trees and node size of 50; (ii) stratified ECDF; (iii) Cox PH model on $C$; (iv) Cox PH model on $C^{*}$; and (v) the true CDF.}\label{fig: smean_compre_method_sim12}
\end{figure}

\newpage
The Figure \ref{fig: smean_compre_method_sim12} compares the estimated coefficients obtained using two approaches, while the censoring distribution is estimated using different methods alongside their true values.

\begin{figure}[H]
    \centering
    \includegraphics[width=1\textwidth]{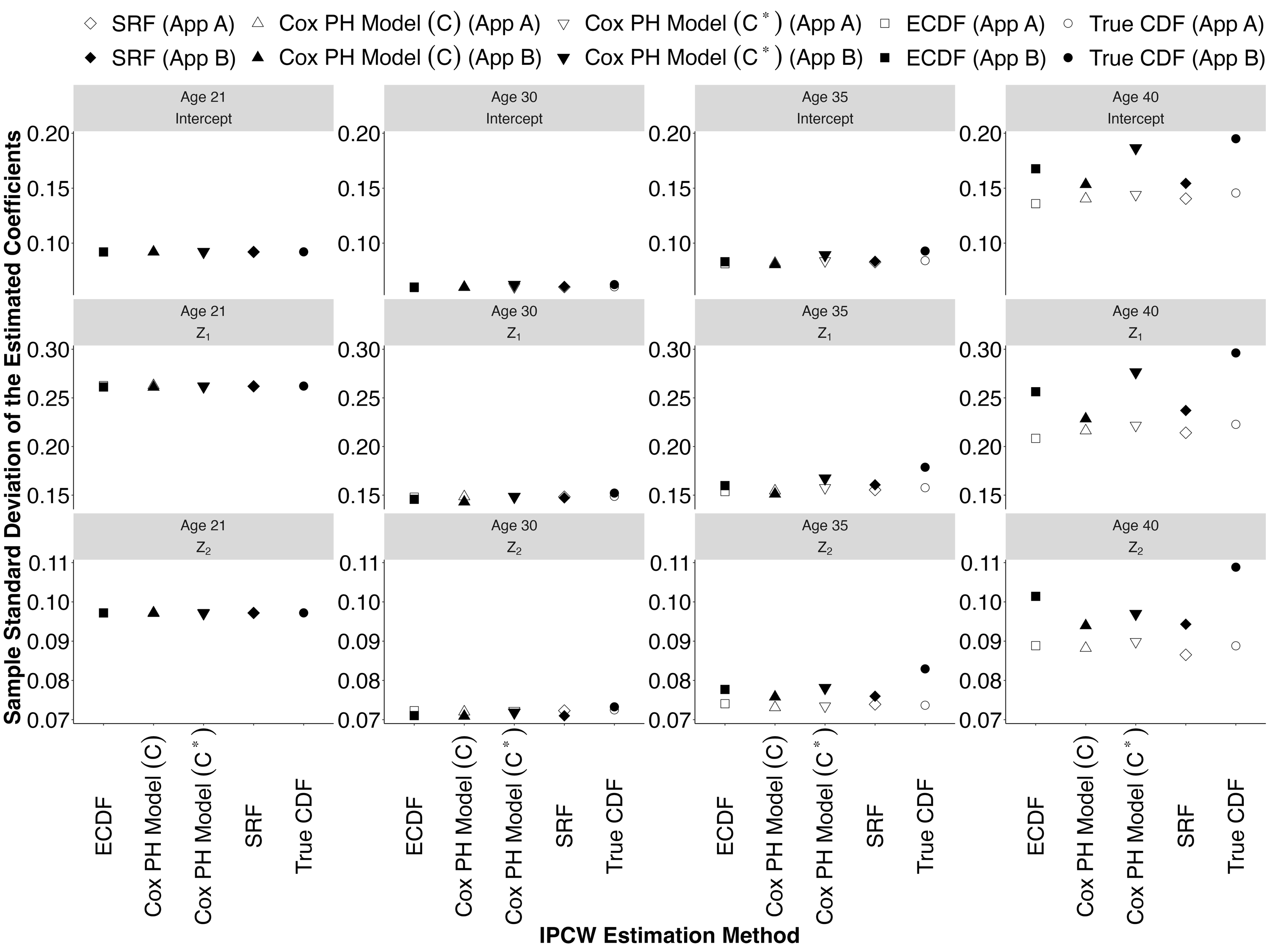}
    \caption{Comparison of sample standard deviation of the estimates (SSD) for coefficient estimates in Simulation 1.1. Results are shown for two approaches across five $\hat{G}(\cdot \mid \mathbf{Z})$ which is obtained using: (i) SRF with 100 trees and node size of 50; (ii) stratified ECDF; (iii) Cox PH model on $C$; (iv) Cox PH model on $C^{*}$; and (v) the true CDF.}\label{fig: ssd_compre_method_sim12}
\end{figure}

\newpage
The Figure \ref{fig: ssd_compre_method_sim12} compares the sample standard deviation of the estimated coefficients using both approaches when the censoring distribution is estimated using different methods.

\begin{figure}[H]
    \centering
    \includegraphics[width=1\textwidth]{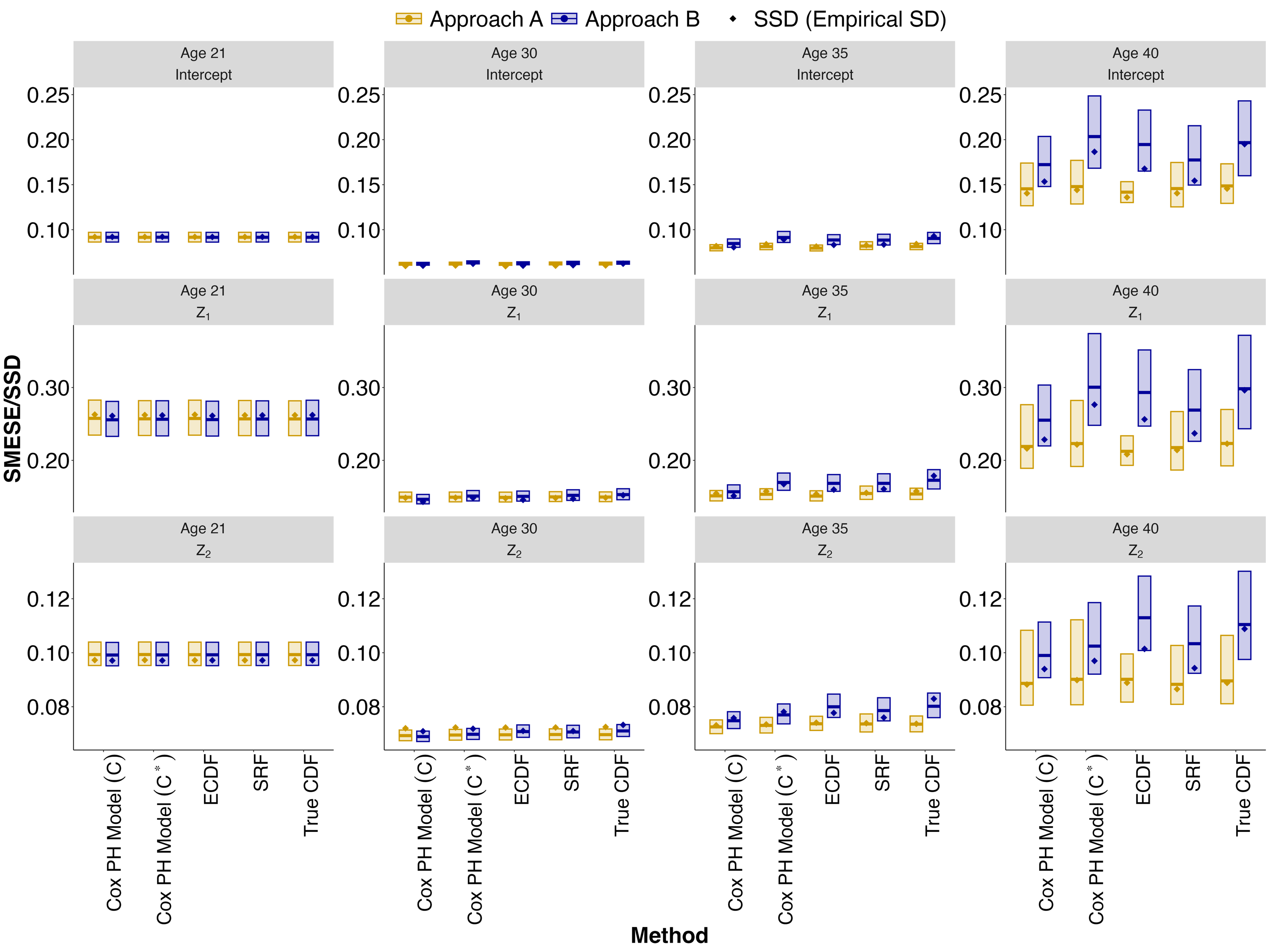}
    \caption{Evaluation of standard error (SE) estimator performance in Simulation 1.1. The plot displays the sample mean of the estimated SEs (SMESE; horizontal lines) and the corresponding 2.5\%–97.5\% empirical quantiles (shaded regions) relative to the sample standard deviation of the coefficient estimates (SSD; diamonds). Results are shown for two approaches across five $\hat{G}(\cdot \mid \mathbf{Z})$ which is obtained using: (i) SRF with 100 trees and node size of 50; (ii) stratified ECDF; (iii) Cox PH model on $C$; (iv) Cox PH model on $C^{*}$; and (v) the true CDF.}\label{fig: smese_ssd_compre_method_sim12}
\end{figure}

\begin{figure}
    \centering
    \includegraphics[width=0.55\textwidth]{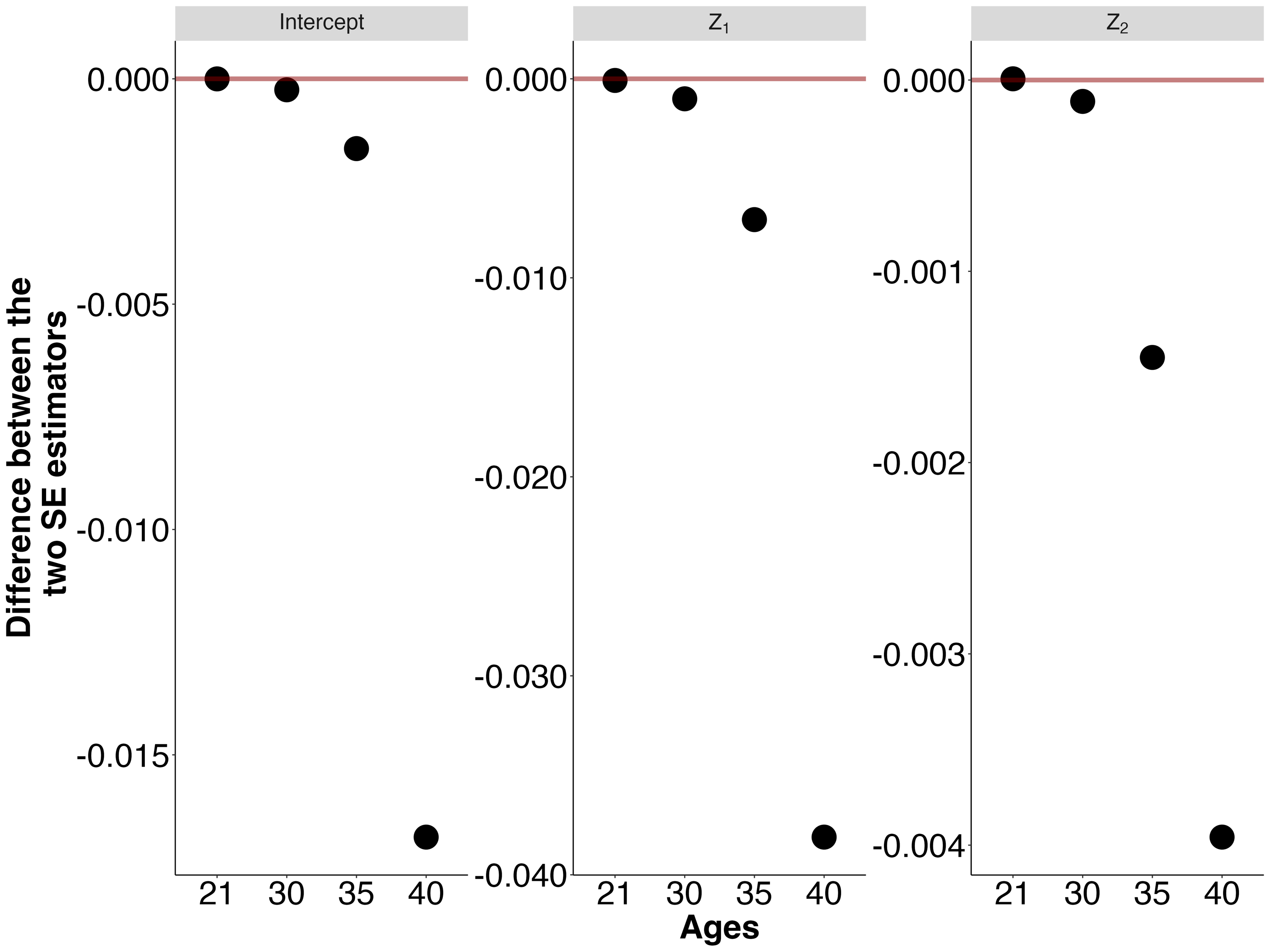}
    \captionof{figure}{Comparison of differences in estimated variances when $G(\cdot|\mathbf{Z})$ is known in Simulation 1.2. The red line in each plot represents a difference of 0.}
    \label{fig:sim12_combined_settings}
\end{figure}

\newpage
The Figure \ref{fig: smese_ssd_compre_method_sim12} compares the SMESE to the SSD for assessing the standard error estimation

\newpage

\newpage

%
%
%
%
\section{Numerical Result of Simulation 2}\label{supp: restul_of_simulation_set2}
The simulation 2.1 is conducted to show the necessity of considering risk set adjustment when the analysis is done before age 21 (Table \ref{table: sim2:result}). In addition, the simulation 2.2 is conducted to show robustness of the approaches. The results in Figure~\ref{fig: sim22a_srf_ci} to Figure~\ref{fig: sim22a_tcdf_ci} highlight that while both approaches perform well when the censoring model is correctly specified.
%
%
%
%
\subsection{Simulation 2.1: Regarding Risk Set Adjustment}\label{supp: restul_of_simulation_set21}
The Table \ref{table: sim2:result} summarizes the estimates of $\alpha_{1}(t)$ and $\bm{\beta}(t)$ (which equals $\bm{\beta}$) in simulation 2, using both approaches and all five methods to obtain $\hat{G}(\cdot|\mathbf{Z})$. We set both the number of trees and the node size equal to 100.
\begin{table}[H]
\centering
\fontsize{4pt}{5pt}\selectfont 

\setlength{\tabcolsep}{1.8pt} 
\renewcommand{\arraystretch}{1.4} 

\caption{The result of simulation 2.1. We report the sample mean of the estimates (SMEAN), the sample standard deviation of the estimates (SSD), the sample mean of the estimated standard errors (SMESE), and the root sample mean squared error of the estimates (RSMSE) of each approach. In SRF, both the number of trees and the node size are set to 100.}\label{table: sim2:result}
\resizebox{1\textwidth}{!}{%
\begin{tabular}{|c|c|ccccc|ccccc|ccccc|}
\hline
 &  & \multicolumn{5}{c|}{$\alpha_{1}(13) = -2.400 $} & \multicolumn{5}{c|}{$\beta_{1}(13) = -6.300$} & \multicolumn{5}{c|}{$\beta_{2}(13) = 1.000$} \\ \hline
Age 13 &  & SRF & Cox $C$ & Cox $C^{*}$ & ECDF & True & SRF & Cox $C$ & Cox $C^{*}$ & ECDF & True & SRF & Cox $C$ & Cox $C^{*}$ & ECDF & True \\ \hline
\multirow{4}{*}{\begin{tabular}[c]{@{}c@{}}Approach used\\ by Im et al\end{tabular}} 
 & SMEAN & -2.298 & -2.290 & -2.298 & -2.294 & -2.298 & -6.955 & -6.978 & -6.955 & -6.968 & -6.954 & 1.021 & 1.022 & 1.021 & 1.021 & 1.020 \\
 & SSD   & 0.227 & 0.227 & 0.227 & 0.227 & 0.227 & 0.617 & 0.619 & 0.617 & 0.619 & 0.618 & 0.214 & 0.214 & 0.214 & 0.214 & 0.214 \\
 & SMESE & 0.221 & 0.221 & 0.221 & 0.221 & 0.221 & 0.600 & 0.602 & 0.600 & 0.602 & 0.600 & 0.214 & 0.214 & 0.214 & 0.214 & 0.214 \\
 & RSMSE & 0.249 & 0.252 & 0.249 & 0.251 & 0.249 & 0.900 & 0.918 & 0.900 & 0.910 & 0.900 & 0.215 & 0.215 & 0.215 & 0.215 & 0.215 \\ \hline
\multirow{4}{*}{Approach A} 
 & SMEAN & -2.425 & -2.417 & -2.425 & -2.421 & -2.425 & -6.305 & -6.330 & -6.304 & -6.318 & -6.303 & 1.018 & 1.019 & 1.018 & 1.018 & 1.017 \\
 & SSD   & 0.248 & 0.248 & 0.248 & 0.248 & 0.248 & 0.750 & 0.752 & 0.750 & 0.751 & 0.750 & 0.213 & 0.213 & 0.213 & 0.213 & 0.213 \\
 & SMESE & 0.242 & 0.242 & 0.242 & 0.242 & 0.242 & 0.728 & 0.730 & 0.728 & 0.730 & 0.728 & 0.213 & 0.213 & 0.213 & 0.213 & 0.213 \\
 & RSMSE & 0.249 & 0.249 & 0.249 & 0.249 & 0.249 & 0.749 & 0.752 & 0.749 & 0.751 & 0.749 & 0.214 & 0.214 & 0.214 & 0.214 & 0.214 \\ \hline
\multirow{4}{*}{Approach B} 
 & SMEAN & -2.425 & -2.430 & -2.425 & -2.428 & -2.425 & -6.304 & -6.289 & -6.303 & -6.295 & -6.303 & 1.017 & 1.017 & 1.017 & 1.017 & 1.017 \\
 & SSD   & 0.248 & 0.248 & 0.248 & 0.248 & 0.248 & 0.750 & 0.748 & 0.749 & 0.748 & 0.749 & 0.213 & 0.213 & 0.213 & 0.213 & 0.213 \\
 & SMESE & 0.242 & 0.242 & 0.242 & 0.242 & 0.242 & 0.728 & 0.726 & 0.728 & 0.727 & 0.728 & 0.213 & 0.213 & 0.213 & 0.213 & 0.213 \\
 & RSMSE & 0.249 & 0.249 & 0.249 & 0.249 & 0.249 & 0.749 & 0.747 & 0.749 & 0.747 & 0.749 & 0.214 & 0.214 & 0.214 & 0.214 & 0.214 \\ \hline
 &  & \multicolumn{5}{c|}{$\alpha_{1}(14) = -2.100 $} & \multicolumn{5}{c|}{$\beta_{1}(14) = -6.300$} & \multicolumn{5}{c|}{$\beta_{2}(14) = 1.000$} \\ \hline
Age 14 &  & SRF & Cox $C$ & Cox $C^{*}$ & ECDF & True & SRF & Cox $C$ & Cox $C^{*}$ & ECDF & True & SRF & Cox $C$ & Cox $C^{*}$ & ECDF & True \\ \hline
\multirow{4}{*}{\begin{tabular}[c]{@{}c@{}}Approach used\\ by Im et al\end{tabular}} 
 & SMEAN & -2.015 & -2.004 & -2.016 & -2.010 & -2.016 & -6.769 & -6.800 & -6.768 & -6.784 & -6.767 & 1.008 & 1.009 & 1.008 & 1.008 & 1.007 \\
 & SSD   & 0.198 & 0.198 & 0.198 & 0.198 & 0.198 & 0.534 & 0.537 & 0.534 & 0.536 & 0.534 & 0.183 & 0.183 & 0.183 & 0.183 & 0.183 \\
 & SMESE & 0.194 & 0.194 & 0.194 & 0.194 & 0.194 & 0.527 & 0.529 & 0.527 & 0.528 & 0.526 & 0.185 & 0.185 & 0.185 & 0.185 & 0.185 \\
 & RSMSE & 0.215 & 0.220 & 0.215 & 0.217 & 0.215 & 0.710 & 0.734 & 0.710 & 0.722 & 0.709 & 0.183 & 0.183 & 0.183 & 0.183 & 0.183 \\ \hline
\multirow{4}{*}{Approach A} 
 & SMEAN & -2.106 & -2.095 & -2.107 & -2.101 & -2.107 & -6.321 & -6.355 & -6.319 & -6.337 & -6.318 & 1.005 & 1.006 & 1.005 & 1.005 & 1.005 \\
 & SSD   & 0.212 & 0.212 & 0.212 & 0.212 & 0.212 & 0.615 & 0.618 & 0.616 & 0.617 & 0.615 & 0.183 & 0.183 & 0.183 & 0.183 & 0.183 \\
 & SMESE & 0.208 & 0.208 & 0.208 & 0.208 & 0.208 & 0.608 & 0.610 & 0.608 & 0.610 & 0.608 & 0.184 & 0.184 & 0.184 & 0.184 & 0.184 \\
 & RSMSE & 0.212 & 0.212 & 0.212 & 0.212 & 0.212 & 0.615 & 0.620 & 0.616 & 0.618 & 0.615 & 0.183 & 0.183 & 0.183 & 0.183 & 0.183 \\ \hline
\multirow{4}{*}{Approach B} 
 & SMEAN & -2.106 & -2.114 & -2.106 & -2.110 & -2.106 & -6.319 & -6.295 & -6.318 & -6.306 & -6.319 & 1.004 & 1.003 & 1.004 & 1.004 & 1.005 \\
 & SSD   & 0.212 & 0.211 & 0.212 & 0.211 & 0.212 & 0.615 & 0.612 & 0.614 & 0.613 & 0.614 & 0.183 & 0.183 & 0.183 & 0.183 & 0.183 \\
 & SMESE & 0.208 & 0.207 & 0.208 & 0.207 & 0.208 & 0.608 & 0.606 & 0.608 & 0.606 & 0.608 & 0.184 & 0.184 & 0.184 & 0.184 & 0.184 \\
 & RSMSE & 0.212 & 0.212 & 0.212 & 0.211 & 0.212 & 0.615 & 0.612 & 0.614 & 0.613 & 0.614 & 0.183 & 0.183 & 0.183 & 0.183 & 0.183 \\ \hline
 &  & \multicolumn{5}{c|}{$\alpha_{1}(15) = -1.800 $} & \multicolumn{5}{c|}{$\beta_{1}(15) = -6.300$} & \multicolumn{5}{c|}{$\beta_{2}(15) = 1.000$} \\ \hline
Age 15 &  & SRF & Cox $C$ & Cox $C^{*}$ & ECDF & True & SRF & Cox $C$ & Cox $C^{*}$ & ECDF & True & SRF & Cox $C$ & Cox $C^{*}$ & ECDF & True \\ \hline
\multirow{4}{*}{\begin{tabular}[c]{@{}c@{}}Approach used\\ by Im et al\end{tabular}} 
 & SMEAN & -1.746 & -1.731 & -1.747 & -1.740 & -1.747 & -6.618 & -6.660 & -6.615 & -6.636 & -6.614 & 1.010 & 1.012 & 1.010 & 1.010 & 1.009 \\
 & SSD   & 0.172 & 0.172 & 0.172 & 0.172 & 0.172 & 0.459 & 0.462 & 0.460 & 0.461 & 0.460 & 0.165 & 0.165 & 0.166 & 0.165 & 0.166 \\
 & SMESE & 0.171 & 0.172 & 0.171 & 0.172 & 0.171 & 0.465 & 0.467 & 0.465 & 0.466 & 0.465 & 0.162 & 0.162 & 0.162 & 0.161 & 0.162 \\
 & RSMSE & 0.180 & 0.185 & 0.180 & 0.182 & 0.180 & 0.558 & 0.585 & 0.557 & 0.570 & 0.556 & 0.166 & 0.166 & 0.166 & 0.166 & 0.166 \\ \hline
\multirow{4}{*}{Approach A} 
 & SMEAN & -1.809 & -1.793 & -1.810 & -1.803 & -1.810 & -6.319 & -6.363 & -6.316 & -6.337 & -6.315 & 1.007 & 1.009 & 1.008 & 1.007 & 1.007 \\
 & SSD   & 0.181 & 0.181 & 0.181 & 0.181 & 0.181 & 0.509 & 0.512 & 0.509 & 0.511 & 0.509 & 0.165 & 0.165 & 0.165 & 0.165 & 0.165 \\
 & SMESE & 0.180 & 0.180 & 0.180 & 0.180 & 0.180 & 0.515 & 0.517 & 0.515 & 0.517 & 0.515 & 0.161 & 0.161 & 0.161 & 0.161 & 0.161 \\
 & RSMSE & 0.181 & 0.181 & 0.181 & 0.181 & 0.181 & 0.509 & 0.515 & 0.509 & 0.512 & 0.509 & 0.165 & 0.165 & 0.165 & 0.165 & 0.165 \\ \hline
\multirow{4}{*}{Approach B} 
 & SMEAN & -1.809 & -1.821 & -1.809 & -1.815 & -1.809 & -6.316 & -6.281 & -6.315 & -6.298 & -6.316 & 1.006 & 1.005 & 1.006 & 1.006 & 1.007 \\
 & SSD   & 0.181 & 0.180 & 0.180 & 0.180 & 0.180 & 0.509 & 0.505 & 0.508 & 0.506 & 0.508 & 0.165 & 0.165 & 0.165 & 0.165 & 0.165 \\
 & SMESE & 0.180 & 0.180 & 0.180 & 0.180 & 0.180 & 0.515 & 0.512 & 0.515 & 0.513 & 0.515 & 0.161 & 0.161 & 0.161 & 0.161 & 0.161 \\
 & RSMSE & 0.181 & 0.181 & 0.180 & 0.180 & 0.180 & 0.509 & 0.505 & 0.508 & 0.506 & 0.508 & 0.165 & 0.165 & 0.165 & 0.165 & 0.165 \\ \hline
\end{tabular}%
}
\end{table}
%
%
%
%
\subsection{Simulation 2.2: Regarding Performance of Different Methods and Approaches}\label{supp: restul_of_simulation_set22}
\begin{figure}[H]
  \centering
    \includegraphics[width=1\textwidth]{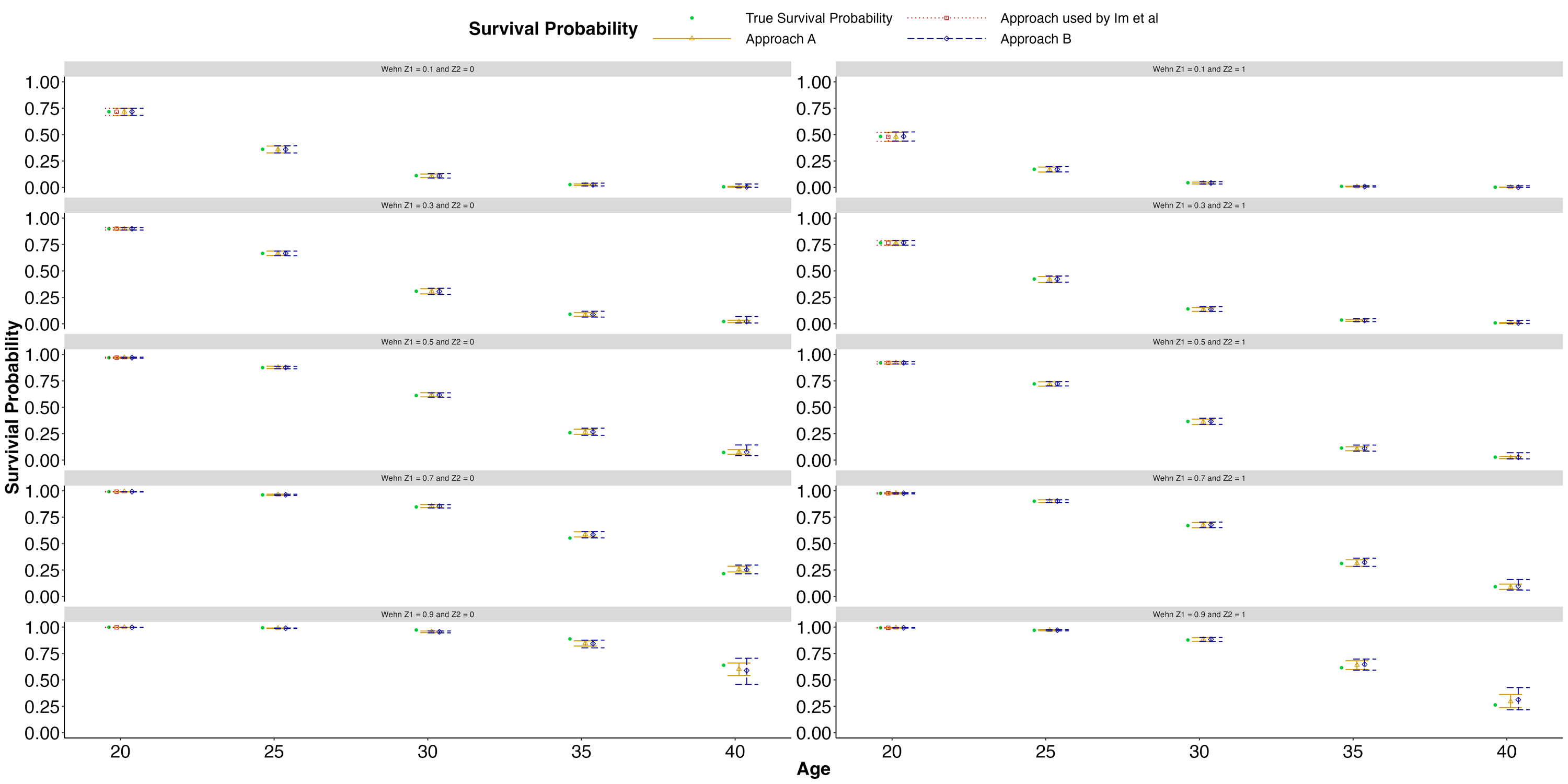}
    \caption{Compare the average estimated survival probability across all approaches with the average true survival probability in simulation 2.2. We use SRF to estimate the survival function of censoring time. In SRF, the number of trees is set to 100, and the node size is set to 500. The standard error of the estimated survival probability is calculated by finding the standard error of the linear predictor (log-odds) using the covariance matrix of estimated coefficients, and then transforming that error to the probability scale.}
     \label{fig: sim22a_srf_ci}
\end{figure}

\begin{figure}[H]
  \centering
    \includegraphics[width=1\textwidth]{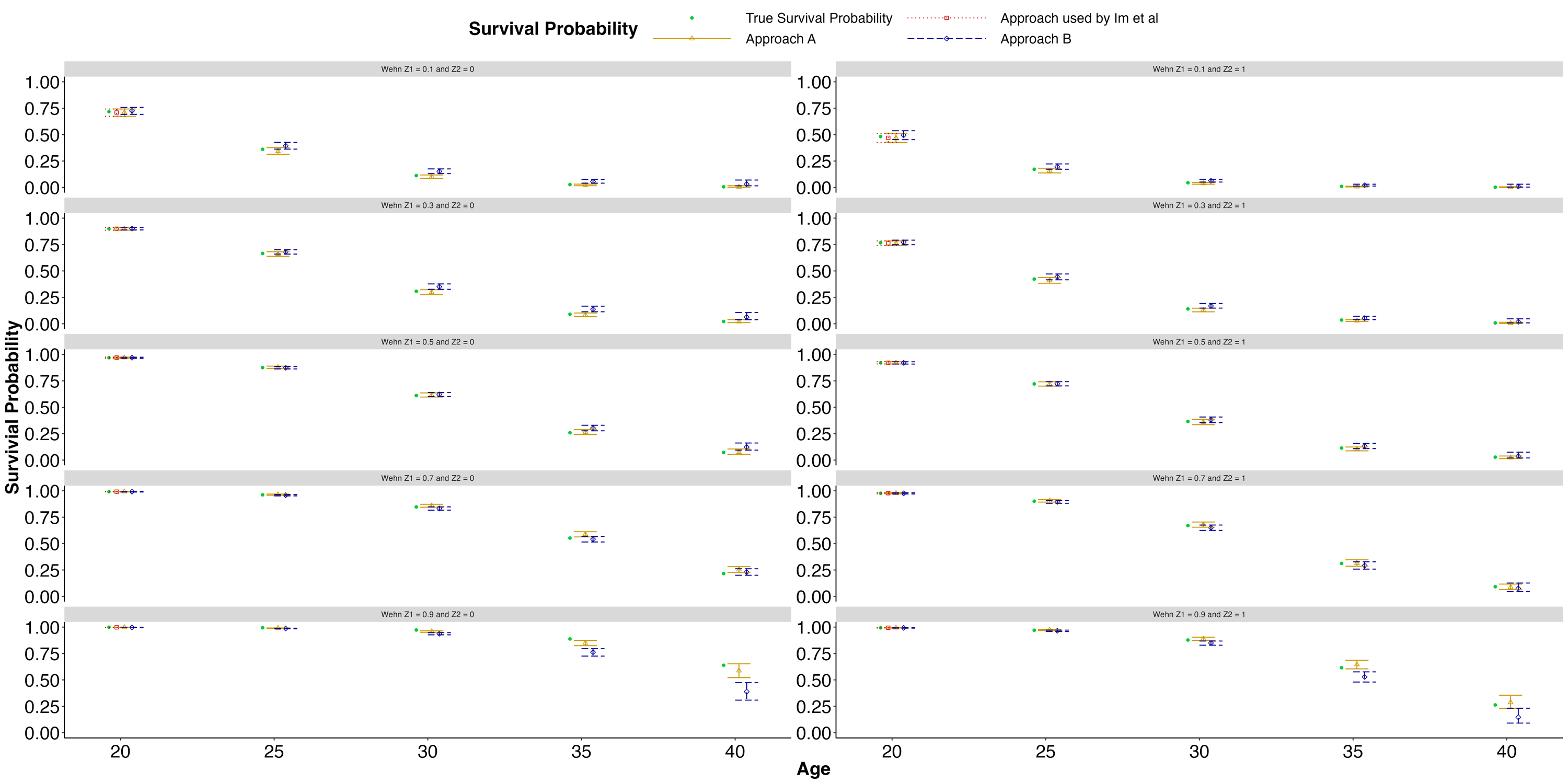}
     \caption{Compare the average estimated survival probability across all approaches with the average true survival probability in simulation 2.2. We use Cox PH Model to estimate the survival function of censoring time. The standard error of estimated survival probability is calculated by finding the standard error of the linear predictor (log-odds) using the covariance matrix of estimated coefficients, and then transforming that error to the probability scale.}
     \label{fig: sim22a_cox_ci}
\end{figure}

\begin{figure}[H]
  \centering
    \includegraphics[width=1\textwidth]{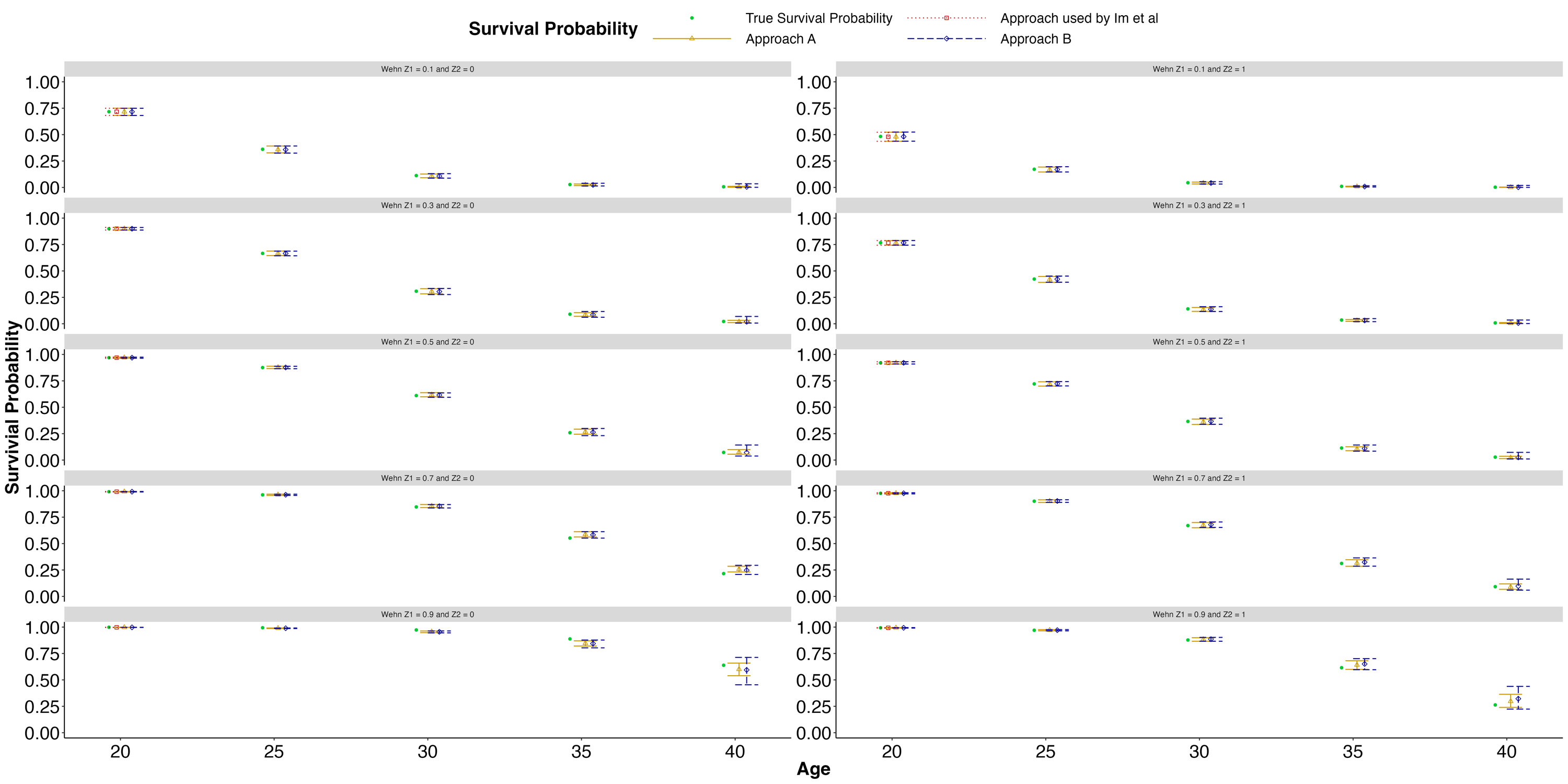}
     \caption{Compare the average estimated survival probability across all approaches with the average true survival probability in simulation 2.2. We use Cox PH Model to estimate the survival function of $C^{*}$,  which equals to $C-V-5$. The standard error of estimated survival probability is calculated by finding the standard error of the linear predictor (log-odds) using the covariance matrix of estimated coefficients, and then transforming that error to the probability scale.}
     \label{fig: sim22a_coxcplus_ci}
\end{figure}

\begin{figure}[H]
  \centering
    \includegraphics[width=1\textwidth]{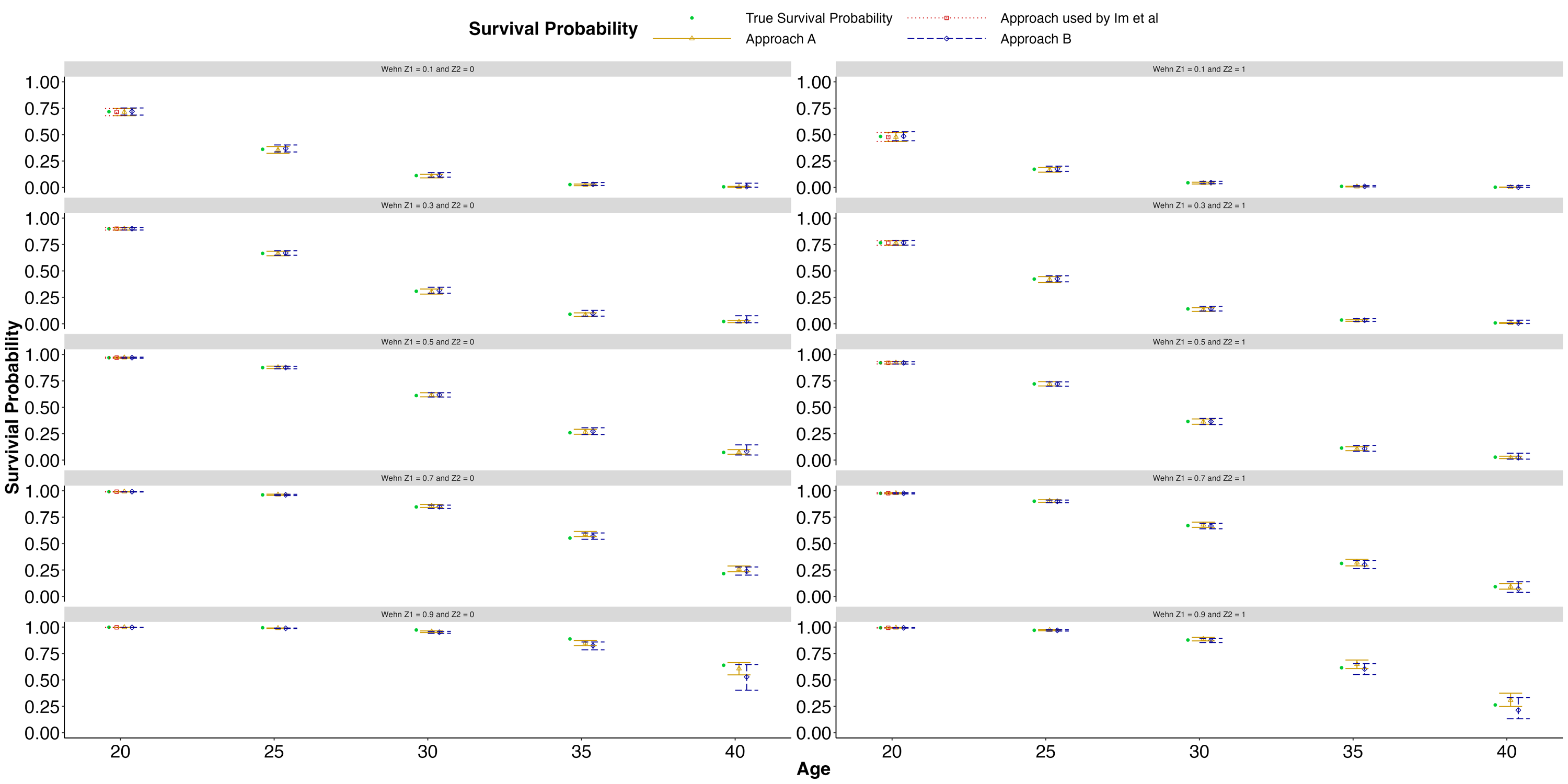}
     \caption{Compare the average estimated survival probability across all approaches with the average true survival probability in simulation 2.2. We use stratified ECDF to estimate the survival function of censoring time. The standard error of estimated survival probability is calculated by finding the standard error of the linear predictor (log-odds) using the covariance matrix of estimated coefficients, and then transforming that error to the probability scale.}
     \label{fig: sim22a_ecdf_ci}
\end{figure}

\begin{figure}[H]
  \centering
    \includegraphics[width=1 \textwidth]{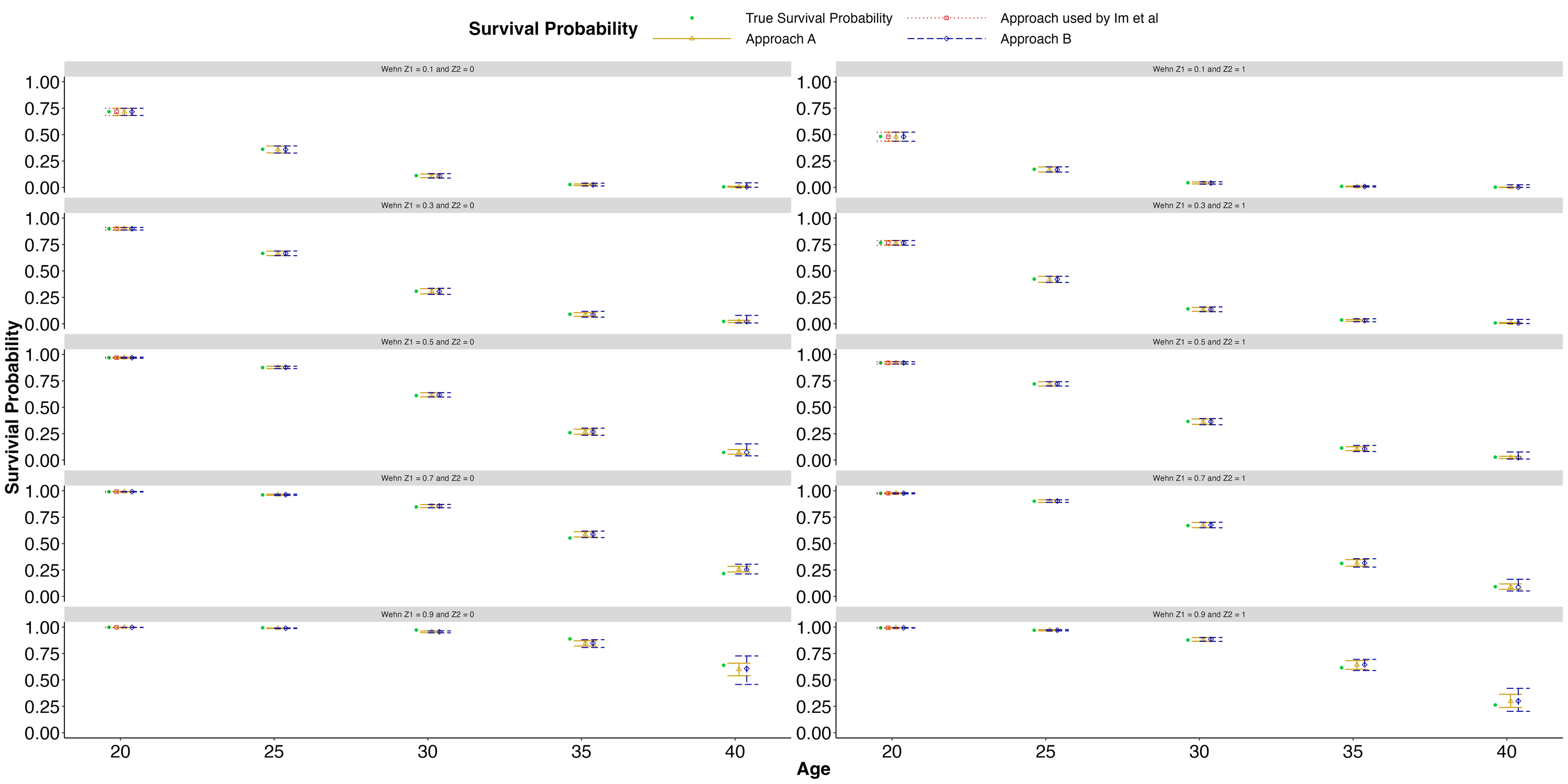}
     \caption{Compare the average estimated survival probability across all approaches with the average true survival probability in simulation 2.2. We use true CDF of censoring time. The standard error of estimated survival probability is calculated by finding the standard error of the linear predictor (log-odds) using the covariance matrix of estimated coefficients, and then transforming that error to the probability scale.}
     \label{fig: sim22a_tcdf_ci}
\end{figure}

In Figure \ref{fig: sim22a_srf_ci} to \ref{fig: sim22a_tcdf_ci}, we report the estimated survival probability using each method for both approaches and compare it with the true survival probability based on the model in simulation 2. We set the number of trees equal to 100 and the node size equal to 500 when using SRF.

\section{Numerical Result of Simulation 3}\label{supp: restul_of_simulation_set3}
\begin{figure}[H]
  \centering
    \includegraphics[width=1\textwidth]{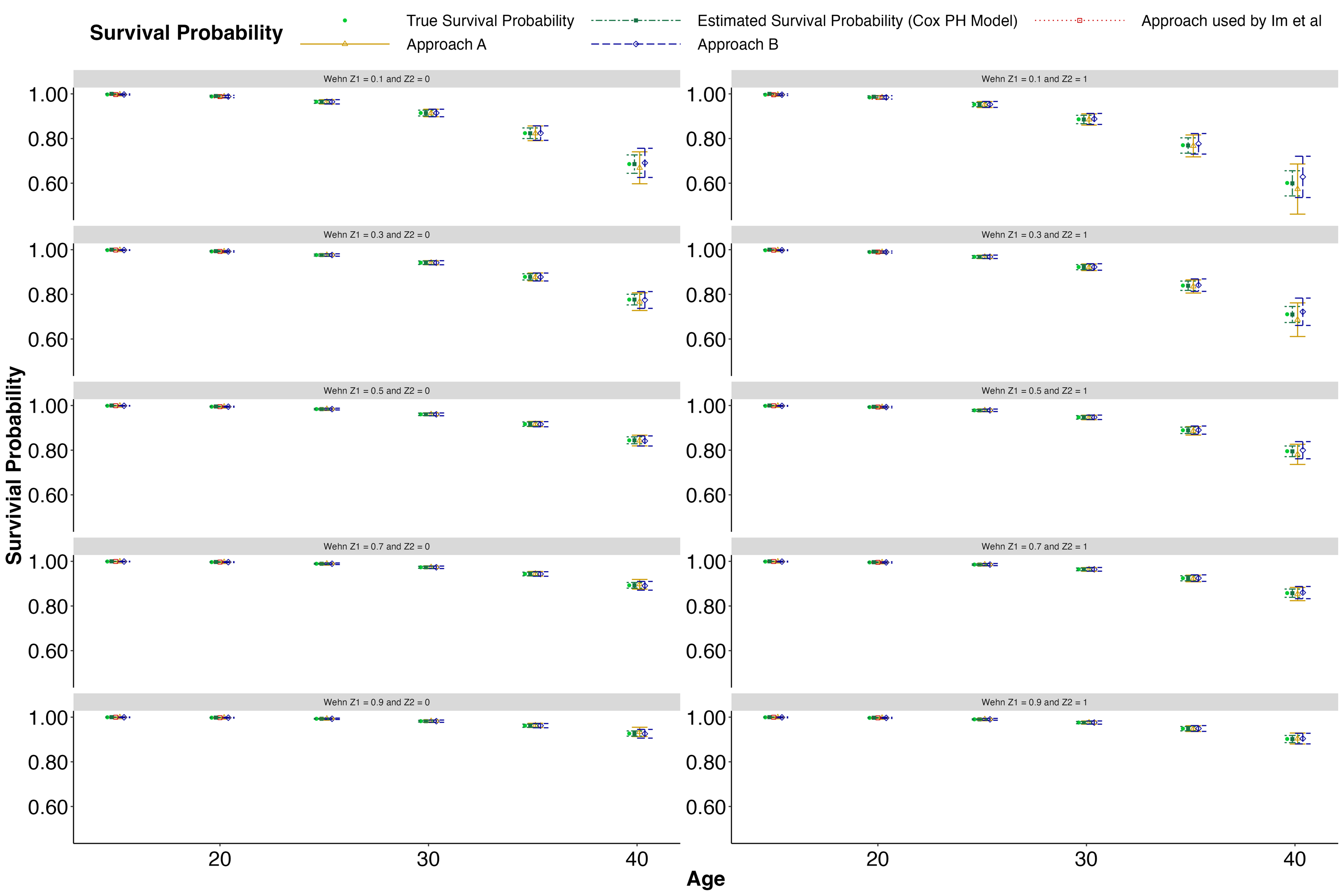}
     \caption{Compare the average estimated survival probability across all approaches with the average true survival probability in simulation 3. We use SRF to estimate the survival function of censoring time. In SRF, the number of trees is set to 100, and the node size are set to 500. The standard error of estimated survival probability is calculated by finding the standard error of the linear predictor (log-odds) using the covariance matrix of estimated coefficients, and then transforming that error to the probability scale. }
     \label{fig: sim31_srf_ci}
\end{figure}

\begin{figure}[H]
  \centering
    \includegraphics[width=1\textwidth]{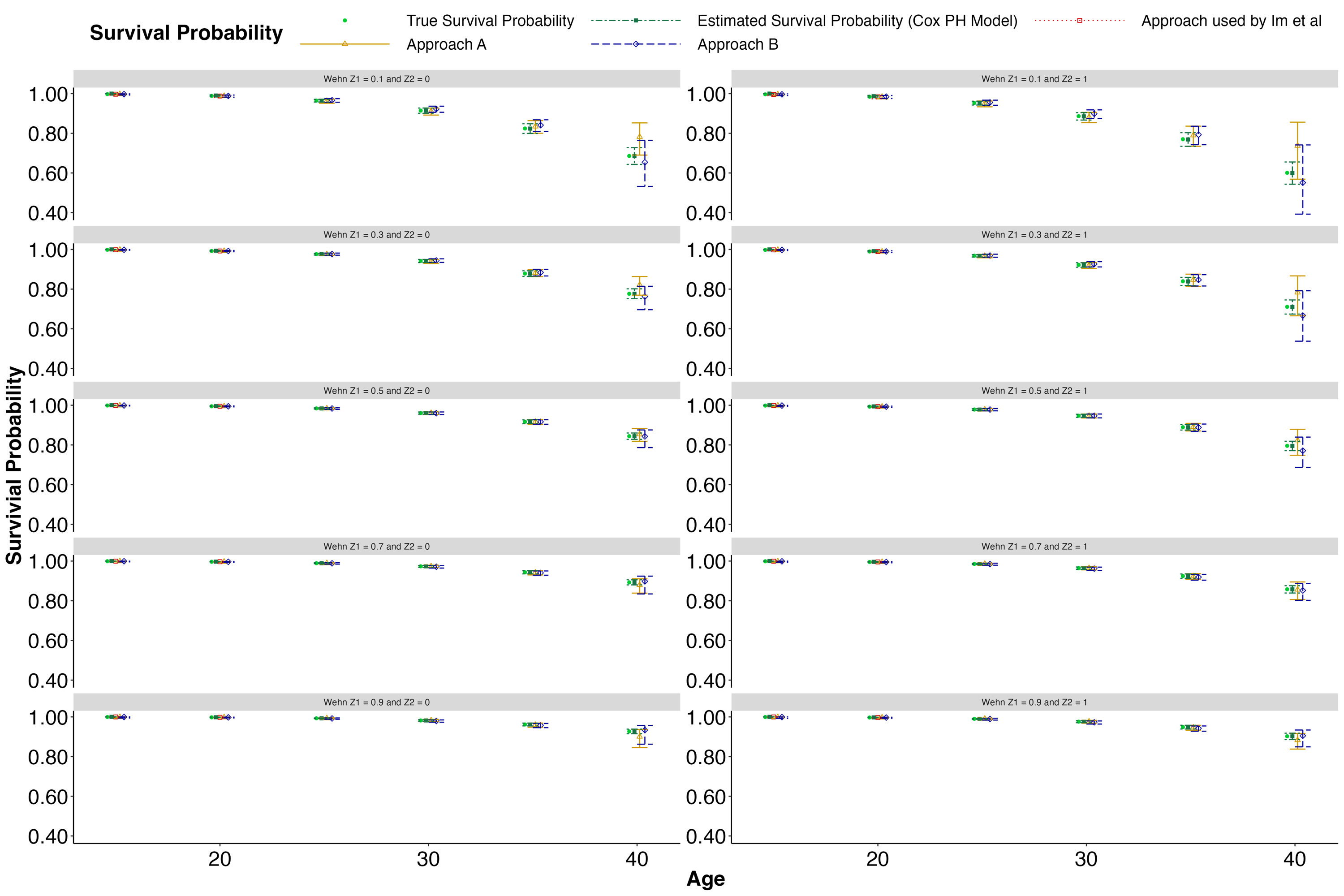}
     \caption{Compare the average estimated survival probability across all approaches with the average true survival probability in simulation 3. We use Cox PH Model to estimate the survival function of censoring time. TThe standard error of estimated survival probability is calculated by finding the standard error of the linear predictor (log-odds) using the covariance matrix of estimated coefficients, and then transforming that error to the probability scale. }
     \label{fig: sim31_cox_ci}
\end{figure}

\begin{figure}[H]
  \centering
    \includegraphics[width=1\textwidth]{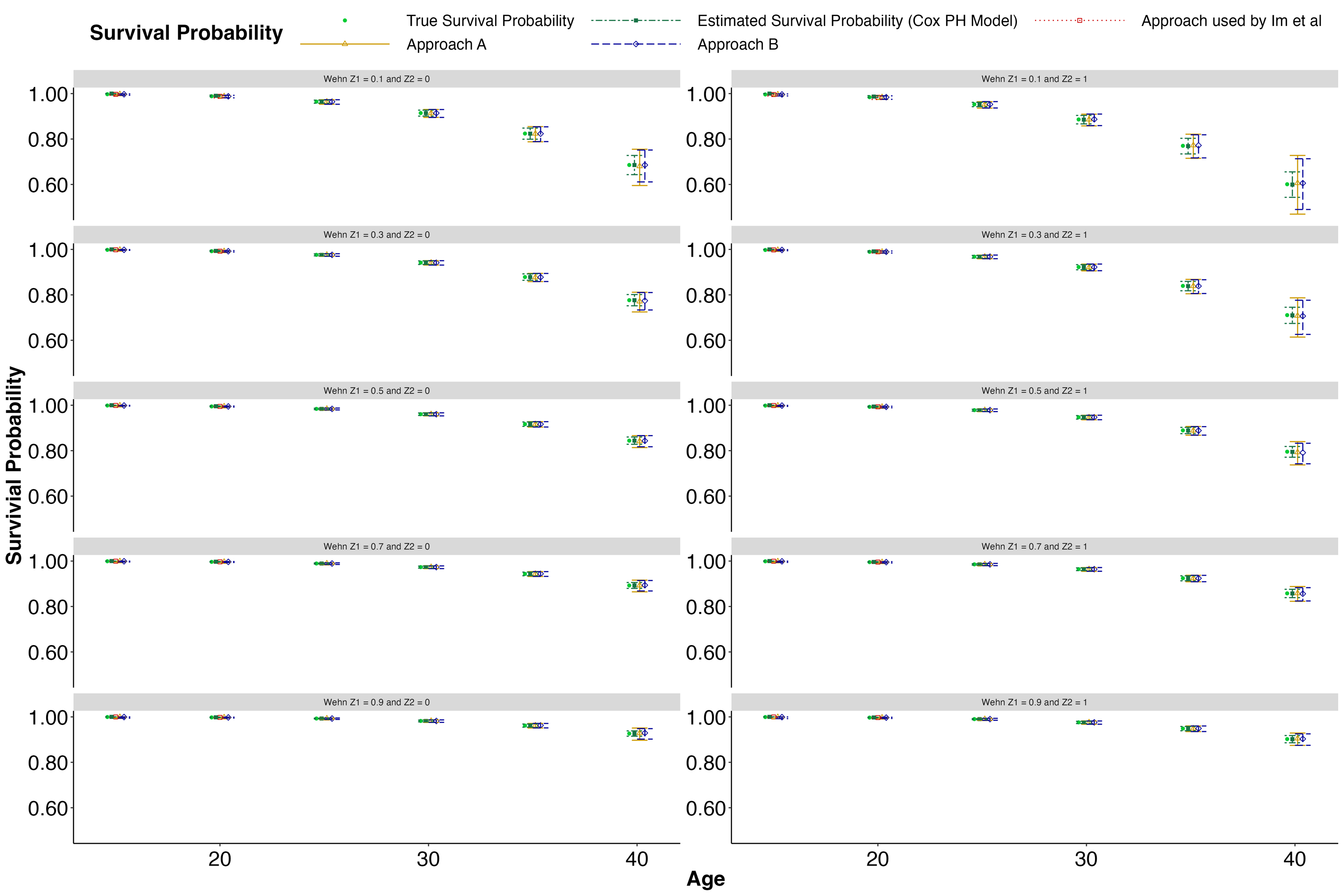}
     \caption{Compare the average estimated survival probability across all approaches with the average true survival probability in simulation 3. We use Cox PH Model to estimate the survival function of $C^{*}$,  which equals to $C-V-5$. The standard error of estimated survival probability is calculated by finding the standard error of the linear predictor (log-odds) using the covariance matrix of estimated coefficients, and then transforming that error to the probability scale. }
     \label{fig: sim31_coxcplus_ci}
\end{figure}

\begin{figure}[H]
  \centering
    \includegraphics[width=1 \textwidth]{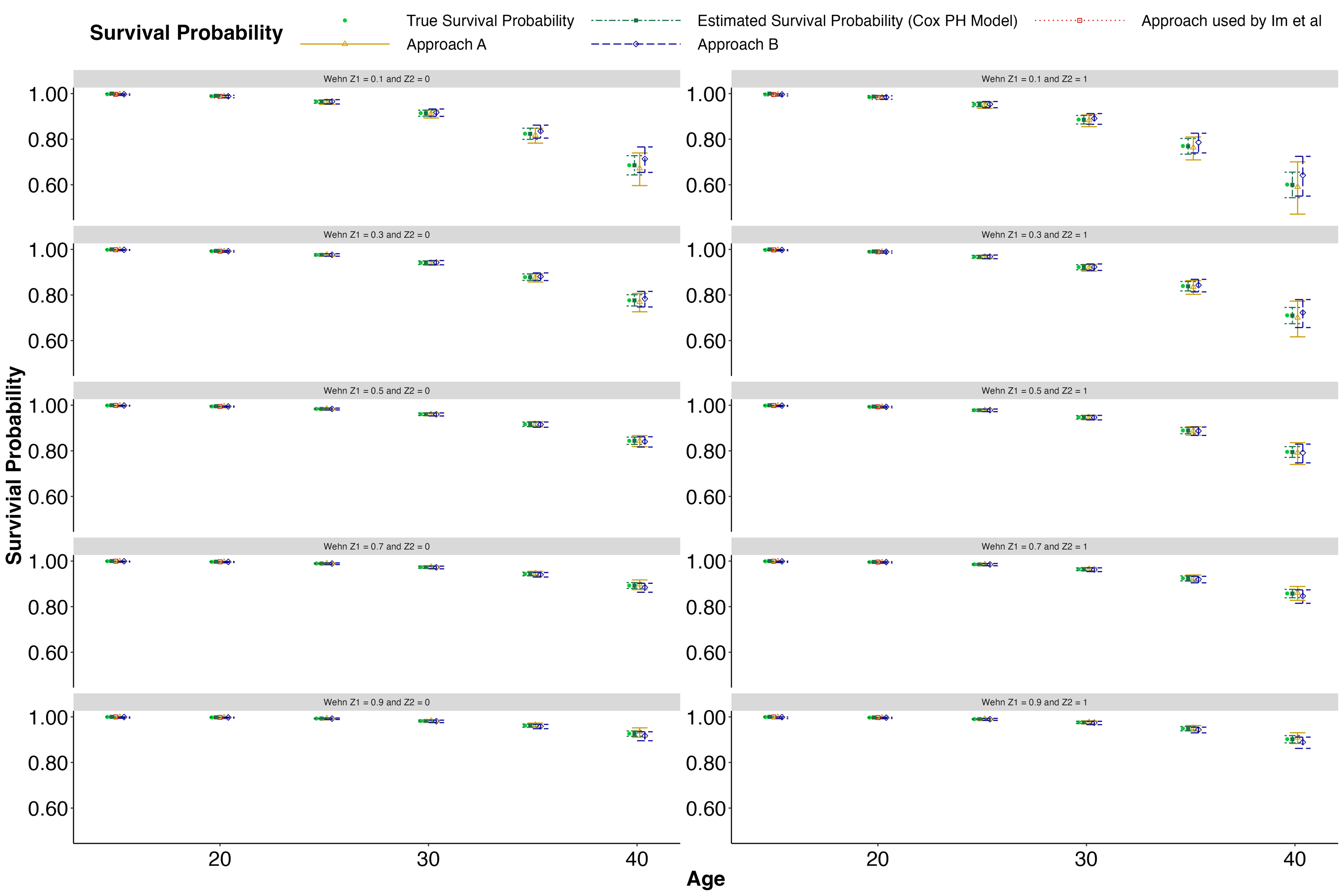}
     \caption{Compare the average estimated survival probability across all approaches with the average true survival probability in simulation 3. We use stratified ECDF to estimate the survival function of censoring time. The standard error of estimated survival probability is calculated by finding the standard error of the linear predictor (log-odds) using the covariance matrix of estimated coefficients, and then transforming that error to the probability scale. }
     \label{fig: sim31_ecdf_ci}
\end{figure}

\begin{figure}[H]
  \centering
    \includegraphics[width=1 \textwidth]{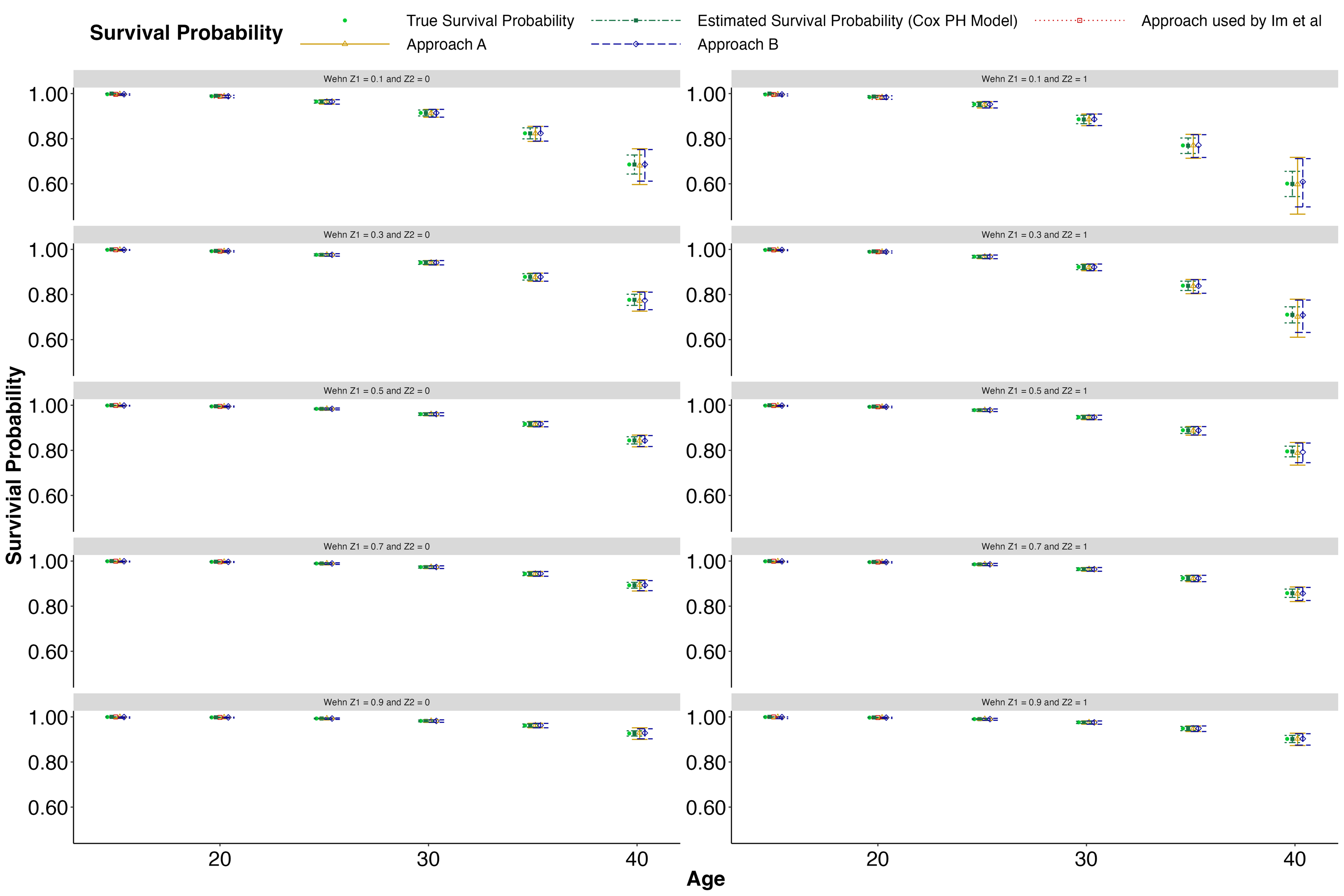}
     \caption{Compare the average estimated survival probability across all approaches with the average true survival probability in simulation 3. We use true CDF of censoring time.The standard error of estimated survival probability is calculated by finding the standard error of the linear predictor (log-odds) using the covariance matrix of estimated coefficients, and then transforming that error to the probability scale. }
     \label{fig: sim31_tcdf_ci}
\end{figure}

In Figure \ref{fig: sim31_srf_ci} to \ref{fig: sim31_tcdf_ci}, we report the estimated survival probability using each method for both approaches. We compare it with the true survival probability based on the model in simulation 3 and an estimated survival probability using the Cox PH model with an unspecified baseline hazard function. We set the number of trees equal to 100 and the node size equal to 500 when using SRF.

\newpage

%
%
%
%
\section{Validation of Data Generation Process}\label{supp: check_alrotihm}
In the simulation studies,the current data generation algorithm only generates data at each selected $t_{0}$. Also, the true event time is only generated for subjects who have $I(T_{i} \le t_{0})=1$. To ensure that this data generation algorithm works, we consider another algorithm that generates $T_{i}$ for all subjects, as shown below.
\begin{algorithm}
    \caption{Another Data Simulation Process of subjects in repetition $k$, where $1\le k \le K$ }\label{alg: another_data_sim}
    \begin{flushleft}
        \textbf{Initialize:} Set parameters for the true model(s) of $T|Z, T > V$ and distribution parameters. For notational convenience, we omit the subscript $k$ in the following steps\\
        \textbf{For} subject $i = 1, 2, \ldots, n$ \textbf{do:}
        \vspace{-10pt}
        \begin{enumerate}
            \item Generate $Z_{1i}$ from Beta$(a_{1},a_{2})$
            \item Generate $Z_{2i}$ from Bernoulli$(p)$
            \item Generate $V_{i}$ from mixture of truncated normal distributions:
            
            \item Generate $T_{i}|\mathbf{Z}_{i}, T_{i} > V_{i}$:
            \begin{itemize}
                \item Sample $B_{i} \sim \text{Uniform}(0,1)$
                \item If $B_{i} < 0.005$, set $B_{i} = 0.005$
                \item Obtain $T_{i}$ using inverse transform sampling
            \end{itemize}
            \item Generate censoring time $C_{i}$:
            \begin{itemize}
                \item Sample $C^{*} \sim \text{Weibull}(\psi_{3i}, \psi_{4i})$
                \item Set $C = C^{*} + V_{i} + 5$
            \end{itemize}
            \item Compute $U_{i} = \min\{T_{i}, C_{i}\}$ and $\delta_{i} = \mathbb{I}(T_{i} \leq C_{i})$
            \item Store observed data: $O_{i} = \{U_{i}, \delta_{i}, V_{i}, C_{i},\mathbf{Z}_{i}\}$
        \end{enumerate}
    \end{flushleft}
\end{algorithm}

\begin{figure}[H]
  \centering
    \includegraphics[width=1 \textwidth]{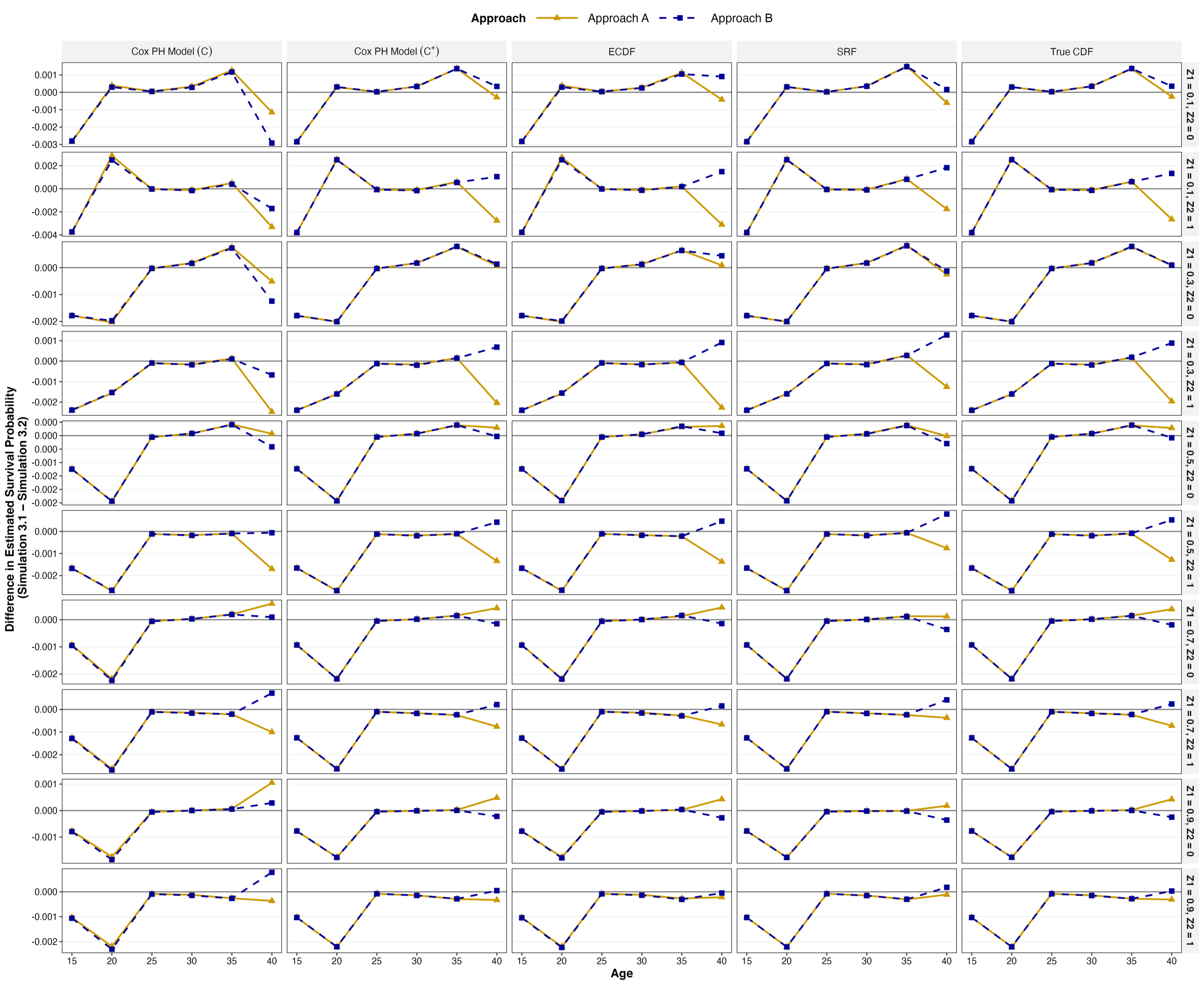}
     \caption{The difference between two average estimated survival probability using both approaches on two different data generation in simulation 3.}
     \label{fig: sim32_comapre}
\end{figure}

We subsequently employ the proposed data-generating algorithm to replicate the datasets from Simulation 3, computing estimated survival probabilities for all methods under both approaches. As illustrated in Figure \ref{fig: sim32_comapre}, the resulting differences in these estimates are smaller.
